\documentclass[aps,prr,reprint,superscriptaddress,twocolumn]{revtex4-2}
\usepackage[dvipdfmx]{graphicx}
\usepackage{graphicx}
\usepackage{color}
\usepackage[dvipsnames]{xcolor}
\definecolor{DarkGreen}{rgb}{0.0, 0.5, 0.0}
\usepackage[colorlinks,citecolor=DarkGreen,linkcolor=blue,linktocpage]{hyperref}

\usepackage[T1]{fontenc}
\usepackage{textcomp}
\usepackage{bm}
\usepackage{amsmath,amssymb,amsthm}
\usepackage{comment}
\usepackage{grffile}
\usepackage{cancel}
\usepackage{algpseudocode}

\newcommand{\cket}[1]{\left|#1\right\rangle}
\newcommand{\bra}[1]{\left\langle#1\right|}
\newcommand{\bracket}[2]{\left\langle#1|#2\right\rangle}

\begin{document}

%\preprint{}

\title{Systematic construction of asymptotic quantum many-body scar states and their relation to supersymmetric quantum mechanics}

\author{Masaya Kunimi}
\email{kunimi@rs.tus.ac.jp}
\affiliation{Department of Physics, Tokyo University of Science, 1-3 Kagurazaka, Tokyo 162-8601,  Japan}
\author{Yusuke Kato}
\email{yusuke@phys.c.u-tokyo.ac.jp}
\affiliation{Department of Basic Science, The University of Tokyo, 3-8-1 Komaba, Tokyo 153-8902, Japan}
\affiliation{Quantum Research Center for Chirality, Institute for Molecular Science, Okazaki, Aichi 444-8585, Japan}
\author{Hosho Katsura}
\email{katsura@phys.s.u-tokyo.ac.jp}
\affiliation{Department of Physics, The University of Tokyo, 7-3-1 Hongo, Tokyo 113-0033, Japan}
\affiliation{Institute for Physics of Intelligence, The University of Tokyo, 7-3-1 Hongo, Tokyo 113-0033, Japan}
\affiliation{Trans-Scale Quantum Science Institute, The University of Tokyo, 7-3-1 Hongo, Tokyo 113-0033, Japan}

%\thanks{}

\date{\today}

%%%%%%%%%%%%%%%%%%%%%%%%%%%%%%%%%%%%%%%%%%%%%%%%%%%%
\begin{abstract}
We develop a systematic method for constructing asymptotic quantum many-body scar (AQMBS) states. While AQMBS states are closely related to quantum many-body scar (QMBS) states, they exhibit key differences. Unlike QMBS states, AQMBS states are not energy eigenstates of the Hamiltonian, making their construction more challenging. We demonstrate that, under appropriate conditions, AQMBS states can be obtained as low-lying gapless excited states of a parent Hamiltonian, which has a QMBS state as its ground state. Furthermore, our formalism reveals a connection between QMBS and supersymmetric (SUSY) quantum mechanics. The QMBS state can be interpreted as a SUSY-unbroken ground state.
\end{abstract}
%%%%%%%%%%%%%%%%%%%%%%%%%%%%%%%%%%%%%%%%%%%%%%%%%%%%
\maketitle
%%%%%%%%%%%%%%%%%%%%%%%%%%%%%%%%%%%%%%%%%%%%%%%%%%%%%%%%%%%%%%
%%%%%%%%%%%%%%%%%%%%%%%%%%%%%%%%%
\section{Introduction}\label{sec:Introduction}
%%%%%%%%%%
%%%%%%%%%%%%%%%%%%%%%%%%%%%%%%%%%
Thermalization in isolated quantum many-body systems is a fundamental issue in statistical physics. A central concept in this context is the eigenstate thermalization hypothesis (ETH)~\cite{Deutsch1991,Srednicki1994,Rigol2008}. If a system satisfies the strong version of the ETH, the long-time average of the expectation value of a physical quantity is equal to the microcanonical ensemble average, indicating that the system thermalizes after a long-time evolution~\cite{DAlessio2016,Mori2018}. Although there is no rigorous mathematical proof of the ETH, extensive numerical studies suggest that generic nonintegrable systems obey the ETH. However, some nonintegrable systems are known to violate the strong version of the ETH. Such systems are referred to as nonergodic. Examples of nonergodic systems include many-body localized systems~\cite{Gornyi2005,Basko2007,Huse2014}, quantum many-body scarred systems~\cite{Bernien2017,Turner2018,Turner2018prb}, and Hilbert space fragmented systems~\cite{Sala2020,Khemani2020}. 

In this paper, we focus on quantum many-body scar (QMBS) states~\cite{Serbyn2021,Papic2022,Moudgalya2022quantum,Chandran2023quantum}, which are special energy eigenstates of nonintegrable Hamiltonians. QMBS states exhibit several distinctive properties: They are typically high-energy excited states, while their entanglement entropy (EE) follows a subvolume law scaling. The energy spectrum of QMBS states is (approximately) equally spaced, a feature referred to as tower of states~\cite{Serbyn2021,Papic2022,Moudgalya2022quantum,Chandran2023quantum}. Since the experimental discovery of QMBS states in Rydberg atom quantum simulators~\cite{Bernien2017}, they have been extensively studied both theoretically~\cite{Choi2019,Ho2019,Lin2019,Schecter2019,Shiraishi2019,Ok2019,Iadecola2020,Bull2020,Chattopadhyay2020,Lee2020,Alhambra2020,Lin2020,Mark2020eta,Mark2020unified,Moudgalya2020eta,Moudgalya2020large,Michailidis2020,ODea2020,Shibata2020,Pakrouski2020,Michailidis2020,Kuno2020,Mizuta2020,Hudomal2020,Surace2020,Sugiura2021,Pakrouski2021,Mondragon2021,chertkov2021motif,Ren2021quasisymmetry,Ren2022deformed,Langlett2022,Tamura2022,Desaules2022,Gotta2022,Tang2022,wildeboer2022quantum,Omiya2023,Omiya2023quantum,Desaules2023,Desaules2023prominent,Sanada2023,Iversen2023,Lalimeh2023,Kaneko2024,Wang2024,Kunimi2024,Matsui2024,teretenkov2024duality,Nakagawa2024,Moudgalya2024exhaustive,Moudgalya2024symmetries,Sanada2024a,Imai2025,Calajo2025} and experimentally~\cite{Bluvstein2021controlling,Zhang2023,Su2023,Zhao2025}.

Certain types of QMBS states can be characterized by algebraic structures known as restricted spectrum generating algebra (RSGA)~\cite{Mark2020unified,Moudgalya2020eta}. When a system satisfies the RSGA conditions, the existence of tower spectra in QMBS states and perfect revival behavior can be analytically demonstrated. The RSGA implies that symmetry emerges within a specific Hilbert subspace, referred to as the scar subspace $\mathcal{H}_{\rm scar}$. We note that several alternative formalisms have also been proposed to characterize QMBS states based on symmetry~\cite{ODea2020,Pakrouski2020,Pakrouski2021,Ren2021quasisymmetry,Ren2022deformed}.

Recently, a class of nonergodic quantum many-body states, dubbed asymptotic quantum many-body scar (AQMBS) states, has been proposed~\cite{Gotta2023}. Although AQMBS states are closely related to QMBS states, they exhibit distinct properties:
\begin{enumerate}
\item[(1)] Orthogonality to any exact QMBS state.
\item[(2)] Low entanglement.
\item[(3)] Vanishing energy variance in the thermodynamic limit.
\end{enumerate}
Among these, property (3) is the most significant. According to the Mandelstam-Tamm bound~\cite{Mandelstam1945,Gong2022}, the fidelity relaxation time is given by $T_{\rm relax}=\pi\hbar/(2\Delta E)$, where $\Delta E$ is the square root of the energy variance. This result implies that, in the thermodynamic limit, the fidelity relaxation time diverges when the AQMBS state is chosen as the initial condition. AQMBS states in other models and their properties have also been investigated~\cite{Kunimi2024,Ren2024,Moudgalya2024symmetries,wei2025spectroscopic}. In particular, Ren {\it et al}.~\cite{Ren2024} pointed out a connection between AQMBS states and Nambu-Goldstone modes. This finding suggests a direct relationship between thermalization in isolated quantum systems and spontaneous symmetry breaking.

Although AQMBS states exhibit intriguing properties, as mentioned above, a systematic method for constructing them remains unknown. All AQMBS states studied in previous works~\cite{Gotta2023,Kunimi2024,Ren2024} were found in a heuristic manner. Unlike QMBS states, AQMBS states are not energy eigenstates of the Hamiltonian, which makes their construction challenging. To address this issue, we propose a systematic method for constructing AQMBS states. The key idea of our formulation is to introduce a Hilbert subspace $\mathcal{H}_P$, which includes the scar subspace $\mathcal{H}_{\rm scar}$. AQMBS states can be obtained as low-lying excited eigenstates of the parent Hamiltonian of QMBS states, defined within the subspace $\mathcal{H}_P$. Our formulation encompasses a wide range of scarred models, including the spin-$1$ XY model~\cite{Schecter2019}, the Fermi-Hubbard models~\cite{Mark2020eta,Moudgalya2020eta}, the DH model~\cite{Kodama2023,Kunimi2024}, the domain-wall-conserving model~\cite{Iadecola2020}, Onsager scar model~\cite{Shibata2020,Tamura2022}, and the nonmaximal spin scar model~\cite{ODea2020}. Furthermore, based on our formalism, we find that QMBS is directly related to supersymmetric (SUSY) quantum mechanics~\cite{Witten1981,Witten1982,Fendley2003lattice,Fendley2003lattice_fermion}. Our results suggest that a QMBS state can be interpreted as a SUSY unbroken ground state.

This paper is organized as follows: In Sec.~\ref{sec:framework}, we present our formalism for constructing AQMBS states. In Sec.~\ref{sec:construction_of_AQMBS}, we demonstrate that our formulation applies to several models that exhibit QMBS states. In Sec.~\ref{sec:summary}, we summarize our results and discuss the implications of our formalism. Some technical details are relegated to the appendices.

%%%%%%%%%%%%%%%%%%%%%%%%%%%%%%%%%%%%%%%%%%%%%%%%%%%%%%%%%%%%%%
%%%%%%%%%%%%%%%%%%%%%%%%%%%%%%%%%
\section{General framework}\label{sec:framework}
%%%%%%%%%%%%%%%%%%%%%%%%%%%%%%%%%%%%%%%%%%%%%%%%%%%%%%%%%%%%%%
%%%%%%%%%%%%%%%%%%%%%%%%%%%%%%%%%

%%%%%%%%%%%%%%%%%%%%%%%%%%%%%%%%%%%%%%%%%%%%%%%%%%%%%%%%%%%%%%
\subsection{Restricted spectrum generating algebra}\label{subsec:RSGA}
%%%%%%%%%%%%%%%%%%%%%%%%%%%%%%%%%%%%%%%%%%%%%%%%%%%%%%%%%%%%%%
In our formalism, we assume that the system satisfies the RSGA \cite{Mark2020unified,Moudgalya2020eta}. For completeness, we recall the definition of an RSGA of order $m$ (RSGA-$m$). 

Let us consider a nonintegrable Hamiltonian $\hat{H}$ and an operator $\hat{Q}^{\dagger}$ such that $\hat{Q}^{\dagger}\cket{S_0}\not=0$ for some normalized state $\cket{S_0}$, i.e., $\bracket{S_0}{S_0}=1)$. We further assume that $\cket{S_0}$ cannot be written in the form $\hat{Q}^{\dagger}\cket{\psi}$ for any eigenstate $\cket{\psi}$ of $\hat{H}$. Under these assumptions, we say that the Hamiltonian $\hat{H}$ exhibits an RSGA-$m$ if it satisfies the following conditions:
\begin{align}
\hat{H}\cket{S_0}&=E_0\cket{S_0},\label{eq:RSGA-n_condition1}\\
[\hat{H}, \hat{Q}^{\dagger}]\cket{S_0}&=\mathcal{E}\hat{Q}^{\dagger}\cket{S_0},\label{eq:RSGA-n_condition2}\\
\underbrace{[[[\hat{H},\hat{Q}^{\dagger}],\hat{Q}^{\dagger}],\ldots],\ldots]}_{r\text{ times}}\cket{S_0}&=0,\quad \text{for } 2\le r \le m,\label{eq:RSGA-n_condition3}\\
\underbrace{[[[\hat{H},\hat{Q}^{\dagger}],\hat{Q}^{\dagger}],\ldots],\ldots]}_{m+1\text{ times}}&=0,\label{eq:RSGA-n_condition4}
\end{align}
where $E_0$ and $\mathcal{E}$ are real numbers. If the above conditions hold, we can obtain the following relations:
\begin{align}
\hat{H}|\tilde{S}_n\rangle&=(E_0+n\mathcal{E})|\tilde{S}_n\rangle,\label{eq:eigenstate_Sn}\\
|\tilde{S}_n\rangle&\equiv 
\begin{cases}
(\hat{Q}^{\dagger})^n\cket{S_0},\quad&\text{for }n\le N_{\rm max},\\
0, &\text{for }n>N_{\rm max},
\end{cases}
\label{eq:definition_of_un_normalized_Sn}
\end{align}
where $|\tilde{S}_n\rangle$ is an unnormalized QMBS state and $N_{\rm max}$ is a positive integer. The normalized QMBS state is defined by $\cket{S_n}\equiv |\tilde{S}_n\rangle/\sqrt{\langle\tilde{S}_n|\tilde{S}_n\rangle}$.

%%%%%%%%%%%%%%%%%%%%%%%%%%%%%%%%%%%%%%%%%%%%%%%%%%%%%%%%%%%%%%
\subsection{Hamiltonian}\label{subsec:Hamiltonian}
%%%%%%%%%%%%%%%%%%%%%%%%%%%%%%%%%%%%%%%%%%%%%%%%%%%%%%%%%%%%%%
In addition to assuming the RSGA, we impose a specific structure of the Hamiltonian based on the symmetry-based formalism~\cite{ODea2020}. The advantage of this formalism is that it makes the structure of the Hamiltonian explicit, which is otherwise unclear in the RSGA formalism alone. Let us consider the following form of the Hamiltonian: 
\begin{align}
\hat{H}=\hat{H}_{\rm A}+\hat{H}_{\rm SG}+\hat{H}_{\rm sym}.\label{eq:Hamiltonian_interest}
\end{align}
Here, $\hat{H}_{\rm A}$ annihilates all QMBS states: 
\begin{align}
\hat{H}_{\rm A}\cket{S_n}=0.\label{eq:definition_of_H_A}
\end{align}
The second term of the Hamiltonian, $\hat{H}_{\rm SG}$, is defined as a linear combination of $\hat{Q}^z$ and $\hat{C}$. Here, $\hat{Q}^z\equiv \sum_{j}\hat{q}_j^z$ is a generator of the onsite symmetry acting independently on each lattice site of the system, i.e., $\hat{q}^z_j$ is hermitian and ${\rm supp}(\hat{q}_j^z)=\{j\}$. We assume that $\hat{Q}^{\dagger}$ can be written as the sum of local operators: $\hat{Q}^{\dagger}=\sum_j \hat{q}_j^{\dagger}$, where each $\hat{q}_j^{\dagger}$ has support only in the vicinity of site $j$. Furthermore, $\hat{Q}^z$ and $\hat{Q}^{\dagger}$ satisfy the following commutation relation:
\begin{align}
[\hat{Q}^z, \hat{Q}^{\dagger}]=\hat{Q}^{\dagger}.\label{eq:commutation_relation_Qz_and_Qdagger}
\end{align}
We note that Eq.~(\ref{eq:commutation_relation_Qz_and_Qdagger}) is an operator identity in contrast to the RSGA. Typically, $\hat{Q}^{z}$ corresponds to the total magnetization operator in quantum spin systems or the total particle number operator in Hubbard-like systems. However, as pointed out in Sec.~III C of Ref.~\cite{ODea2020}, the operators $\hat{Q}^z$ and $\hat{Q}^{\dagger}$ are not necessarily associated with a Lie algebra. For example, the domain-wall-conserving model~\cite{Iadecola2020} and the DH model~\cite{Kunimi2024} do not exhibit any Lie algebra structure. The operator $\hat{C}$ satisfies the relations $[\hat{Q}^z, \hat{C}]=0$ and $[\hat{Q}^{\dagger}, \hat{C}]\not=0$. The third term of the Hamiltonian, $\hat{H}_{\rm sym}$, satisfies the commutation relations $[\hat{H}_{\rm sym}, \hat{Q}^{\dagger}]=0$ and $[\hat{H}_{\rm sym}, \hat{Q}^z]=0$. We note that setting $\hat{H}_{\rm sym}=0$ is allowed within the symmetry-based formalism.

Furthermore, we assume that $\cket{S_0}$ is an eigenstate of $\hat{Q}^z$: $\hat{Q}^z\cket{S_0}=Q^z_0\cket{S_0}$, where $Q^z_0$ is a real number. This assumption ensures the orthogonality of $\cket{S_n}$ even when $\mathcal{E}=0$.

%%%%%%%%%%%%%%%%%%%%%%%%%%%%%%%%%%%%%%%%%%%%%%%%%%%%%%%%%%%%%%
\subsection{Subspace}\label{subsec:Subspace}
%%%%%%%%%%%%%%%%%%%%%%%%%%%%%%%%%%%%%%%%%%%%%%%%%%%%%%%%%%%%%%
In this subsection, we introduce the subspace $\mathcal{H}_{P}$, which plays a crucial role in our formalism. 

First, we define the total Hilbert space $\mathcal{H}$, assuming that the dimension of $\mathcal{H}$ is finite. We also define the scar subspace $\mathcal{H}_{\rm scar}$, which is spanned by all QMBS states generated by $\hat{Q}^{\dagger}$, as
\begin{align}
\mathcal{H}_{\rm scar}\equiv {\rm Span}\{\cket{S_0}, \cket{S_1},\ldots, \cket{S_{N_{\rm max}}}\}.\label{eq:definition_of_scar_subspace}
\end{align}
The total Hilbert space can be expressed as a direct sum: $\mathcal{H}=\mathcal{H}_{\rm thermal}\oplus\mathcal{H}_{\rm scar}$, where $\mathcal{H}_{\rm thermal}$ denotes a complement of $\mathcal{H}_{\rm scar}$ and is referred to as the thermal subspace~\cite{Serbyn2021}. The dimension of the scar subspace is typically of the order of the system size, whereas the dimension of the thermal subspace scales exponentially with the system size. 

\begin{figure}[t]
\centering
\includegraphics[width=8.6cm,clip]{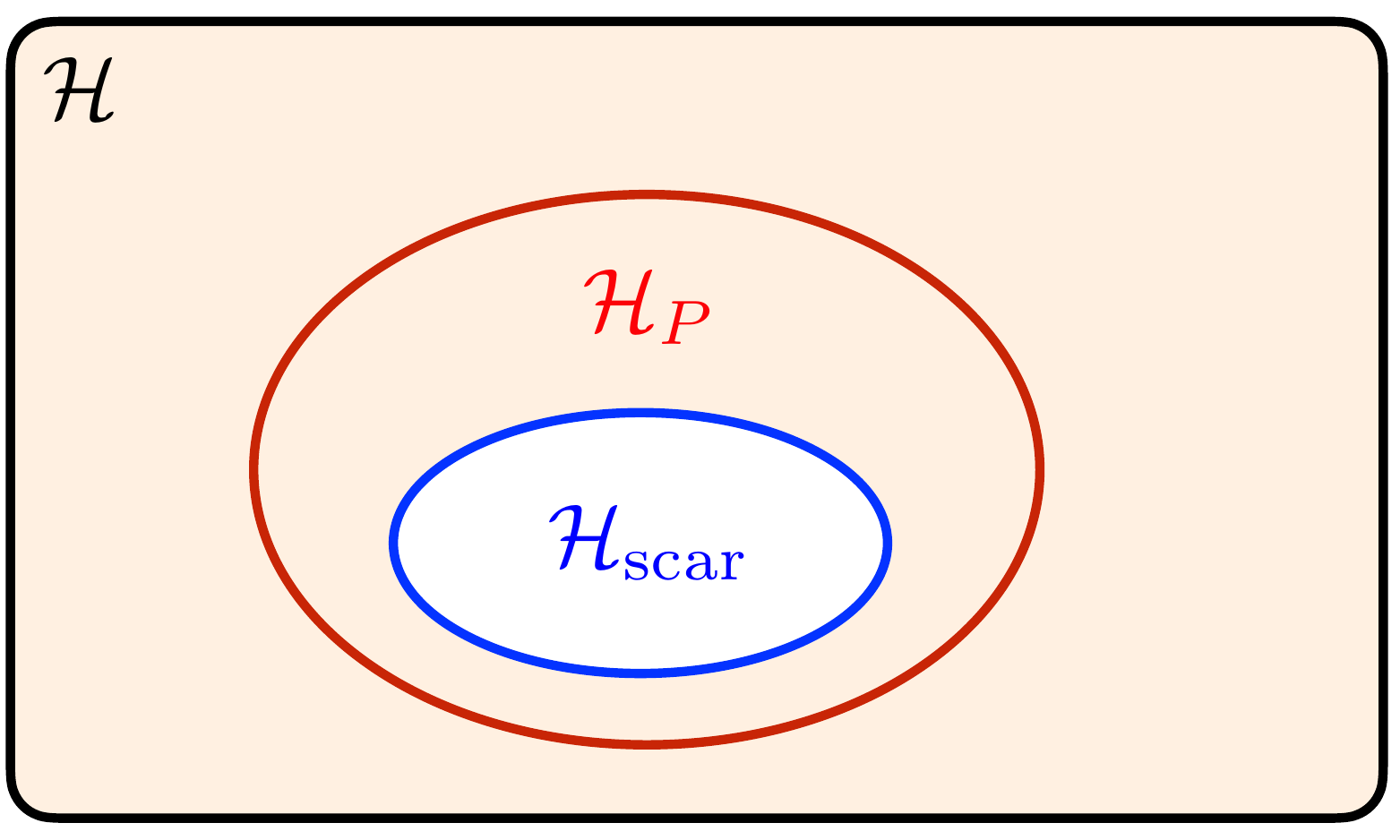}
\caption{Schematic of Hilbert subspaces introduced in this paper. The region surrounded by the red solid and the blue solid line represents $\mathcal{H}_P$ and $\mathcal{H}_{\rm scar}$, respectively. The colored region represents $\mathcal{H}_{\rm thermal}=\mathcal{H}_{\rm scar}^{\perp}$. The AQMBS states belong to $\mathcal{H}_P\setminus \mathcal{H}_{\rm scar}$.
}
\label{fig:schematic_Hilbert_space}
\vspace{-0.75em}
\end{figure}

Then, we introduce the Hilbert subspace $\mathcal{H}_P$, defined as
\begin{align}
\mathcal{H}_P&\equiv \bigoplus_{n=0}^{N_{\rm max}}\mathcal{H}_{P_n},\label{eq:definition_of_H_P}\\
\mathcal{H}_{P_n}&\equiv {\rm Span}\{\cket{\bm{m}}\in \mathcal{H} \;|\; \bracket{\bm{m}}{S_n}\not=0\},\label{eq:definition_of_H_Pn}
\end{align}
where $\cket{\bm{m}}$ is a product state and an eigenstate of $\hat{Q}^z$. Since $\hat{Q}^z$ is a generator of the onsite symmetry, we can choose a product basis that diagonalizes $\hat{Q}^z$. Given the assumed properties of $\hat{Q}^z$ and $\hat{Q}^{\dagger}$, all states in $\mathcal{H}_{P_n}$ are orthogonal to those in $\mathcal{H}_{P_m}$ for $n\not=m$, because the eigenvalues of $\hat{Q}^z$ for the states in $\mathcal{H}_{P_n}$ and $\mathcal{H}_{P_m}$ are $Q^z_0 + n$ and $Q^z_0 + m$, respectively. In general, the following relation holds: $\mathcal{H}_{\rm scar}\subset \mathcal{H}_P\subset \mathcal{H}$. This hierarchy results from the fact that QMBS states are formed as superpositions of states within a specific portion of the total Hilbert space. Additionally, we define the complement of $\mathcal{H}_P$ as $\mathcal{H}_Q$ and introduce the projection operators onto $\mathcal{H}_{P_n}$, $\mathcal{H}_P$, and $\mathcal{H}_Q$ as $\hat{\mathcal{P}}_n$, $\hat{\mathcal{P}}\equiv \sum_{n=0}^{N_{\rm max}}\hat{\mathcal{P}}_n$, and $\hat{\mathcal{Q}}\equiv \hat{1}-\hat{\mathcal{P}}$, respectively, where $\hat{1}$ is the unit operator in $\mathcal{H}$. We note that these projection operators are generally nonlocal. Figure~\ref{fig:schematic_Hilbert_space} schematically shows the relation between these subspaces.

We further assume that the term $\hat{H}_{\rm A}$ can be decomposed into two parts:
\begin{align}
\hat{H}_{\rm A}&\equiv \hat{H}_0+\hat{H}_{\rm p}',\label{eq:decomposition_of_H_A}\\
\hat{H}_0&\equiv \sum_j\hat{h}_j,\label{eq:definition_of_H0}\\
\hat{H}_{\rm p}'&\equiv \sum_{n=0}^{N_{\rm max}}c_n\hat{\mathcal{P}}_n\hat{H}_0^2\hat{\mathcal{P}}_n,\label{eq:definition_of_H_p'}
\end{align}
where $\hat{H}_0(\not=0)$ satisfies $\hat{H}_0\cket{S_n}=0$,  $\hat{h}_j$ is a local operator, which has support only in the vicinity of site $j$, satisfying $\hat{\mathcal{P}}\hat{h}_j\hat{\mathcal{P}}=0$ for all $j$, and $c_n$ is a real constant with the dimension of the inverse of the energy (note that setting all $c_n=0$ is allowed).

As we will discuss in Sec.~\ref{subsec:energy_variance}, the term
\begin{align}
\hat{H}_{\rm p}\equiv \hat{\mathcal{P}}\hat{H}_0^2\hat{\mathcal{P}},\label{eq:definition_of_parent_hamiltonian}
\end{align}
is a parent Hamiltonian of the QMBS state $\{\cket{S_n}\}$, which means that the QMBS state $\{\cket{S_n}\}$ is a ground state of $\hat{\mathcal{P}}\hat{H}_0^2\hat{\mathcal{P}}$ as we show below. By definition, the operator $\hat{\mathcal{P}}\hat{H}_0^2\hat{\mathcal{P}}$ is positive semidefinite ($\hat{\mathcal{P}}\hat{H}_0^2\hat{\mathcal{P}}\ge 0$). It then suffices to show that $\hat{H}_0\cket{S_n}=0$. From the relation $\hat{H}_{\rm A}\cket{S_n}=\hat{H}_0\cket{S_n}+\hat{H}_{\rm p}' \cket{S_n}=0$, we see that the first term $\hat{H}_0 \cket{S_n}$ belongs to $\mathcal{H}_Q$ (since $\hat{\mathcal{P}}\hat{h}_j\hat{\mathcal{P}}=0$), while the second term $\hat{H}_{\rm p}' \cket{S_n} =c_n \hat{\mathcal{P}}_n\hat{H}_0^2\hat{\mathcal{P}}_n\cket{S_n}$ belongs to $\mathcal{H}_{P_n}$. Thus, $\hat{H}_0 \cket{S_n}$ and $\hat{H}_{\rm p}' \cket{S_n}$ are orthogonal. Since their sum is zero, it follows that $\hat{H}_0\cket{S_n}=0$.

Here, we remark on the locality of the parent Hamiltonian. The parent Hamiltonian seems to be nonlocal due to the term $\hat{H}_0^2$. In Appendix~\ref{app:remark_locality}, we show that the parent Hamiltonian is written as a sum of local operators if $\mathcal{H}_P$ has a tensor product structure.

Finally, we add an assumption on the off-diagonal elements of the parent Hamiltonian. From Eq.~(\ref{eq:definition_of_parent_hamiltonian}), the parent Hamiltonian reduces to
\begin{align}
\hat{H}_{\rm p}=\sum_{n,m}\hat{\mathcal{P}}_n\hat{H}_0^2\hat{\mathcal{P}}_m.\label{eq:parent_Hamiltonian_general_form}
\end{align}
We assume that the off-diagonal elements $\hat{\mathcal{P}}_n\hat{H}_0^2\hat{\mathcal{P}}_m\;(n\not=m)$ vanish. This assumption holds when $\hat{H}_0$ and $\hat{Q}^z$ commute. Even if they do not commute, the off-diagonal elements can still vanish due to other algebraic relations. This will be discussed in Appendix~\ref{app:remark_off-diagonal_element_parent_Hamiltonian}. Under this assumption, the parent Hamiltonian becomes
\begin{align}
\hat{H}_{\rm p}=\sum_n\hat{\mathcal{P}}_n\hat{H}_0^2\hat{\mathcal{P}}_n\equiv \sum_n\hat{H}^{(n)}_{\rm p}.\label{eq:definition_of_H_p^n}
\end{align}

%%%%%%%%%%%%%%%%%%%%%%%%%%%%%%%%%%%%%%%%%%%%%%%%%%%%%%%%%%%%%%
\subsection{Energy variance}\label{subsec:energy_variance}
%%%%%%%%%%%%%%%%%%%%%%%%%%%%%%%%%%%%%%%%%%%%%%%%%%%%%%%%%%%%%%
Here, we discuss the energy variance, which characterizes the AQMBS state. Let us assume that $\cket{\varphi_n}$ is a normalized state in $\mathcal{H}_{P_n}$. The energy variance of $\cket{\varphi_n}$ is defined as
\begin{align}
\Delta E^2_n&\equiv \bra{\varphi_n}\hat{H}^2\cket{\varphi_n}-(\bra{\varphi_n}\hat{H}\cket{\varphi_n})^2\notag \\
&=\bra{\varphi_n}\hat{\mathcal{P}}_n\hat{H}^2\hat{\mathcal{P}}_n\cket{\varphi_n}-(\bra{\varphi_n}\hat{\mathcal{P}}_n\hat{H}\hat{\mathcal{P}}_n\cket{\varphi_n})^2,\label{eq:definition_of_energy_variance}
\end{align}
where we used $\hat{\mathcal{P}}_n\cket{\varphi_n}=\cket{\varphi_n}$. Using the relations $\hat{\mathcal{P}}_n\hat{H}_0\hat{\mathcal{P}}_n=0$~\footnote{This relation can be obtained by the assumption $\hat{\mathcal{P}}\hat{h}_j\hat{\mathcal{P}}=0$. From the definition of $\hat{\mathcal{P}}$, we have $\hat{\mathcal{P}}=\sum_{\cket{\bm{n}}\in\mathcal{H}_P}\cket{\bm{n}}\bra{\bm{n}}$, where $\cket{\bm{n}}$ is a direct product state. Then, we obtain $\hat{\mathcal{P}}\hat{h}_j\hat{\mathcal{P}}=\sum_{\cket{\bm{n}},\cket{\bm{m}}\in\mathcal{H}_P}\bra{\bm{n}}\hat{h}_j\cket{\bm{m}}\cket{\bm{n}}\bra{\bm{m}}=0$. For this equation to hold, it is necessary that $\bra{\bm{n}}\hat{h}_j\cket{\bm{m}}=0$ for any $\cket{\bm{n}}$ and $\cket{\bm{m}}\in\mathcal{H}_P$. As a special case, we obtain $\hat{\mathcal{P}}_n\hat{h}_j\hat{\mathcal{P}}_n=0$.}, $\hat{\mathcal{P}}_n(\hat{H}_{\rm p}'+\hat{H}_{\rm SG}+\hat{H}_{\rm sym})\hat{\mathcal{Q}}=0$, and $\hat{\mathcal{Q}}(\hat{H}_{\rm p}'+\hat{H}_{\rm SG}+\hat{H}_{\rm sym})\hat{\mathcal{P}}_n=0$ \footnote{The relation $\hat{\mathcal{P}}_n(\hat{H}_{\rm SG}+\hat{H}_{\rm sym})\hat{\mathcal{Q}}=0$ can be proven as follows: Since $\hat{Q}^z$ is the generator of the onsite symmetry, we can show that $\hat{Q}^z\cket{\bm{m}}=Q^z_n\cket{\bm{m}}$ holds for any product state $\cket{\bm{m}}$ in $\mathcal{H}_{P_n}$, where $Q^z_n$ is an eigenvalue of $\hat{Q}^z$ in $\mathcal{H}_{P_n}$. Furthermore, we also obtain $\hat{H}_{\rm sym}\cket{\bm{m}}=\mathcal{E}_0\cket{\bm{m}}$, where $\mathcal{E}_0$ is a real number. From these relations, $\hat{Q}^z$ and $\hat{H}_{\rm sym}$ commute with $\hat{\mathcal{P}}_n$. Therefore, we obtain $\hat{\mathcal{P}}_n(\hat{H}_{\rm SG}+\hat{H}_{\rm sym})\hat{\mathcal{Q}}=(\hat{H}_{\rm SG}+\hat{H}_{\rm sym})\hat{\mathcal{P}}_n\hat{\mathcal{Q}}=0$.}, we obtain{\small 
\begin{align}
\Delta E^2_n&=\bra{\varphi_n}\hat{\mathcal{P}}_n\hat{H}^2_0\hat{\mathcal{P}}_n\cket{\varphi_n}\notag \\
&+c_n^2\left[\bra{\varphi_n}(\hat{\mathcal{P}}_n\hat{H}_0^2\hat{\mathcal{P}}_n)^2\cket{\varphi_n}-(\bra{\varphi_n}\hat{\mathcal{P}}_n\hat{H}_0^2\hat{\mathcal{P}}_n\cket{\varphi_n})^2\right].\label{eq:energy_variance_calculation}
\end{align}
}If we choose $\cket{\varphi_n}$ as a normalized eigenstate of $\hat{\mathcal{P}}_n\hat{H}_0^2\hat{\mathcal{P}}_n$, the second and third terms in Eq.~(\ref{eq:energy_variance_calculation}) vanish. Therefore, the energy variance reduces to
\begin{align}
\Delta E^2_n&=\bra{AS_n}\hat{\mathcal{P}}_n\hat{H}^2_0\hat{\mathcal{P}}_n\cket{AS_n},\label{eq:energy_variance_final_form}
\end{align}
where $\cket{AS_n}$ is a normalized eigenstate of $\hat{\mathcal{P}}_n\hat{H}_0^2\hat{\mathcal{P}}_n$.

Here, we regard $\hat{H}_{\rm p}^{(n)}=\hat{\mathcal{P}}_n\hat{H}_0^2\hat{\mathcal{P}}_n$ as a Hamiltonian of a system defined in $\mathcal{H}_{P_n}$. As discussed in Sec.~\ref{subsec:Subspace}, the QMBS state $\cket{S_n}$ is a ground state of $\hat{H}_{\rm p}^{(n)}$. If $\hat{H}_{\rm p}^{(n)}$ has gapless excited states, we can systematically construct the AQMBS states by obtaining the gapless excited state of $\hat{H}_{\rm p}^{(n)}$.

We discuss how the state $\cket{AS_n}$ constructed in this manner satisfies the conditions of AQMBS (1), (2), and (3) mentioned in Sec.~\ref{sec:Introduction}. Condition (1) is automatically satisfied because $\cket{S_n}$ and $\cket{AS_m}$ are eigenstates of $\hat{H}_{\rm p}$ with different eigenenergies. Therefore, $\bracket{S_n}{AS_m}=0$ always holds. For condition (2), we expect that the entanglement entropy of $\cket{AS_n}$ does not obey the volume-law scaling. The reasons are as follows: First, the QMBS state $\cket{S_n}$ typically obeys subvolume law scaling, except in the rainbow scar~\cite{Langlett2022,wildeboer2022quantum}. Second, low-energy gapless excited states of a local many-body Hamiltonian can be well described by the single-mode approximation in many cases; i.e., an excited state is given by $\hat{E}\cket{S_n}$, where $\hat{E}$ is an operator that creates the excitation. According to Ref.~\cite{moudgalya2018entanglement}, the bond dimension of the matrix product representation of the operator $\hat{E}$ scales polynomially with the system size. Combining these facts, we expect that the state $\cket{AS_n}$ obeys subvolume law scaling. Condition (3) is also satisfied due to the assumption that $\cket{AS_n}$ is a gapless excited state. From these results, we conclude that $\cket{AS_n}$ can be regarded as an AQMBS state. 

We note that the \emph{optimal} AQMBS can be defined from Eq.~(\ref{eq:energy_variance_final_form}) based on the variational principle, where \emph{optimal} means that the energy variance is minimized. As pointed out in Ref.~\cite{Kunimi2024}, the AQMBS states are not unique; several states can satisfy conditions (1), (2), and (3). By applying the variational principle, the AQMBS state can be adjusted to achieve a lower energy variance when we cannot obtain the exact eigenstate of $\hat{H}_{\rm p}$. This aspect will be discussed further in Secs.~\ref{subsec:spin-1_XY} and \ref{subsec:DH_model}.

At the end of this subsection, we comment on the special case of a one-dimensional system with periodic boundary conditions. In our formalism, it is essential that $\hat{H}_{\rm p}$ is gapless. In general, determining whether a given many-body Hamiltonian is gapless is a difficult problem. However, for a translationally invariant local Hamiltonian in one dimension, the Lieb-Schultz-Mattis theorem~\cite{lieb1961two} can be applied and, under certain conditions, imposes strong constraints on the nature of low-energy excitations~\cite{tasaki2020physics}.

%%%%%%%%%%%%%%%%%%%%%%%%%%%%%%%%%%%%%%%%%%%%%%%%%%%%%%%%%%%%%%
\subsection{Relation to supersymmetric quantum mechanics}\label{subsec:relation_SUSY}
%%%%%%%%%%%%%%%%%%%%%%%%%%%%%%%%%%%%%%%%%%%%%%%%%%%%%%%%%%%%%%
Here, we show that our formalism, discussed so far, is directly related to SUSY quantum mechanics. For completeness, we first summarize SUSY quantum mechanics \cite{Witten1981,Witten1982,Fendley2003lattice,Fendley2003lattice_fermion} before showing our results.

In $\mathcal{N}=2$ SUSY quantum mechanics, we consider two supercharge operators $\hat{\tilde{Q}}$ and $\hat{\tilde{Q}}^{\dagger}$, which satisfy the conditions $(\hat{\tilde{Q}})^2=0$ and $(\hat{\tilde{Q}}^{\dagger})^2=0$. The SUSY Hamiltonian is defined as the anticommutator of these operators:
\begin{align}
\hat{H}_{\rm SUSY}\equiv \{\hat{\tilde{Q}}, \hat{\tilde{Q}}^{\dagger}\}.\label{eq:definition_of_SUSY_Hamiltonian}
\end{align}
By definition, the supercharges commute with the SUSY Hamiltonian:
\begin{align}
[\hat{H}_{\rm SUSY}, \hat{\tilde{Q}}]=0,\quad [\hat{H}_{\rm SUSY}, \hat{\tilde{Q}}^{\dagger}]=0.\label{eq:commutation_relation_SUSY_Hamiltonian_and_supercharge}
\end{align}
We also introduce the total fermion number operator $\hat{F}$. In $\mathcal{N}=2$ SUSY quantum mechanics, the following relations are assumed:
\begin{align}
\{(-1)^{\hat{F}}, \hat{\tilde{Q}}\}=0,\quad \{(-1)^{\hat{F}}, \hat{\tilde{Q}}^{\dagger}\}=0.\label{eq:anticommutator_fermion_parity_and_supercharge}
\end{align}
Equations~(\ref{eq:definition_of_SUSY_Hamiltonian}) and (\ref{eq:anticommutator_fermion_parity_and_supercharge}) lead to $\mathbb{Z}_2$ symmetry:
\begin{align}
[\hat{H}_{\rm SUSY}, (-1)^{\hat{F}}]=0.\label{eq:Z2_symmetry_SUSY_quantum_mechanics}
\end{align}

Then, we discuss several important properties that follow from these algebraic relations~\cite{Fendley2003lattice}. By definition, we can show that $\hat{H}_{\rm SUSY}$ is positive semidefinite. Let $\cket{\psi}$ be an eigenstate of $\hat{H}_{\rm SUSY}$ with eigenvalue $E$. It follows that
\begin{align}
E=\bra{\psi}(\hat{\tilde{Q}}\hat{\tilde{Q}}^{\dagger}+\hat{\tilde{Q}}^{\dagger}\hat{\tilde{Q}})\cket{\psi}=\|\hat{\tilde{Q}}\cket{\psi} \|^2+\|\hat{\tilde{Q}}^{\dagger}\cket{\psi} \|^2\ge0,\label{eq:proof_positive_semidefinite_ness_SUSY}
\end{align}
where $\|\cdots\|$ denotes the norm $\sqrt{\bracket{\cdots}{\cdots}}$. For any positive eigenenergy $E>0$, there exists a corresponding eigenstate $\cket{\psi}$ such that $\hat{\tilde{Q}}\cket{\psi}=0$.
From the properties of the supercharge, it follows that $\hat{\tilde{Q}}^{\dagger}\cket{\psi}$ is also an eigenstate of $\hat{H}_{\rm SUSY}$ with eigenvalue $E$. This implies that eigenstates with positive eigenvalues are always at least twofold degenerate~\footnote{We prove that every positive energy level is at least twofold degenerate. Let $|\psi\rangle$ be a simultaneous eigenstate of $\hat {H}_{\rm SUSY}$ and $(-1)^{\hat{F}}$ with eigenvalues $E>0$ and $p=\pm 1$, respectively. If $\hat{{\tilde Q}}|\psi\rangle=0$, then $|\phi\rangle = \hat{{\tilde Q}}^\dagger |\psi\rangle$ is an eigenstate of $\hat{H}_{\rm SUSY}$ with the same eigenvalue $E$. This can be seen as follows: We first note that $|\phi\rangle \ne 0$, since otherwise we would have $\hat{{\tilde Q}}|\psi\rangle=\hat{{\tilde Q}}^\dagger|\psi\rangle=0$, which implies $E=0$, contradicting our assumption that $E>0$. We then see that $|\psi\rangle$ and $|\phi\rangle$ must be orthogonal, since they have opposite fermionic parity. Thus, it follows from Eq. \eqref{eq:commutation_relation_SUSY_Hamiltonian_and_supercharge} that $|\psi\rangle$ and $|\phi\rangle$ form a degenerate pair. If, on the other hand, $\hat{{\tilde Q}}|\psi\rangle \ne 0$, then we define $|\chi\rangle = \hat{{\tilde Q}} |\psi\rangle$, which is an eigenstate of $\hat{H}_{\rm SUSY}$ with eigenvalue $E$. By the same reasoning as before, $|\chi\rangle$ must be nonzero and orthogonal to $|\psi\rangle$. Thus, in both cases, we obtain the desired result.}. For eigenstates with $E=0$ ($\cket{\psi_0}$), we obtain $\hat{\tilde{Q}}\cket{\psi_0}=0$ and $\hat{\tilde{Q}}^{\dagger}\cket{\psi_0}=0$ because of Eq.~(\ref{eq:proof_positive_semidefinite_ness_SUSY}). If a zero-energy eigenstate of $\hat{H}_{\rm SUSY}$ exists, SUSY remains unbroken. 

Here, we demonstrate the connection between our formalism for AQMBS and SUSY quantum mechanics. We adopt the same assumptions as in Secs.~\ref{subsec:RSGA}, \ref{subsec:Hamiltonian}, \ref{subsec:Subspace}, and \ref{subsec:energy_variance}. The matrix representation of $\hat{H}_0$ is given by
\begin{align}
\hat{H}_0\to 
\begin{array}{l}
\vspace{0.3em}\hspace{0.7em}\mathcal{H}_P \hspace{0.2em}\mathcal{H}_{Q_1}\hspace{0.1em}\mathcal{H}_{Q_2} \\
\left[
\begin{array}{ccc}
0 & A  & 0 \\ 
A^{\dagger} & C & B\\
0 & B^{\dagger} & D
\end{array}
\right]
\begin{array}{l}
\} \mathcal{H}_P \\
\} \mathcal{H}_{Q_1} \\
\} \mathcal{H}_{Q_2}
\end{array}
\end{array}
,\label{eq:matrix_representation_of_H0}
\end{align}
where $A$, $B$, $C$, $D$ are nonzero matrices with dimensions ${\rm dim}\mathcal{H}_P\times {\rm dim}\mathcal{H}_{Q_1}$, ${\rm dim}\mathcal{H}_{Q_1}\times {\rm dim}\mathcal{H}_{Q_2}$, ${\rm dim}\mathcal{H}_{Q_1}\times {\rm dim}\mathcal{H}_{Q_1}$, and ${\rm dim}\mathcal{H}_{Q_2}\times{\rm dim}\mathcal{H}_{Q_2}$, respectively. Here, the subspace $\mathcal{H}_Q$ is decomposed as a direct sum: $\mathcal{H}_Q=\mathcal{H}_{Q_1}\oplus\mathcal{H}_{Q_2}$. The subspace $\mathcal{H}_{Q_1}$ is defined by
\begin{align}
\mathcal{H}_{Q_1}&={\rm Span}\{\hat{H}_0\cket{\bm{m}}\not=0 \;|\;\cket{\bm{m}}\in\mathcal{H}_P\},\label{eq:definition_of_subspace_Q1}
\end{align}
where $\cket{\bm{m}}$ is a direct product state in $\mathcal{H}_{P}$. The subspace $\mathcal{H}_{Q_2}$ is defined by the complementary subspace of $\mathcal{H}_{Q_1}$ in $\mathcal{H}_Q$. The matrix representation of the projection operator onto $\mathcal{H}_P$ is given by
\begin{align}
\hat{\mathcal{P}}\to 
\begin{array}{l}
\vspace{0.3em}\hspace{0.2em}{\text{{\footnotesize$\mathcal{H}_P$}}} \hspace{0.1em}{\text{{\footnotesize$\mathcal{H}_{Q_1}$}}}\hspace{0.0em}{\text{{\footnotesize$\mathcal{H}_{Q_2}$}}} \\
\left[
\begin{array}{ccc}
I & 0  & 0 \\ 
0 & 0 & 0 \\
0 & 0 & 0
\end{array}
\right]
\begin{array}{l}
\} \mathcal{H}_P \\
\} \mathcal{H}_{Q_1} \\
\} \mathcal{H}_{Q_2}
\end{array}
\end{array}
,\label{eq:matrix_representation_of_P}
\end{align}
where $I$ represents the unit matrix. From Eqs.~(\ref{eq:matrix_representation_of_H0}) and (\ref{eq:matrix_representation_of_P}), we define $\hat{\tilde{Q}}$ and $\hat{\tilde{Q}}^{\dagger}$ as
\begin{align}
\hat{\tilde{Q}}&\equiv \hat{\mathcal{P}}\hat{H}_0\to 
\begin{array}{l}
\vspace{0.3em}\hspace{0.2em}{\text{{\footnotesize$\mathcal{H}_P$}}} \hspace{0.1em}{\text{{\footnotesize$\mathcal{H}_{Q_1}$}}}\hspace{0.0em}{\text{{\footnotesize$\mathcal{H}_{Q_2}$}}} \\
\left[
\begin{array}{ccc}
0 & A  & 0 \\ 
0 & 0 & 0 \\
0 & 0 & 0
\end{array}
\right]
\begin{array}{l}
\} \mathcal{H}_P \\
\} \mathcal{H}_{Q_1} \\
\} \mathcal{H}_{Q_2}
\end{array}
\end{array}
,\label{eq:matrix_representation_of_PH}\\
\hat{\tilde{Q}}^{\dagger}&\equiv\hat{H}_0\hat{\mathcal{P}}\to 
\begin{array}{l}
\vspace{0.3em}\hspace{0.2em}{\text{{\footnotesize$\mathcal{H}_P$}}} \hspace{0.1em}{\text{{\footnotesize$\mathcal{H}_{Q_1}$}}}\hspace{0.0em}{\text{{\footnotesize$\mathcal{H}_{Q_2}$}}} \\
\left[
\begin{array}{ccc}
0 & 0  & 0 \\ 
A^{\dagger} & 0 & 0 \\
0 & 0 & 0
\end{array}
\right]
\begin{array}{l}
\} \mathcal{H}_P \\
\} \mathcal{H}_{Q_1} \\
\} \mathcal{H}_{Q_2}
\end{array}
\end{array}
.\label{eq:matrix_representation_of_HP}
\end{align}
The operators $\hat{\tilde{Q}}$ and $\hat{\tilde{Q}}^{\dagger}$ satisfy the conditions $(\hat{\tilde{Q}})^2=0$ and $(\hat{\tilde{Q}}^{\dagger})^2=0$ due to the assumption $\hat{\mathcal{P}}\hat{H}_0\hat{\mathcal{P}}=0$. We can construct the SUSY Hamiltonian as
\begin{align}
\hat{H}_{\rm SUSY}&\equiv \{\hat{\tilde{Q}}, \hat{\tilde{Q}}^{\dagger}\} \to
\begin{array}{l}
\vspace{0.3em}\hspace{1em}\mathcal{H}_P \hspace{1.0em}\mathcal{H}_{Q_1}\hspace{0.1em}\mathcal{H}_{Q_2} \\
\left[
\begin{array}{ccc}
AA^{\dagger} & 0  & 0 \\ 
0 & A^{\dagger}A & 0 \\
0 & 0 & 0
\end{array}
\right]
\begin{array}{l}
\} \mathcal{H}_P \\
\} \mathcal{H}_{Q_1} \\
\} \mathcal{H}_{Q_2}
\end{array}
\end{array}
.\label{eq:matrix_representation_SUSY_Hamiltonian}
\end{align}
We introduce the fermionic parity operator as
\begin{align}
(-1)^{\hat{F}}&\to
\begin{array}{l}
\vspace{0.3em}\hspace{0.2em}{\text{{\footnotesize$\mathcal{H}_P$}}} \hspace{0.1em}{\text{{\footnotesize$\mathcal{H}_{Q_1}$}}}\hspace{0.0em}{\text{{\footnotesize$\mathcal{H}_{Q_2}$}}} \\
\left[
\begin{array}{ccc}
I & 0  & 0 \\ 
0 & -I & 0 \\
0 & 0 & 0
\end{array}
\right]
\begin{array}{l}
\} \mathcal{H}_P \\
\} \mathcal{H}_{Q_1} \\
\} \mathcal{H}_{Q_2}
\end{array}
\end{array}
.\label{eq:fermionc_parity_operator_for_SUSY}
\end{align}
We confirm that the system exhibits $\mathbb{Z}_2$ symmetry. Therefore, SUSY quantum mechanics naturally emerges in the Hilbert subspace $\mathcal{H}_P\oplus\mathcal{H}_{Q_1}$ within our formalism.

Next, we discuss the ground state of the SUSY Hamiltonian (\ref{eq:matrix_representation_SUSY_Hamiltonian}), which can be written as
\begin{align}
\hat{H}_{\rm SUSY}&=\hat{\tilde{Q}}\hat{\tilde{Q}}^{\dagger}+\hat{\tilde{Q}}^{\dagger}\hat{\tilde{Q}}\notag \\
&=\hat{\mathcal{P}}\hat{H}_0^2\hat{\mathcal{P}}+\hat{H}_0\hat{\mathcal{P}}\hat{H}_0.\label{eq:SUSY_Hamiltonian_explict_form}
\end{align}
The first term in Eq.~(\ref{eq:SUSY_Hamiltonian_explict_form}) is nothing but the parent Hamiltonian introduced in Sec.~\ref{subsec:energy_variance}, which is a local Hamiltonian under some conditions and acts on the subspace $\mathcal{H}_P$ as shown in Appendix~\ref{app:remark_locality}. This term corresponds to $AA^{\dagger}$ in Eq.~(\ref{eq:matrix_representation_SUSY_Hamiltonian}). In contrast, the second term in Eq.~(\ref{eq:SUSY_Hamiltonian_explict_form}), which acts on the subspace $\mathcal{H}_{Q_1}$ and corresponds to $A^{\dagger}A$ in Eq.~(\ref{eq:matrix_representation_SUSY_Hamiltonian}), generically includes nonlocal terms. While these terms have different locality, the spectral structure is exactly the same as shown in this section. Using the relation $\hat{H}_0\cket{S_n}=0$, we establish that the QMBS state $\cket{S_n}$ is a zero-energy eigenstate of the SUSY Hamiltonian. Consequently, the QMBS state $\cket{S_n}$ can be interpreted as a SUSY unbroken ground state.

%%%%%%%%%%%%%%%%%%%%%%%%%%%%%%%%%%%%%%%%%%%%%%%%%%%%%%%%%%%%%%
\subsection{Summary of our formalism}\label{subsec:summary_of_formalism}
%%%%%%%%%%%%%%%%%%%%%%%%%%%%%%%%%%%%%%%%%%%%%%%%%%%%%%%%%%%%%%

Here, we summarize our formalism for the systematic construction of AQMBS states. Our assumptions are as follows: 
\begin{enumerate}
\item[(1)] The system satisfies the RSGA.
\item[(2)] The Hamiltonian can be written as $\hat{H}=\hat{H}_{\rm A}+\hat{H}_{\rm SG}+\hat{H}_{\rm sym}$.
\item[(3)]The annihilation operator $\hat{H}_{\rm A}$ can be written as $\hat{H}_{\rm A}=\hat{H}_0+\hat{H}_{\rm p}'$. 
\item[(4)]$\hat{H}_0=\sum_j\hat{h}_j$ is a local Hamiltonian and satisfies $\hat{\mathcal{P}}\hat{h}_j\hat{\mathcal{P}}=0$.
\item[(5)]  The off-diagonal elements of the parent Hamiltonian are zero: $\hat{\mathcal{P}}_n\hat{H}_0^2\hat{\mathcal{P}}_m=0\;(n\not=m).$
\end{enumerate}
Under these assumptions, we can obtain the AQMBS states as low-energy gapless excited states of the parent Hamiltonian $\hat{H}_{\rm p}=\hat{\mathcal{P}}\hat{H}_0^2\hat{\mathcal{P}}$.

%%%%%%%%%%%%%%%%%%%%%%%%%%%%%%%%%%%%%%%%%%%%%%%%%%%%
%%%%%%%%%%%%%%%%%%%%%%%%%%%%%%%%%%%%%%%%%%%%%%%%%%%%
%%%%%%%%%%%%%%%%%%%%%%%%%%%%%%%%%%%%%%%%%%%%%%%%%%%%
\section{Construction of asymptotic quantum many-body scar states}\label{sec:construction_of_AQMBS}
%%%%%%%%%%%%%%%%%%%%%%%%%%%%%%%%%%%%%%%%%%%%%%%%%%%%
%%%%%%%%%%%%%%%%%%%%%%%%%%%%%%%%%%%%%%%%%%%%%%%%%%%%
%%%%%%%%%%%%%%%%%%%%%%%%%%%%%%%%%%%%%%%%%%%%%%%%%%%%
In this section, we apply our formulation of the AQMBS states to various systems. Throughout this section, we consider one-dimensional open chains with a total number of lattice sites $M$. For simplicity, we assume that $M$ is even. Although we present results in these specific situations, we emphasize that our formulation does not depend on dimensionality or boundary conditions, as long as the assumptions discussed in the previous section are satisfied. We note that the notation of the parameters in the Hamiltonian is independent across the following subsections, as we consider various models.

In our formalism, the parent Hamiltonian plays a key role in constructing the AQMBS states, as discussed in the previous section. In all models considered in this paper, the parent Hamiltonian is closely related to the spin-$1/2$ ferromagnetic Heisenberg model, likely due to the SU(2)-like structure of the $\hat{Q}^\dagger$ operator assumed in this work.

%%%%%%%%%%%%%%%%%%%%%%%%%%%%%%%%%%%%%%%%%%%%%%%%%%%%%%%%%%%%%%
\subsection{Spin-1 XY model}\label{subsec:spin-1_XY}
%%%%%%%%%%%%%%%%%%%%%%%%%%%%%%%%%%%%%%%%%%%%%%%%%%%%%%%%%%%%%%
Here, we consider the spin-1 XY model. The Hamiltonian is defined by
\begin{align}
\hat{H}&=\hat{H}_{XY}^{S=1}+h\sum_{j=1}^M\hat{\tau}_j^z+D\sum_{j=1}^M(\hat{\tau}_j^z)^2,\label{eq:definition_of_S=1_XY_Hamiltonian}\\
\hat{H}_{XY}^{S=1}&\equiv \sum_{k=1,3,\ldots, M-1}J_k\sum_{j=1}^{M-k}(\hat{\tau}_j^x\hat{\tau}_{j+k}^x+\hat{\tau}_j^y\hat{\tau}_{j+k}^y),\label{eq:definition_of_XY_interaction_S=1}
\end{align}
where $\hat{\tau}_j^{\mu}\;(\mu=x,y,z)$ is the spin-1 operator at site $j$, $J_k$ is the XY interaction strength of $k$th neighbor, $h$ is the strength of the linear Zeeman term, and $D$ is the strength of the quadratic Zeeman term. We note that the $k\ge 3$-th neighbor interaction terms break the hidden SU(2) symmetry for the spin-1 XY model with the nearest-neighbor interaction with open boundary conditions~\cite{Kitazawa2003}.

As shown in Refs.~\cite{Schecter2019,Gotta2023}, this model has the QMBS and AQMBS states. We define the operator $\hat{Q}^{\dagger}$ and state $\cket{S_0}$ as
\begin{align}
\hat{Q}^{\dagger}&\equiv \frac{1}{2}\sum_{j=1}^M(-1)^j(\hat{\tau}_j^+)^2,\label{eq:definition_of_Qdagger_for_spin-1}\\
\cket{S_0}&\equiv \cket{-_1,-_2,\ldots,-_M},\label{eq:definition_of_S0_Spin-1_XY_model}
\end{align}
where $\hat{\tau}^{\pm}_j\equiv \hat{\tau}_j^x\pm i\hat{\tau}_j^y$. Here, $\cket{+_j}, \cket{0_j}$, and $\cket{-_j}$ are the eigenstates of $\hat{\tau}_j^z$ with eigenvalues $+1$, $0$, and $-1$, respectively. We can show that $\hat{H}_{XY}^{S=1}$, $\hat{Q}^{\dagger}$, and $\cket{S_0}$ satisfy the RSGA-1. The QMBS state is given by
\begin{align}
\cket{S_n}&=\frac{1}{n!\sqrt{\mathcal{N}_{XY}(M,n)}}(\hat{Q}^{\dagger})^n\cket{S_0},\label{eq:explicit_expression_scar_spin1XY}\\
\mathcal{N}_{XY}(M,n)&\equiv \binom{M}{n},\label{eq:nomralization_constant_spin1XY_scar_state}
\end{align}
where $n=0,1,\ldots, M$. The QMBS state $\cket{S_n}$ satisfies
\begin{align}
\hat{H}\cket{S_n}&=(-Mh+2n h+MD)\cket{S_n}.\label{eq:eigenvalue_equation_S=1_XY_model}
\end{align}

Here, we apply our formalism to the spin-1 XY model. In this case, $\hat{H}_{\rm A}=\hat{H}_0=\hat{H}_{XY}^{S=1}$, $\hat{H}_{\rm SG}=h\hat{Q}^z=h\sum_{j=1}^M\hat{\tau}_j^z$, and $\hat{H}_{\rm sym}=D\sum_{j=1}^M(\hat{\tau}_j^z)^2$.Here, $\hat{H}_{\rm p}'=0$. The subspaces $\mathcal{H}_{P}$ and $\mathcal{H}_{P_n}$ are given by
\begin{align}
\mathcal{H}_{P}&={\rm Span}\{\cket{\bm{m}}\in\mathcal{H} \;|\; m_j=+_j,-_j \},\label{eq:definition_of_H_P_spin-1_XY}\\
\mathcal{H}_{P_n}&={\rm Span}\{\cket{\bm{m}}\in\mathcal{H}_P \;|\; \hat{\tau}^z_{\rm tot}\cket{\bm{m}}=(2n-M)\cket{\bm{m}}\},\label{eq:definition_of_H_Pn_spin-1_XY}
\end{align}
where $\hat{\tau}^z_{\rm tot}\equiv \sum_{j=1}^M\hat{\tau}_j^z$. We can find that $\mathcal{H}_P$ is the same as the Hilbert space of the $S=1/2$ systems with lattice sites $M$. Here, we introduce the spin-1/2 operator at site $j$ as $\hat{S}_j^{\mu}\;(\mu=x,y,z)$. Let $\cket{\uparrow_j}$ and $\cket{\downarrow_j}$ be eigenstates of $\hat{S}_j^z$ with eigenvalues $+1/2$ and $-1/2$, respectively. We assign $\cket{\uparrow_j}\to \cket{+_j}$ and $\cket{\downarrow_j}\to\cket{-_j}$. For simplicity, we consider only the nearest-neighbor terms $(J_{k\ge 3}=0)$. Direct calculations show that the parent Hamiltonian becomes
\begin{align}
\hat{H}_{\rm p}=2J_1^2\sum_{j=1}^{M-1}\left(\frac{1}{4}+\hat{S}_j^x\hat{S}_{j+1}^x+\hat{S}_j^y\hat{S}_{j+1}^y-\hat{S}_j^z\hat{S}_{j+1}^z\right).\label{eq:parent_Hamlitonian_for_S=1_XY_model}
\end{align}
This parent Hamiltonian acts on the subspace $\mathcal{H}_P$. We note that the previous work has derived this parent Hamiltonian in a different context~\cite{Kaneko2024}. The Hamiltonian (\ref{eq:parent_Hamlitonian_for_S=1_XY_model}) is transformed to the ferromagnetic Heisenberg model by the sublattice spin rotation:
\begin{align}
\hat{U}^{\dagger}_{\rm rot}\hat{H}_{\rm p}\hat{U}_{\rm rot}&=2J_1^2\sum_{j=1}^{M-1}\left(\frac{1}{4}-\hat{\bm{S}}_j\cdot\hat{\bm{S}}_{j+1}\right),\label{eq:parent_Hamiltonian_uintary_transformation_S=1_XY}\\
\hat{U}_{\rm rot}&\equiv\prod_{j=1,3,\ldots,M-1}e^{-i\hat{S}_j^z\pi}.\label{eq:rotation_operator_sublattice}
\end{align}
It is well known that low-lying excitations of the ferromagnetic Heisenberg model are given by magnon excitations. The magnon excitation energy for ferromagnetic Heisenberg with the open boundary conditions is given by (see the details in Appendix~\ref{app:Magnon_excitation_ferro_Heisenberg_model})
\begin{align}
\Delta E^2_n&=2J_1^2\left[1-\cos\left(\frac{\pi l}{M}\right)\right],\quad l=0,1,\ldots, M-1.\label{eq:one-magnon_excitation_ferromagnetic_Heisenberg_model}
\end{align}
Here, $l$ represents the number of nodes in the wave function of the magnon excitation with open boundary conditions. For fixed $l=O(1)(\not=0)$, the energy variance goes to zero in the limit $M\to \infty$.

Then, we discuss the EE of the magnon excited state. Using the matrix product state (MPS) representation, we can show that the bond dimension of the magnon excited state is at most $2n$. Because the entanglement entropy is bounded by the logarithm of the bond dimension, the entanglement entropy of the magnon excited state obeys a subvolume law scaling. The details are discussed in Appendix~\ref{app:Magnon_excitation_ferro_Heisenberg_model}.  Therefore, we obtain the AQMBS states of the spin-1 XY model as one-magnon excitations of the ferromagnetic spin-1/2 Heisenberg model. 

Here, we compare our results with those of the previous work. According to Ref.~\cite{Gotta2023}, they proposed the AQMBS state of the spin-1 XY model as
\begin{align}
|\widetilde{AS_n}\rangle&\equiv\hat{Q}^{\dagger}(l)(\hat{Q}^{\dagger})^{n-1}\cket{S_0},\label{eq:AQMBS_Gotta}\\
\hat{Q}^{\dagger}(l)&\equiv \frac{1}{2}\sum_{j=1}^{M}(-1)^je^{2\pi i l j/M}(\hat{\tau}_j^+)^2,\label{eq:definition_of_Qdagger_for_spin1_XY}
\end{align}
where $l=0,1,2,\ldots, M-1$. The energy variance of the state for the open boundary conditions becomes (see the Supplemental Material of Ref.~\cite{Gotta2023})
\begin{align}
\widetilde{\Delta E^2_n}&=2J_1^2\left(1-\frac{1}{M}\right)\left[1-\cos\left(\frac{2\pi l}{M}\right)\right].\label{eq:energy_variance_previous_work}
\end{align}
For $l=O(1)\not=0$, we show that the energy variance derived by the parent Hamiltonian (\ref{eq:one-magnon_excitation_ferromagnetic_Heisenberg_model}) is smaller than that of the previous work~(\ref{eq:energy_variance_previous_work}). This is a direct consequence of the variational principle discussed in Sec.~\ref{subsec:energy_variance}. The state (\ref{eq:AQMBS_Gotta}) proposed in the previous work \cite{Gotta2023} is not an eigenstate of the parent Hamiltonian $\hat{H}_{\rm p}$. Therefore, we successfully modify the energy variance of the AQMBS by considering the parent Hamiltonian. 

It is worth noting that when $J_k$ is constant for all odd $k$, we find that the parent Hamiltonian becomes gapped. We also prove that the parent Hamiltonian becomes gapped when $J_k\propto 1/k^{\alpha}$ for $\alpha\le 1/2$. The details are discussed in Appendix~\ref{app:spin-1_XY_with_long-range}.

Then, we consider the SUSY Hamiltonian. We define the supercharges $\hat{\tilde{Q}}$ and $\hat{\tilde{Q}}^{\dagger}$ as
\begin{align}
\hat{\tilde{Q}}&\equiv \hat{\mathcal{P}}\hat{H}_{XY}^{S=1},\quad \hat{\tilde{Q}}^{\dagger}\equiv \hat{H}_{XY}^{S=1}\hat{\mathcal{P}}.\label{eq:definition_of_supercharge_S=1_XY}
\end{align}
We also define the fermionic parity operator $(-1)^{\hat{F}}$ as
\begin{align}
(-1)^{\hat{F}}&\equiv \prod_{j=1,3,\ldots,M-1}e^{-i\hat{\tau}_j^z\pi}.\label{eq:definition_of_fermionic_parity_operator_S=1_XY}
\end{align}
We can check that these operators satisfy the conditions of SUSY quantum mechanics. The SUSY Hamiltonian is given by
\begin{align}
\hat{H}_{\rm SUSY}&=\hat{\mathcal{P}}(\hat{H}_{XY}^{S=1})^2\hat{\mathcal{P}}+\hat{H}_{XY}^{S=1}\hat{\mathcal{P}}\hat{H}_{XY}^{S=1}.\label{eq:SUSY_Hamiltonian_S=1_XY}
\end{align}
The first term of Eq.~(\ref{eq:SUSY_Hamiltonian_S=1_XY}) is the parent Hamiltonian derived in this section. The second term of Eq.~(\ref{eq:SUSY_Hamiltonian_S=1_XY}) is a nonlocal Hamiltonian acting on the subspace $\mathcal{H}_{Q_1}$, which is defined by
\begin{align}
\mathcal{H}_{Q_1}&\equiv {\rm Span}\{\hat{H}_{XY}^{S=1}\cket{\bm{m}}\not=0 \; | \; \cket{\bm{m}}\in\mathcal{H}_P \}.\label{eq:definition_of_H_Q1_S=1_XY}
\end{align}
The states in $\mathcal{H}_{Q_1}$ include one pair of $\cket{0_j0_k}$, where $j$ and $k$ belong to different sublattices, and the others consist of $\cket{+_s}$ or $\cket{-_s}\;(s\not=j, k)$. 

%%%%%%%%%%%%%%%%%%%%%%%%%%%%%%%%%%%%%%%%%%%%%%%%%%%%%%%%%%%%%%
\subsection{Fermi-Hubbard model}\label{subsec:Fermi-Hubbard}
%%%%%%%%%%%%%%%%%%%%%%%%%%%%%%%%%%%%%%%%%%%%%%%%%%%%%%%%%%%%%%
Here, we consider the Fermi-Hubbard model with correlated hopping terms. The Hamiltonian is defined by
\begin{align}
\hat{H}&\equiv \hat{H}_{\rm hop}+\hat{H}_{\rm c}+\hat{H}_{\mu}+\hat{H}_U,\label{eq:definition_of_Fermi_Hubbard_Hamiltonian}
\end{align}
\begin{subequations}   
\begin{align}
\hat{H}_{\rm hop}&\equiv -J\sum_{j=1}^{M-1}\sum_{\sigma=\uparrow,\downarrow}(\hat{c}^{\dagger}_{j+1,\sigma}\hat{c}_{j,\sigma}+\hat{c}^{\dagger}_{j,\sigma}\hat{c}_{j+1,\sigma}),\label{eq:definition_of_hopping_term_Fermi-Hubbard}\\
\hat{H}_{\rm c}&\equiv J_{\rm c}\sum_{j,\sigma}(\hat{n}_{j,-\sigma}+\hat{n}_{j+1,-\sigma})(\hat{c}^{\dagger}_{j+1,\sigma}\hat{c}_{j,\sigma}+\hat{c}^{\dagger}_{j,\sigma}\hat{c}_{j+1,\sigma}),\label{eq:definitionof_correlated_hopping}\\
\hat{H}_{\mu}&\equiv-\mu\sum_{j=1}^M\hat{n}_j,\label{eq:defiition_of_H_mu}\\
\hat{H}_U&\equiv U\sum_{j=1}^M\hat{n}_{j,\uparrow}\hat{n}_{j,\downarrow},\label{eq:definition_of_H_U}
\end{align}
\end{subequations}
where $\hat{c}_{j,\sigma}(\hat{c}^{\dagger}_{j,\sigma})$ is the Fermion annihilation (creation) operator at site $j$ with spin $\sigma(=\uparrow,\downarrow)$, $\hat{n}_{j,\sigma}\equiv \hat{c}^{\dagger}_{j,\sigma}\hat{c}_{j,\sigma}$, $\hat{n}_j\equiv \hat{n}_{j,\uparrow}+\hat{n}_{j,\downarrow}$, $J$ is the strength of the hopping, $J_{\rm c}$ is the strength of the correlated hopping, $\mu$ is the chemical potential, and $U$ is the strength of the onsite interaction. Here, $-\sigma$ represents the opposite spin direction of $\sigma$. The pure Fermi-Hubbard model ($J_{\rm c}=0$) has exact eigenstates called the $\eta$-pairing state~\cite{Yang1989}. The $\eta$-pairing state originates from the $\eta$-SU(2) symmetry. As shown in Refs.~\cite{Mark2020eta,Moudgalya2020eta}, the correlated hopping terms (or density-dependent hopping terms)~\cite{Hirsch1989} break $\eta$-SU(2) symmetry, while $\eta$-pairing states are still eigenstates of the Hamiltonian. In this situation, the $\eta$-pairing states can be regarded as QMBS states. We do not consider the case $J=J_{\rm c}$ because Hamiltonian (\ref{eq:definition_of_Fermi_Hubbard_Hamiltonian}) becomes integrable at this point~\cite{Arrachea1994,Schadschneider1995}.

The Fermi-Hubbard model with the correlated hopping terms satisfies the RSGA-1. The operator $\hat{Q}^{\dagger}$ and the state $\cket{S_0}$ are defined by
\begin{align}
\hat{Q}^{\dagger}&\equiv \sum_{j=1}^M(-1)^j\hat{c}^{\dagger}_{j,\uparrow}\hat{c}_{j,\downarrow}^{\dagger},\label{eq:definition_of_Qdagger_Fermi-Hubbard}\\
\cket{S_0}&\equiv \cket{\rm vac},\label{eq:definition_of_vac_Fermi-Hubbard}
\end{align}
where $\cket{\rm vac}$ is the vacuum state, i.e., the normalized state such that ${\hat c}_{j,\sigma} \cket{\rm vac}=0$ for all $j$ and $\sigma$. The QMBS state of the Fermi-Hubbard model is given by
\begin{align}
\cket{S_n}&=\frac{1}{n!\sqrt{\mathcal{N}_{\rm FH}(M,n)}}(\hat{Q}^{\dagger})^n\cket{S_0},\label{eq:definition_of_Sn_Fermi-Hubbard}\\
\mathcal{N}_{\rm FH}(M,n)&\equiv \binom{M}{n},\label{eq:definition_of_normalization_constant_FH}
\end{align}
where $n=0,1,\ldots, M$.
The QMBS state (\ref{eq:definition_of_Sn_Fermi-Hubbard}) satisfies
\begin{align}
\hat{H}\cket{S_n}&=(U-2\mu)n\cket{S_n}.\label{eq:eigenvalue_equation_QMBS_FH}
\end{align}

Here, we apply our formalism to the Fermi-Hubbard model. In this case, $\hat{H}_0=\hat{H}_{\rm hop}+\hat{H}_{\rm c}$, $\hat{H}_{\rm p}'=0$, $\hat{H}_{\rm SG}=-2\mu\hat{Q}^z+\hat{C}=\hat{H}_{\mu}+\hat{H}_{U}$, $\hat{Q}^z\equiv \frac{1}{2}\sum_{j=1}^M\hat{n}_j$, and $\hat{H}_{\rm sym}=0$. The subspace $\mathcal{H}_P$ is spanned by the states containing doubly occupied sites (doublons $\cket{\uparrow\downarrow_j})$ or empty sites (holons $\cket{0_j}$). There are no singly occupied sites (singlons, $\cket{\uparrow_j}$ or $\cket{\downarrow_j}$) in $\mathcal{H}_P$. Direct calculations show that the parent Hamiltonian becomes{\small 
\begin{align}
\hat{H}_{\rm p}&=4(-J+J_{\rm c})^2\sum_{j=1}^{M-1}\left(\frac{1}{4}+\hat{S}_j^x\hat{S}_{j+1}^x+\hat{S}_j^y\hat{S}_{j+1}^y-\hat{S}_j^z\hat{S}_{j+1}^z\right),\label{eq:parent_Hamiltonian_Fermi_Hubbard}
\end{align}
}where we assign a doublon to an up spin and a holon to a down spin. We note that this parent Hamiltonian acts on the subspace $\mathcal{H}_P$. Therefore, the parent Hamiltonian of the Fermi-Hubbard model is identical to that of the spin-1 XY model, up to an overall constant. The AQMBS state can be obtained by the same procedure as the spin-1 XY model. The energy variance is then given by
\begin{align}
\Delta E^2_n&=4(-J+J_{\rm c})^2\left[1-\cos\left(\frac{\pi l}{M}\right)\right],\label{eq:energy_variance_FH}
\end{align}
where $l=0,1,\ldots, M-1$. Here, $l$ represents the number of nodes in the wave function of the magnon excitation. 

It is worth noting that the parent Hamiltonian (\ref{eq:parent_Hamiltonian_Fermi_Hubbard}) vanishes at $J=J_{\rm c}$. This is because $\hat{H}\cket{\psi}=-\hat{H}_{\rm c}\cket{\psi}$ for $\cket{\psi}\in \mathcal{H}_P$. In this case, off-diagonal elements of $\hat{H}_{\rm hop}+\hat{H}_{\rm c}$ vanish.

Now, we discuss the EE of the Fermi-Hubbard model. As mentioned above, the parent Hamiltonian of the Fermi-Hubbard model is the same as that of the spin-1 XY model up to an overall constant. Therefore, the EE obeys subvolume law scaling in the Fermi-Hubbard model.

Then, we consider the SUSY Hamiltonian. We define the supercharges and fermionic parity operator as
\begin{align}
\hat{\tilde{Q}}&\equiv \hat{\mathcal{P}}(\hat{H}_{\rm hop}+\hat{H}_{\rm c}),\quad \hat{\tilde{Q}}^{\dagger}\equiv (\hat{H}_{\rm hop}+\hat{H}_{\rm c})\hat{\mathcal{P}},\label{eq:supercharge_operator_FH}\\
(-1)^{\hat{F}}&\equiv \prod_{j=1,3,\ldots,M-1}e^{-i\hat{n}_j\pi}.\label{eq:fermionic_parity_operator_FH}
\end{align}
The SUSY Hamiltonian of the Fermi-Hubbard model is given by{\small
\begin{align}
\hat{H}_{\rm SUSY}&=\hat{\mathcal{P}}(\hat{H}_{\rm hop}+\hat{H}_{\rm c})^2\hat{\mathcal{P}}+(\hat{H}_{\rm hop}+\hat{H}_{\rm c})\hat{\mathcal{P}}(\hat{H}_{\rm hop}+\hat{H}_{\rm c}).\label{eq:SUSY_Hamiltonian_FH}
\end{align}
}The subspace $\mathcal{H}_{Q_1}$ is spanned by states containing two singlons $\cket{\uparrow_j,\downarrow_{j+1}}$ or $\cket{\downarrow_j, \uparrow_{j+1}}$.

%%%%%%%%%%%%%%%%%%%%%%%%%%%%%%%%%%%%%%%%%%%%%%%%%%%%%%%%%%%%%%
\subsection{DH model}\label{subsec:DH_model}
%%%%%%%%%%%%%%%%%%%%%%%%%%%%%%%%%%%%%%%%%%%%%%%%%%%%%%%%%%%%%%
Here, we consider the DH model~\cite{Kodama2023}, which is defined by
{\small \begin{align}
\hat{H}&=D\sum_{j=1}^{M-1}(\hat{S}_j^z\hat{S}_{j+1}^x-\hat{S}_j^x\hat{S}_{j+1}^z)-h\sum_{j=1}^M\hat{S}_j^z-\frac{D}{2}(\hat{S}_1^x-\hat{S}_M^x)\notag \\
&=D\sum_{j=1}^M\hat{S}_j^x(\hat{S}_{j-1}^z-\hat{S}_{j+1}^z)-h\sum_{j=1}^M\hat{S}_j^z\notag \\
&\equiv \hat{H}_{\rm DM}+\hat{H}_z,\label{eq:definition_of_DH_model}
\end{align}
}where $D$ is the strength of the Dzyaloshinskii-Moriya (DM) interaction, $h$ is the strength of the Zeeman field, and $\hat{S}_0^z=\hat{S}_{M+1}^z=-1/2$. This model was introduced in the context of quantum chiral magnetism. Our previous work showed that the DH model has the QMBS and AQMBS states~\cite{Kunimi2024}. In the following, we show that our previous work can be reproduced and modified by the present formalism.

The DH model satisfies the RSGA-1. We define $\hat{Q}^{\dagger}$ operator and $\cket{S_0}$ as
\begin{align}
\hat{Q}^{\dagger}&\equiv \sum_{j=1}^{M}\hat{P}_{j-1}\hat{S}_j^+\hat{P}_{j+1},\label{eq:definition_of_Qdagger_for_DH_model}\\
\cket{S_0}&\equiv \cket{\downarrow_1,\downarrow_2,\ldots,\downarrow_M},\label{eq:definition_of_S0_for_DH_model}
\end{align}
where $\hat{P}_j\equiv 1/2-\hat{S}_j^z$ is the projection operator onto the down spin state and $\hat{P}_0\equiv 1$ and $\hat{P}_{M+1}\equiv1$. The QMBS state of the DH model is given by
\begin{align}
\cket{S_n}&\equiv \frac{1}{n!\sqrt{\mathcal{N}_{\rm DH}(M,n)}} (\hat{Q}^{\dagger})^n\cket{S_0},\label{eq:definition_of_Scar_state_DH_model}\\
\mathcal{N}_{\rm DH}(M,n)&\equiv \binom{M-n+1}{n}.\label{eq:definition_of_normalization_constant_QMBS_DH}
\end{align}
The QMBS state (\ref{eq:definition_of_Scar_state_DH_model}) satisfies
\begin{align}
\hat{H}\cket{S_n}&=\left(\frac{M}{2}-n\right)h\cket{S_n},\label{eq:eigenstate_scar_DH}
\end{align}
where $n=0,1,\ldots, M/2$.

Here, we apply our formalism to the DH model. In this case, $\hat{H}_0=\hat{H}_{\rm DM}$ and $\hat{H}_{\rm SG}=-h\hat{Q}^z=\hat{H}_z$. Now, $\hat{H}_{\rm p}'=0$ and $\hat{H}_{\rm sym}=0$. We introduce the subspace $\mathcal{H}_{P_n}$ and $\mathcal{H}_P$ followed by Eqs.~(\ref{eq:definition_of_H_P}) and (\ref{eq:definition_of_H_Pn}). In the DH model, the states in $\mathcal{H}_P$ do not include any configurations of $\uparrow_j\uparrow_{j+1}$. The direct calculations show that the parent Hamiltonian of the DH model is given by
\begin{align}
\hat{H}_{\rm p}&=\frac{D^2}{2}\sum_{j=1}^{M-1}\hat{P}_{j-1}\left(\frac{1}{4}-\hat{\bm{S}}_j\cdot\hat{\bm{S}}_{j+1}\right)\hat{P}_{j+2}.\label{eq:parent_Hamiltonian_for_DH_model}
\end{align}
See the details of the derivation of the parent Hamiltonian in Appendix~\ref{app:derivation_of_parent_Hamiltonian_DH_model}. We note that $\hat{H}_{\rm p}$ acts on the subspace $\mathcal{H}_P$. According to the previous works \cite{Gomez1993,Cheong2009}, the Hamiltonian (\ref{eq:parent_Hamiltonian_for_DH_model}) in $\mathcal{H}_{P_n}$ can be exactly mapped to a pure ferromagnetic Heisenberg with $M-n+1$ lattice sites, where $n~(1\le n\le M/2)$ is the number of up spins. The mapping rule is to delete the site to the right of up spins except the rightmost up spin. For example, we obtain the case for $M=10$ and $n=3$:
\begin{align}
&\cket{\downarrow\uparrow\cancel{\downarrow}\downarrow\uparrow\cancel{\downarrow}\downarrow\downarrow\uparrow\downarrow}\rightarrow\cket{\downarrow\uparrow\downarrow\uparrow\downarrow\downarrow\uparrow\downarrow}.\label{eq:mapping_rule_DH_model}
\end{align}
The parent Hamiltonian projected onto the subspace $\mathcal{H}_{P_n}$ is mapped onto
\begin{align}
\hat{\tilde{H}}_{\rm p}^{(n)}=\frac{D^2}{2}\sum_{n=1}^{M-n}\left(\frac{1}{4}-\hat{\tilde{\bm{S}}}_j\cdot\hat{\tilde{\bm{S}}}_{j+1}\right),\label{eq:mapped_parent_Hamiltonian_DH_model}
\end{align}
where $\hat{\tilde{\bm{S}}}_j$ is the spin-1/2 operator after mapping. Because the above mapping is exact, the energy spectra of $\hat{H}_{\rm p}^{(n)}$ and $\hat{\tilde{H}}_{\rm p}^{(n)}$ are the same. Therefore, the low-energy gapless excitation is given by the magnon excitation of the pure ferromagnetic Heisenberg model. The energy variance in $\mathcal{H}_{P_n}$ can be written as
\begin{align}
\Delta E^2_n&=\frac{D^2}{2}\left[1-\cos\left(\frac{\pi l}{M-n+1}\right)\right],\label{eq:energy_variance_AQMBS_DH_model}
\end{align}
where $l=0,1,\ldots, M-n$ represents the number of nodes in the wave function of the magnon excitation. The energy variance goes to zero in the limit of $M\to \infty$ for fixed $l=O(1)\not=0$. 

\begin{figure}[t]
\centering
\includegraphics[width=8.6cm,clip]{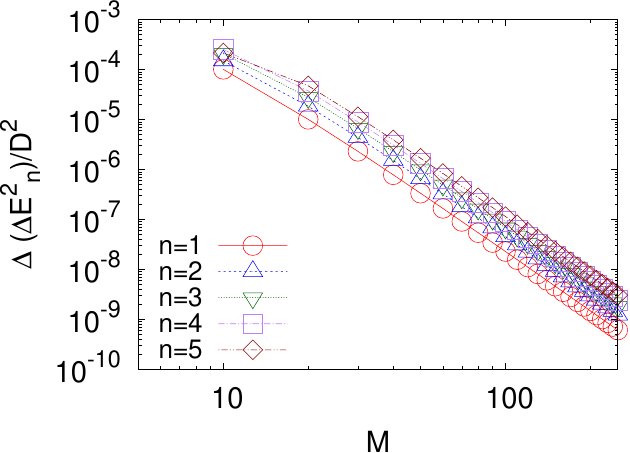}
\caption{Difference of the energy variance $\Delta (\Delta E^2_n)=\widetilde{\Delta E^2_n}-\Delta E_n^2$ as a function of the system size for several $n$. Here, we fix $l=1$ in Eq.~(\ref{eq:energy_variance_AQMBS_DH_model}).}
\label{fig:difference_energy_variance}
\vspace{-0.75em}
\end{figure}

Here, we compare the above results with those from our previous work~\cite{Kunimi2024}. In Ref.~\cite{Kunimi2024}, we assumed that the AQMBS state is given by
\begin{align}
|\widetilde{AS_n}\rangle&\equiv \hat{A}^{\dagger}(\hat{Q}^{\dagger})^{n-1}\cket{S_0},\label{eq:definition_of_ASn_previous_study}\\
\hat{A}^{\dagger}&\equiv \sum_{j=1}^Mf_j\hat{P}_{j-1}\hat{S}_j^+\hat{P}_{j+1},\label{eq:definition_of_Adagger_previous_work}\\
f_j&\equiv \cos\left(\frac{\pi j}{M+1}\right).\label{eq:defintion_of_fj_for_AS_n}
\end{align}
We calculate the difference in the energy variance $\Delta (\Delta E^2_n)\equiv \widetilde{\Delta E_n^2}-\Delta E_n^2$, where $\widetilde{\Delta E_n^2}$ is the energy variance calculated from Eq.~(\ref{eq:definition_of_ASn_previous_study}). The results are shown in Fig.~\ref{fig:difference_energy_variance}. We find that $\Delta(\Delta E^2_n)$ is always positive. This means that the energy variance presented in this paper is smaller than that in our previous work. Therefore, we can obtain the AQMBS of the DH model in our formalism and improve the energy variance presented in our previous work. 

\begin{figure}[t]
\centering
\includegraphics[width=8.6cm,clip]{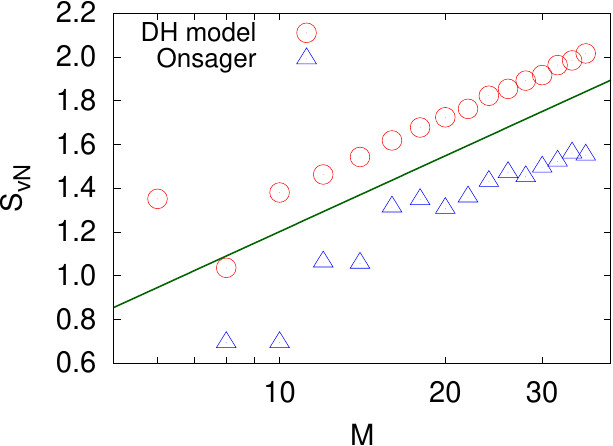}
\caption{Half-chain von Neumann EE $S_{\rm vN}$ of AQMBS in the DH model (red circles) and the Onsager scar model (blue triangles). Here, we fix $l = 1$ for both models and choose $n = \lfloor M/4 \rfloor + 1$ for the DH model and $n = \lfloor M/4 \rfloor - 1$ for the Onsager scar model, where $\lfloor \cdot \rfloor$ denotes the floor function. The solid green line represents $(1/2)\ln(M) + 1/20$ for a guide to the eye.}
\label{fig:ee_dh_and_onsager}
\vspace{-0.75em}
\end{figure}

Now, we discuss the half-chain EE of the state $\cket{AS_n}$. Although the excitation energy of the parent Hamiltonian can be obtained analytically via the mapping, it is difficult to derive an analytical expression for the EE because the mapping alters the wave function. However, we can construct the wave function numerically, since the magnon excitation wave function of the parent Hamiltonian is known (see Appendix~\ref{app:Magnon_excitation_ferro_Heisenberg_model}) and the mapping rules are well defined. Figure~\ref{fig:ee_dh_and_onsager} shows the system-size dependence of the half-chain EE for $\cket{AS_n}$ with $l=1$, where $n=\lfloor M/4\rfloor+1$. We find that the EE exhibits subvolume law scaling.

We also find that the DH model exhibits perfect revival originating from the AQMBS state. The details of this behavior are discussed in Appendix~\ref{app:perfect_revival_DH}.

Then, we discuss the SUSY Hamiltonian of the DH model. The supercharges and fermionic parity operator are defined by
\begin{align}
\hat{\tilde{Q}}&\equiv \hat{\mathcal{P}}\hat{H}_{\rm DM},\quad \hat{\tilde{Q}}^{\dagger}\equiv \hat{H}_{\rm DM}\hat{\mathcal{P}},\label{eq:defnition_of_supercharge_DH_model}\\
(-1)^{\hat{F}}&\equiv \prod_{j=1}^{M}(2\hat{S}_j^z),\label{eq:definition_of_fermionic_parity_operator_DH_model}
\end{align}
where $\hat{\mathcal{P}}$ is the projection operator onto $\mathcal{H}_P$. We can easily verify that the conditions for SUSY quantum mechanics are satisfied. The SUSY Hamiltonian is given by
\begin{align}
\hat{H}_{\rm SUSY}=\hat{\mathcal{P}}\hat{H}_{\rm DM}^2\hat{\mathcal{P}}+\hat{H}_{\rm DM}\hat{\mathcal{P}}\hat{H}_{\rm DM}.\label{eq:SUSY_Hamiltonian_for_DH_model}
\end{align}
The subspace $\mathcal{H}_{Q_1}$ of the DH model is defined by the states containing one $\uparrow_j\uparrow_{j+1}$ configuration because the DM interaction term always creates one $\uparrow_j\uparrow_{j+1}$ configuration when acting on the states in $\mathcal{H}_P$.

%%%%%%%%%%%%%%%%%%%%%%%%%%%%%%%%%%%%%%%%%%%%%%%%%%%%%%%%%%%%%%
\subsection{Domain-wall conserving model}\label{subsec:Domain_wall_conserving_model}
%%%%%%%%%%%%%%%%%%%%%%%%%%%%%%%%%%%%%%%%%%%%%%%%%%%%%%%%%%%%%%
The domain-wall conserving (DWC) model has been proposed by Iadecola and Schecter~\cite{Iadecola2020}. The AQMBS of the DWC model with periodic boundary conditions has been shown in Ref.~\cite{Ren2024}. 

The Hamiltonian of the DWC model with open boundary conditions is given by{\small 
\begin{align}
\hat{H}&=J\sum_{j=1}^M(\hat{S}_j^x-4\hat{S}_{j-1}^z\hat{S}_j^x\hat{S}_{j+1}^z)-h\sum_{j=1}^M\hat{S}_j^z+J_z\sum_{j=0}^M\hat{S}_j^z\hat{S}_{j+1}^z\notag \\
&\equiv \hat{H}_{\rm int}+\hat{H}_z+\hat{H}_{\rm Ising},\label{eq:Definition_of_Hamiltonian_DWC_model}
\end{align}
}where $J$, $h$, and $J_z$ are the strength of the three-spin interaction, the Zeeman field term, and the Ising interaction, respectively. As in the case of the DH model, we define $\hat{S}_0^z\equiv-1/2$ and $\hat{S}_{M+1}^z\equiv-1/2$. 

The DWC model satisfies the RSGA-1. The $\hat{Q}^{\dagger}$ operator and $\cket{S_0}$ is defined as
\begin{align}
\hat{Q}^{\dagger}&\equiv \sum_{j=1}^M(-1)^j\hat{P}_{j-1}\hat{S}_j^+\hat{P}_{j+1},\label{eq:definition_of_Qdagger_for_DWC}\\
\cket{S_0}&\equiv\cket{\downarrow_1,\downarrow_2,\ldots,\downarrow_M}.\label{eq:definition_of_S0_DWC_model}
\end{align}
The QMBS state of the DWC model is given by
\begin{align}
\cket{S_n}&=\frac{1}{n!\sqrt{\mathcal{N}_{\rm DWC}(M,n)}}(\hat{Q}^{\dagger})^n\cket{S_0},\label{eq:QMBS_DWC_model}\\
\mathcal{N}_{\rm DWC}(M,n)&\equiv \binom{M-n+1}{n},\label{eq:definition_of_normalization_constant_QMBS_DWC_model}
\end{align}
where $n=0,1,\ldots, M/2$. The QMBS state (\ref{eq:QMBS_DWC_model}) satisfies
\begin{align}
\hat{H}\cket{S_n}&=\left[h\left(\frac{M}{2}-n\right)+J_z\left(\frac{M+1}{4}-n\right)\right]\cket{S_n}.\label{eq:eigenenergy_of_QMBS_DWC_model}
\end{align}
We note that the QMBS states of the DWC model are almost the same as those of the DH model except the phase factor. 

Here, we apply our formalism to the DWC model. In this case, $\hat{H}_0=\hat{H}_{\rm int}$, $\hat{H}_{\rm SG}=-h\hat{Q}^z+\hat{C}=\hat{H}_z+\hat{H}_{\rm Ising}$, $\hat{H}_{\rm p}'=0$, and $\hat{H}_{\rm sym}=0$. The subspace $\mathcal{H}_P$ is the same as that of the DH model. Therefore, the parent Hamiltonian is given by{\small
\begin{align}
&\hat{H}_{\rm p}\notag \\
&=2J^2\sum_{j=1}^{M-1}\hat{P}_{j-1}\left(\frac{1}{4}+\hat{S}_j^x\hat{S}_{j+1}^x+\hat{S}_j^y\hat{S}_{j+1}^y-\hat{S}_j^z\hat{S}_{j+1}^z\right)\hat{P}_{j+2}.\label{eq:parent_Hamiltonian_DWC_model}
\end{align}
}This Hamiltonian has been derived in a different context in Ref.~\cite{Iadecola2020} and acts on the subspace $\mathcal{H}_P$. Applying the unitary transformation $\hat{U}_{\rm rot}\equiv \prod_{j=1,3,\ldots,M-1}e^{-i\hat{S}_j^z\pi}$ yielding the $\pi$ rotation around $z$ axis on one of the sublattices, we obtain
\begin{align}
\hat{\tilde{H}}_{\rm p}&\equiv \hat{U}_{\rm rot}^{\dagger}\hat{H}_{\rm p}\hat{U}_{\rm rot}\notag \\
&=2J^2\sum_{j=1}^{M-1}\hat{P}_{j-1}\left(\frac{1}{4}-\hat{\bm{S}}_j\cdot\hat{\bm{S}}_{j+1}\right)\hat{P}_{j+2}.\label{eq:parent_Hamiltonian_DWC_model_after_transformation}
\end{align}
This transformed Hamiltonian is the same as the parent Hamiltonian of the DH model (\ref{eq:parent_Hamiltonian_for_DH_model}) up to an overall constant. Therefore, repeating the same procedure as in the DH model, we obtain the AQMBS states of the DWC model. 

Now, we discuss the half-chain EE of AQMBS in the DWC model. As shown above, the parent Hamiltonian of the DWC model and DH model is the same up to a local unitary transformation. Since the local unitary transformation does not change the EE, the properties of the EE are identical to those of the DH model. As shown in the previous section, the EE of the DH model obeys subvolume law scaling. Therefore, the EE of the DWC model also obeys subvolume law scaling.

Here, we discuss the SUSY Hamiltonian of the DWC model. We define the supercharges and fermionic parity operator as
\begin{align}
\hat{\tilde{Q}}&\equiv \hat{\mathcal{P}}\hat{H}_{\rm int},\quad \hat{\tilde{Q}}^{\dagger}\equiv \hat{H}_{\rm int}\hat{\mathcal{P}},\label{eq:definition_of_supercharges_DWC_model}\\
(-1)^{\hat{F}}&\equiv \prod_{j=1}^{M}(2\hat{S}_j^z).\label{eq:definition_of_fermionic_parity_operator_DWC_model}
\end{align}
The SUSY Hamiltonian is given by
\begin{align}
\hat{H}_{\rm SUSY}&=\hat{\mathcal{P}}\hat{H}_{\rm int}^2\hat{\mathcal{P}}+\hat{H}_{\rm int}\hat{\mathcal{P}}\hat{H}_{\rm int}.\label{eq:SUSY_Hamiltonian_DWC_model}
\end{align}
The structure of the subspace $\mathcal{H}_{Q_1}$ is the same as that of the DH model.

%%%%%%%%%%%%%%%%%%%%%%%%%%%%%%%%%%%%%%%%%%%%%%%%%%%%%%%%%%%%%%
\subsection{Onsager scar model}\label{subsec:Onsager_scar}
%%%%%%%%%%%%%%%%%%%%%%%%%%%%%%%%%%%%%%%%%%%%%%%%%%%%%%%%%%%%%%
Here, we consider the Onsager scar~\cite{Shibata2020,Tamura2022}, which is related to the Onsager algebra \cite{Onsager1944,vernier2019onsager}. The previous work shows that the spinless fermion model with a density-dependent hopping term is equivalent to the Onsager scar model~\cite{Tamura2022}. In this paper, we map the spinless fermion system to the spin-1/2 system via the Jordan-Wigner transformation. The Hamiltonian is given by
\begin{align}
\hat{H}&\equiv \hat{H}_{XY}^{S=1/2}+\hat{H}_{\rm c}+\hat{H}_z,\label{eq:definition_of_Onsager_scar_Hamiltonian}\\
\hat{H}_{XY}^{S=1/2}&\equiv -2J\sum_{j=1}^{M-1}(\hat{S}_j^x\hat{S}_{j+1}^x+\hat{S}_j^y\hat{S}_{j+1}^y),\label{eq:definition_of_S=1/2_XY}\\
\hat{H}_{\rm c}&\equiv J_{\rm c}\sum_{j=1}^M\hat{h}_{{\rm c}j},\quad \hat{H}_z\equiv -h\sum_{j=1}^M\hat{S}_j^z,\label{eq:definition_of_Hc_and_Hz_terms_Onsgager}
\end{align}
{\small
\begin{align}
\hat{h}_{{\rm c}j}&\equiv \frac{1}{2}\left(\hat{P}_{j-1}\hat{P}'_j+\hat{P}'_j\hat{P}_{j+1}+\hat{S}_{j-1}^-\hat{P}_j'\hat{S}_{j+1}^++\hat{S}_{j-1}^+\hat{P}_j'\hat{S}_{j+1}^-\right),\label{eq:definition_of_correlated_hopping_term}
\end{align}
}where $J$, $J_{\rm c}$, and $h$ are the strength of the XY interaction, three-spin interaction, and Zeeman field, respectively, and we defined $\hat{P}'_j\equiv 1/2+\hat{S}_j^z$ is the projection operator onto the up spin state, and $\hat{P}_0=0$, $\hat{P}_{M+1}=0$, $\hat{S}_0^{\pm}=0$, and $\hat{S}_{M+1}^{\pm}=0$. The bracket form of Eq.~(\ref{eq:definition_of_correlated_hopping_term}) can be written as{\small
\begin{align}
\hat{h}_{{\rm c}j}&=\cket{\downarrow_{j-1}\uparrow_j\downarrow_{j+1}}\bra{\downarrow_{j-1}\uparrow_j\downarrow_{j+1}}\notag \\
&\quad +\frac{1}{2}\left(\cket{\uparrow_{j-1}\uparrow_j\downarrow_{j+1}}+\cket{\downarrow_{j-1}\uparrow_j\uparrow_{j+1}}\right)\notag \\
&\quad \quad \quad \times \left(\bra{\uparrow_{j-1}\uparrow_j\downarrow_{j+1}}+\bra{\downarrow_{j-1}\uparrow_j\uparrow_{j+1}}\right),\label{eq:bracket_form_of_hcj_Onsager_scar}\\
\hat{h}_{{\rm c}1}&=\frac{1}{2}\cket{\uparrow_1\downarrow_2}\bra{\uparrow_1\downarrow_2},\quad \hat{h}_{{\rm c}M}=\frac{1}{2}\cket{\downarrow_{M-1}\uparrow_M}\bra{\downarrow_{M-1}\uparrow_M}.\label{eq:bracket_for_hcj_edge}
\end{align}
}

The Onsager scar Hamiltonian (\ref{eq:definition_of_Onsager_scar_Hamiltonian}) satisfies the RSGA-1. The operator $\hat{Q}^{\dagger}$ and the state $\cket{S_0}$ are defined by
\begin{align}
\hat{Q}^{\dagger}&\equiv \sum_{j=1}^{M-1}(-1)^{j-1}\hat{S}_j^+\hat{S}_{j+1}^+,\label{eq:definition_of_Qdagger_Onsager}\\
\cket{S_0}&\equiv \cket{\downarrow_1,\downarrow_2,\ldots, \downarrow_M}.\label{eq:definition_of_S0_Onsager_scar}
\end{align}
The QMBS state of the Onsager scar model is given by
\begin{align}
\cket{S_n}&=\frac{1}{n!\sqrt{\mathcal{N}_{\rm Ons}(M,n)}}(\hat{Q}^{\dagger})^n\cket{S_0},\label{eq:definition_of_Sn_Onsager}\\
\mathcal{N}_{\rm Ons}(M, n)&\equiv\binom{M-n}{n},\label{eq:definition_of_normalization_constant_Onsager_Scar}
\end{align}
where $n=0,1,\ldots, M/2$. The QMBS state (\ref{eq:definition_of_Sn_Onsager}) satisfies
\begin{align}
\hat{H}\cket{S_n}&=h\left(\frac{M}{2}-2n\right)\cket{S_n}.\label{eq:eigenvalue_equation_QMBS_Onsager_scar}
\end{align}

Here, we apply our formalism to the Onsager scar model. In this case, $\hat{H}_0=\hat{H}_{XY}^{S=1/2}$, $\hat{H}_{\rm SG}=-2h\hat{Q}^z=\hat{H}_z$, $\hat{H}_{\rm p}'=\hat{H}_{\rm c}$, and $\hat{H}_{\rm sym}=0$. The subspace $\mathcal{H}_{P_n}$ is spanned by states containing $2n$ up spins, where the up spins form clusters with an even number of up spins. The examples are as follows:
\begin{align}
|\ldots\downarrow\underbrace{\uparrow\uparrow}_{\text{Size 2 cluster}}\downarrow\downarrow\underbrace{\uparrow\uparrow\uparrow\uparrow}_{\text{Size 4 cluster}}\downarrow\ldots\rangle.\label{eq:example_onsager_scar_state}
\end{align}
This comes from the definition of $\hat{Q}^{\dagger}$. The parent Hamiltonian of the Onsager scar model becomes
\begin{align}
\hat{H}_{\rm p}&=2J^2\sum_{j=1}^{M-1}\left(\frac{1}{4}-\hat{S}_{j}^z\hat{S}_{j+1}^z\right)\notag \\
&\quad +2J^2\sum_{j=2}^{M-1}(\hat{S}_{j-1}^x\hat{P}_j'\hat{S}_{j+1}^x+\hat{S}_{j-1}^y\hat{P}_j'\hat{S}_{j+1}^y).\label{eq:parent_Hamiltonian_Onsager_scar}
\end{align}
We note that this parent Hamiltonian acts on the subspace $\mathcal{H}_P$ (see Appendix~\ref{app:derivation_of_parent_Hamiltonian_Onsager_scar_model} for details of the derivation of the parent Hamiltonian). We find that the parent Hamiltonian (\ref{eq:parent_Hamiltonian_Onsager_scar}) can be mapped to a pure Heisenberg Hamiltonian with $M-n$ lattice sites, where $n$ is the order of the QMBS state. The mapping rule is to remove half of the up spins containing each up spin cluster. For example, we obtain
\begin{align} 
|\underbrace{\ldots\downarrow\uparrow\cancel{\uparrow}\uparrow\cancel{\uparrow}\downarrow\uparrow\cancel{\uparrow}\downarrow\ldots}_M\rangle\to|\underbrace{\ldots\downarrow\uparrow\uparrow\downarrow\uparrow\downarrow\ldots}_{M-n}\rangle.\label{eq:example_mapping_Onsager}
\end{align}
The parent Hamiltonian projected onto the subspace $\mathcal{H}_{P_n}$ is mapped onto
\begin{align}
\hat{\tilde{H}}_{\rm p}^{(n)}&=2J^2\sum_{j=1}^{M-n-1}\left(\frac{1}{4}+\hat{\tilde{S}}_j^x\hat{\tilde{S}}_{j+1}^x+\hat{\tilde{S}}_j^y\hat{\tilde{S}}_{j+1}^y-\hat{\tilde{S}}_j^z\hat{\tilde{S}}_{j+1}^z\right),\label{eq:parent_Hamiltonian_after_mapping_Onsager}
\end{align}
where $\hat{\tilde{S}}_j^{\mu}$ represents the spin-1/2 operator after mapping. We note that the related mapping has been discussed in the context of the Bariev-like model ~\cite{Bariev1991,Chhajlany2016,Pozsgay2021,Borsi2023,borsi2025volume}. Performing the sublattice spin rotation used in Sec.~\ref{subsec:Domain_wall_conserving_model}, we obtain the pure ferromagnetic Heisenberg model. Therefore, the AQMBS state of the Onsager scar model is given by the magnon excitation of the parent Hamiltonian (\ref{eq:parent_Hamiltonian_after_mapping_Onsager}). The energy variance of this state reads
\begin{align}
\Delta E_n^2&=2J^2\left[1-\cos\left(\frac{\pi l}{M-n}\right)\right],\label{eq:energy_variance_Onsager_scar}
\end{align}
where $l=0,1,\ldots, M-n-1$. Here, $l$ represents the number of nodes in the wave function of the magnon excitation.

Now, we discuss the half-chain EE of the state $\cket{AS_n}$. For the same reasons as in the DH model, it is difficult to derive an analytical expression for the EE of the Onsager scar model. Thus, we numerically calculate the half-chain EE. The results are shown in Fig.~\ref{fig:ee_dh_and_onsager}, where we find that the EE exhibits subvolume law scaling.

Then, we discuss the SUSY Hamiltonian of the Onsager scar model. The supercharges and fermionic parity operator are defined by
\begin{align}
\hat{\tilde{Q}}&\equiv \hat{\mathcal{P}}\hat{H}_{XY}^{S=1/2},\quad \hat{\tilde{Q}}^{\dagger}\equiv \hat{H}_{XY}^{S=1/2}\hat{\mathcal{P}},\label{eq:supercharge_operator_Onsager}\\
(-1)^{\hat{F}}&\equiv \prod_{j=1,3,\ldots,M-1}e^{-i\hat{P}_j'\pi},\label{eq:fermionic_parity_operator_Onsager_scar}
\end{align}
where $\hat{\mathcal{P}}$ is the projection operator onto the subspace $\mathcal{H}_P$. Therefore, the SUSY Hamiltonian is given by
\begin{align}
\hat{H}_{\rm SUSY}&=\hat{\mathcal{P}}(\hat{H}_{XY}^{S=1/2})^2\hat{\mathcal{P}}+\hat{H}_{XY}^{S=1/2}\hat{\mathcal{P}}\hat{H}_{XY}^{S=1/2}.\label{eq:SUSY_Hamiltonian_Onsager_Scar}
\end{align}
The subspace $\mathcal{H}_{Q_1}$ of the Onsager scar model is defined by states containing one $\downarrow\uparrow\downarrow$ configuration and an odd up-spin cluster adjacent to the $\downarrow\uparrow\downarrow$ configuration. For example, the following states belong to $\mathcal{H}_{Q_1}$:
\begin{align}
\cket{\ldots\downarrow\uparrow\downarrow\uparrow\uparrow\uparrow\downarrow\ldots},\label{eq:example1_Onsager_scar_HQ1}\\
\cket{\ldots\downarrow\uparrow\uparrow\uparrow\uparrow\uparrow\downarrow\uparrow\downarrow\ldots}.\label{eq:example2_Onsager_scar_HQ1}
\end{align}

%%%%%%%%%%%%%%%%%%%%%%%%%%%%%%%%%%%%%%%%%%%%%%%%%%%%%%%%%%%%%%
\subsection{Nonmaximal spin scar model}\label{subsec:Nonmaximal_spins}
%%%%%%%%%%%%%%%%%%%%%%%%%%%%%%%%%%%%%%%%%%%%%%%%%%%%%%%%%%%%%%
Here, we discuss the nonmaximal spin scar model proposed by O'Dea {\it et al}.~\cite{ODea2020}, which is defined for spin-1 systems. The Hamiltonian of the model is defined by
\begin{align}
\hat{H}&=\hat{H}_{J_1}+\hat{H}_{J_2}+\hat{H}_{B_1}+\hat{H}_{B_2}+\hat{H}_{J_z}\notag \\
&\quad +\hat{H}_D+\hat{H}_z+\hat{H}_{{\rm DM}0},\label{eq:definition_of_nonmaximal_spin_scar}
\end{align}
\begin{subequations}   
\begin{align}   
\hat{H}_{J_1}&\equiv J_1\sum_{j=1}^{M-1}\left[\frac{1}{3}(\hat{\bm{\tau}}_j\cdot\hat{\bm{\tau}}_{j+1})^2-\frac{1}{3}\right]\equiv J_1\sum_{j=1}^{M-1}\hat{P}^{S=0}_{j,j+1},\label{eq:defintion_of_J1_term_nommaximal}\\
\hat{H}_{J_2}&\equiv J_2\sum_{j=1}^{M-2}\left[1-\frac{1}{2}(\hat{\bm{\tau}}_j\cdot\hat{\bm{\tau}}_{j+2})^2-\frac{1}{2}\hat{\bm{\tau}}_j\cdot\hat{\bm{\tau}}_{j+2}\right]\notag \\
&\equiv J_2\sum_{j=1}^{M-2}\hat{P}^{S=1}_{j,j+2},\label{eq:definition_of_J2_term_nommaximal}\\
\hat{H}_{B_1}&\equiv B_1\sum_{j=2}^{M-1}[(\hat{\tau}_{j-1}^z)^2-(\hat{\tau}_{j+1}^z)^2](\hat{\tau}_{j}^x+\hat{\tau}_{j}^y),\label{eq:definition_of_B1_term_nonmaximal}\\
\hat{H}_{B_2}&\equiv B_2\sum_{j=2}^{M-1}[(\hat{\tau}_{j-1}^z)^2-(\hat{\tau}_{j+1}^z)^2]\hat{\tau}_j^z,\label{eq:definition_of_B2_term_nonmaximal}\\
\hat{H}_{J_z}&\equiv J_z\sum_{j=1}^{M-1}\hat{\tau}_j^z\hat{\tau}_{j+1}^z,\label{eq:definition_of_Jz_term_nonmaximal}\\
\hat{H}_D&\equiv D\sum_{j=1}^M(\hat{\tau}_j^z)^2,\label{eq:definition_of_HD_nonmaximal}\\
\hat{H}_z&\equiv h\sum_{j=1}^M\hat{\tau}_j^z,\label{eq:definition_of_Hz_nonmaximal}\\
\hat{H}_{{\rm DM}0}&\equiv D_z\sum_{j=2}^{M-1}(\hat{\bm{\tau}}_{j-1}\times\hat{\bm{\tau}}_{j+1})_z[1-(\hat{\tau}_j^z)^2]\notag \\
&=D_z\sum_{j=2}^{M-1}(\tau_{j-1}^x\hat{\tau}_{j+1}^y-\hat{\tau}_{j-1}^y\hat{\tau}_{j+1}^x)[1-(\hat{\tau}_j^z)^2],\label{eq:definition_of_DM_nonmaximal}
\end{align}
\end{subequations}
where $J_1$, $J_2$, $B_1$, $B_2$, $J_z$, $D$, $h$, and $D_z$ are real parameters, $\hat{P}^{S=0}_{j,j+1}$ and $\hat{P}^{S=1}_{j,j+2}$ are the projectors onto the singlet state and antisymmetric spin-1 states, respectively. The last term $\hat{H}_{{\rm DM}0}$ is absent in the original paper~\cite{ODea2020}. The reason for adding this term will be discussed below. 

The nonmaximal spin scar model (\ref{eq:definition_of_nonmaximal_spin_scar}) satisfies the RSGA-2. The operator $\hat{Q}^{\dagger}$ and state $\cket{S_0}$ are defined by
\begin{align}
\hat{Q}^{\dagger}&\equiv \frac{1}{2}\sum_{j=1}^M(\hat{\tau}_j^+)^2,\label{eq:definition_of_Qdagger_nonmaximal_spin}\\
\cket{S_0}&=\frac{1}{\sqrt{2}}(\cket{0_1,-_2,0_3,-_4,\ldots,0_{M-1},-_M}\notag \\
&\hspace{2.0em}+\cket{-_1,0_2,-_3,0_4,\ldots,-_{M-1},0_M}).\label{eq:root_state_of_nonmaximal_spin}
\end{align}
In contrast to other models discussed in this paper, $\cket{S_0}$ is not a direct product state. The QMBS state of the nonmaximal spin scar model is given by
\begin{align}
\cket{S_n}&=\frac{1}{n!\sqrt{\mathcal{N}_{\rm NMS}(M,n)}}(\hat{Q}^{\dagger})^n\cket{S_0},\label{eq:definition_of_Sn_nonmaximal_spin_scar}\\
\mathcal{N}_{\rm NMS}(M,n)&\equiv \binom{M/2}{n}.\label{eq:normalization_constant_nonmaximal_spin}
\end{align}
The QMBS state (\ref{eq:definition_of_Sn_nonmaximal_spin_scar}) satisfies
\begin{align}
\hat{H}\cket{S_n}&=\left[h\left(2n-\frac{M}{2}\right)+\frac{M}{2}D\right]\cket{S_n},\label{eq:eigenvalue_equation_QMBS_nonmaximal}
\end{align}
where $n=0,1,\ldots, M/2$.

Here, we apply our formalism to the model. The subspace $\mathcal{H}_P$ is spanned by the states $\cket{0_1,n_2,0_3,\ldots, 0_{M-1},n_M}$ and $\cket{n_1,0_2,n_3,\ldots, n_{M-1},0_M}$ for $n_j=+_j$, or $-_j$. In this case, we obtain $\hat{H}_0=\hat{H}_{J_1}+\hat{\mathcal{Q}}\hat{H}_{J_2}\hat{\mathcal{Q}}+\hat{H}_{B_1}+\hat{H}_{B_2}+\hat{H}_{J_z}+\hat{H}_{{\rm DM}0}$, $\hat{H}_{\rm SG}=h\hat{Q}^z=\hat{H}_{z}$, $\hat{H}_{\rm sym}=\hat{H}_{D}$, and $\hat{H}_{\rm p}'=\hat{\mathcal{P}}\hat{H}_{J_2}\hat{\mathcal{P}}$. Here, we used the fact that $\hat{\mathcal{P}}\hat{H}_{J_2}\hat{\mathcal{Q}}=0$. We note that for $\cket{\bm{n}}\in\mathcal{H}_P$, the following relations hold: $\hat{H}_{J_1}\cket{\bm{n}}=0$, $\hat{H}_{B_1}\cket{\bm{n}}=0$, $\hat{H}_{B_2}\cket{\bm{n}}=0$, and $\hat{H}_{J_z}\cket{\bm{n}}=0$. The parent Hamiltonian becomes
\begin{align}
\hat{H}_{\rm p}&=D^2_z\sum_{j=2}^{M-1}[1-(\hat{\tau}_j^z)^2]\notag \\
&\times\left[-\frac{1}{4}(\hat{\tau}_{j-1}^+)^2(\hat{\tau}_{j+1}^-)^2-\frac{1}{4}(\hat{\tau}_{j-1}^-)^2(\hat{\tau}_{j+1}^+)^2\right.\notag \\
&\left. \quad \quad +\frac{1}{2}(\hat{\tau}_{j-1}^z)^2(\hat{\tau}_{j+1}^z)^2-\frac{1}{2}\hat{\tau}_{j-1}^z\hat{\tau}_{j+1}^z\right].\label{eq:parent_Hamiltonian_nonmaximal_spin}
\end{align}
We note that this parent Hamiltonian acts on the subspace $\mathcal{H}_P$ (see Appendix~\ref{app:derivation_of_parent_Hamiltonian_nonmaximal_spin_scar_model} for details of the derivation of the parent Hamiltonian). This parent Hamiltonian can be mapped to the spin-1/2 ferromagnetic Heisenberg on the two independent ladders. Each ladder has $M/2$ sites. The mapping rule is given by
\begin{align}
\cket{0_1,n_2,\ldots, 0_{M-1}, n_{M}} &\to \cket{n_2,n_4,\ldots,n_{M}}\quad (k=1),\label{eq:mapping_upper_ladder}\\
\cket{n_1,0_2,\ldots, n_{M-1}, 0_{M}} &\to \cket{n_1,n_3,\ldots,n_{M-1}}\quad (k=2),\label{eq:mapping_lower_ladder}
\end{align}
where $n_j=\uparrow_j$ or $\downarrow_j$ and $k$ is the index of the ladder. The parent Hamiltonian after mapping becomes
\begin{align}
\hat{\tilde{H}}_{\rm p}&=\frac{D^2_z}{2}\sum_{k=1,2}\sum_{j=1}^{M/2-1}\left(\frac{1}{4}-\hat{\bm{s}}_{j,k}\cdot\hat{\bm{s}}_{j+1,k}\right),\label{eq:parent_Hamiltonian_nonmaximal_spin_after_mapping}
\end{align}
where $\hat{\bm{s}}_{j,k}$ is the spin-1/2 operator at site $j$ of the ladder $k\;(k=1,2)$. Since there is no interaction between the ladders, the spectra of $\hat{\tilde{H}}_{\rm p}$ are identical to those of the one-dimensional ferromagnetic Heisenberg model with lattice sites $M/2$, except that they are doubly degenerate. The degeneracy comes from the symmetry of the exchange of the two ladders. Here, we choose the even-parity sector of the symmetry because $\cket{S_n}$ has even parity under the exchange of the ladders. Therefore, the AQMBS state is given by the magnon excited state of the ferromagnetic Heisenberg model with even parity. The energy variance becomes
\begin{align}
\Delta E^2_n&=\frac{D^2_z}{2}\left[1-\cos\left(\frac{2\pi l}{M}\right)\right],\label{eq:energy_variance_nonmaximal_spin}
\end{align}
where $l=0,1,\ldots,M/2-1$. Here, $l$ represents the number of nodes in the wave function of the magnon excitation.

Now, we discuss the half-chain EE of the state $\cket{AS_n}$. Since the parent Hamiltonian is constructed by the ferromagnetic Heisenberg model on the two independent ladder, we can obtain the analytical expression of the state $\cket{AS_n}$ as
\begin{align}
\cket{AS_n}&=\frac{1}{\sqrt{2}}(\cket{AS_n^{\rm odd}}+\cket{AS_n^{\rm even}}),\label{eq:expression_of_AS_n_nonmaximal_spin_scar}
\end{align}
where $\cket{AS_n^{\rm odd}}$ and $\cket{AS_n^{\rm even}}$ are the normalized states. As shown in Appendix~\ref{app:derivation_of_AQMBS_nonmaximal_spin_scar_model}, the bond dimension of these states is given by at most $2n$. Then, the bond dimension of the state $\cket{AS_n} $ is at most $4n$ because the bond dimension of the sum of two MPS is given by the sum of their bond dimensions~\cite{schollwock2011density}. This means that the half-chain EE of the state $\cket{AS_n}$ obeys subvolume law scaling.

Here, we discuss why we added the term $\hat{H}_{{\rm DM}0}$. When $D_z=0$, we obtain $\hat{H}_{\rm p}=0$. This comes from the properties $\hat{\mathcal{P}}(\hat{H}_{J_1}+\hat{\mathcal{Q}}\hat{H}_{J_2}\hat{\mathcal{Q}}+\hat{H}_{B_1}+\hat{H}_{B_2}+\hat{H}_{J_z})\hat{\mathcal{P}}=0$ and $\hat{\mathcal{Q}}(\hat{H}_{J_1}+\hat{\mathcal{Q}}\hat{H}_{J_2}\hat{\mathcal{Q}}+\hat{H}_{B_1}+\hat{H}_{B_2}+\hat{H}_{J_z})\hat{\mathcal{P}}=0$. In order to obtain the nonzero parent Hamiltonian, we need nonzero off-diagonal elements. The original model does not satisfy this property~\cite{ODea2020}. From the viewpoint of the subspace, $\mathcal{H}_P$ does not have common elements of $\mathcal{H}_{\rm thermal}$ in contrast to the case shown in Fig.~\ref{fig:schematic_Hilbert_space}. Because the term $\hat{H}_{{\rm DM}0}$ has nonzero off-diagonal elements, we can successfully construct the AQMBS states by our formalism. Here, we note that the absence of $\hat{H}_{\rm DM0}$ does not necessarily imply the nonexistence of the AQMBS in the model.

Then, we consider the SUSY Hamiltonian of the nonmaximal spin scar model. The supercharges and fermionic parity operator are defined by
\begin{align}
\hat{\tilde{Q}}&\equiv \hat{\mathcal{P}}\hat{H}_{{\rm DM}0},\quad \hat{\tilde{Q}}^{\dagger}\equiv \hat{H}_{{\rm DM}0}\hat{\mathcal{P}},\label{eq:definition_of_superchage_nonmaximal_spin_scar}\\
(-1)^{\hat{F}}&\equiv \prod_{j=1,2,5,6,9,10,\ldots}e^{-i\hat{\tau}_j^z\pi}.\label{eq:definition_of_fermionic_parity_operator_nonmaximal_spin_scar}
\end{align}
The SUSY Hamiltonian of the nonmaximal spin scar model is given by
\begin{align}
\hat{H}_{\rm SUSY}&=\hat{\mathcal{P}}\hat{H}_{{\rm DM}0}^2\hat{\mathcal{P}}+\hat{H}_{{\rm DM}0}\hat{\mathcal{P}}\hat{H}_{{\rm DM}0}.\label{eq:SUSY_Hamiltonian_nonmaximal_spin_scar}
\end{align}
The subspace $\mathcal{H}_{Q_1}$ of the nonmaximal spin scar model is spanned by states in which a local configuration of three consecutive 0s (i.e., $\cket{0_{j-1},0_j,0_{j+1}}$ for some $j$) appears exactly once.

%%%%%%%%%%%%%%%%%%%%%%%%%%%%%%%%%%%%%%%%%%%%%%%%%%%%%
%%%%%%%%%%%%%%%%%%%%%%%%%%%%%%%%%%%%%%%%%%%%%%%%%%%%%
%%%%%%%%%%%%%%%%%%%%%%%%%%%%%%%%%%%%%%%%%%%%%%%%%%%%%
\section{Summary and discussion}\label{sec:summary}
%%%%%%%%%%%%%%%%%%%%%%%%%%%%%%%%%%%%%%%%%%%%%%%%%%%%%
%%%%%%%%%%%%%%%%%%%%%%%%%%%%%%%%%%%%%%%%%%%%%%%%%%%%%
%%%%%%%%%%%%%%%%%%%%%%%%%%%%%%%%%%%%%%%%%%%%%%%%%%%%%
%%%%%%%%%%%%%%%%%%%%%%%%%%%%%%%%%%%%%%%%%%%%%%%%%%%%%%%%%%%%%%

In this paper, we proposed a method to construct AQMBS states systematically. Our assumptions are as follows: (1) The system satisfies the RSGA. (2) The Hamiltonian can be decomposed into three parts, $\hat{H}_{\rm A}$, $\hat{H}_{\rm SG}$, and $\hat{H}_{\rm sym}$ as discussed in the symmetry-based formalism of the QMBS~\cite{ODea2020}. (3) The annihilation operator $\hat{H}_{\rm A}$ can be written as $\hat{H}_{\rm A}=\hat{H}_0+\hat{H}_{\rm p}'$. (4) $\hat{H}_0=\sum_j\hat{h}_j$ is a local Hamiltonian and satisfies $\hat{\mathcal{P}}\hat{h}_j\hat{\mathcal{P}}=0$. (5) The off-diagonal elements of the parent Hamiltonian are zero: $\hat{\mathcal{P}}_n\hat{H}_0^2\hat{\mathcal{P}}_m=0\;(n\not=m).$ Under these assumptions, we can obtain the AQMBS states as low-energy gapless excited states of the parent Hamiltonian $\hat{\mathcal{P}}\hat{H}_0^2\hat{\mathcal{P}}$.

In the previous section, we showed that the AQMBS states in several models can be obtained using our formalism presented in Sec.~\ref{sec:framework}. However, our formalism cannot be applied to certain scar models. One such example is the PXP model~\cite{Turner2018,Turner2018prb}. Although Ref.~\cite{Ren2024} demonstrated that the PXP model exhibits AQMBS states, our formalism cannot be applied to it. This is because the PXP model does not satisfy the RSGA.

Another example is the Affleck-Kennedy-Lieb-Tasaki (AKLT) model. As shown in Refs.~\cite{Mark2020unified, Moudgalya2020large}, the AKLT model exhibits QMBS and satisfies RSGA-2. Reference~\cite{Ren2024} demonstrated that the AKLT model also possesses AQMBS states. However, the AKLT model does not satisfy the condition $\hat{\mathcal{P}}\hat{h}_j\hat{\mathcal{P}}=0$. Therefore, our formalism cannot be applied to the AKLT model. We note that the parent Hamiltonian of the AKLT scar states has been discussed in a different context~\cite{Moudgalya2024exhaustive}, and it is more complicated than those considered in this paper.

One more example is given by the maximal spin scar model \cite{Mark2020eta,ODea2020} and spin models with scalar spin chirality~\cite{Sanada2023,Sanada2024a}. Although these models satisfy the RSGA, the subspace $\mathcal{H}_P$ is equal to the total Hilbert space $\mathcal{H}$ because $\cket{S_0}=\cket{\downarrow_1,\downarrow_2,\ldots,\downarrow_M}$ and $\hat{Q}^{\dagger}\equiv \hat{S}_{\rm tot}^+=\sum_{j=1}^M\hat{S}_j^+$. Therefore, $\hat{\mathcal{P}}\hat{h}_j\hat{\mathcal{P}}\not=0$ holds in these models.

From these results, our formalism does not cover \emph{all }AQMBS states. Extending our formalism to other AQMBS states remains an open problem.

In our formalism, the parent Hamiltonian plays a crucial role for constructing the AQMBS states. As shown in Sec.~\ref{sec:construction_of_AQMBS}, the parent Hamiltonians of all models presented in this paper are related to the spin-$1/2$ ferromagnetic Heisenberg model. This may be due to the structure of $\hat{Q}^{\dagger}$.  In this paper, we assume the existence of a single $\hat{Q}^{\dagger}$ operator, whose algebraic structure resembles that of the SU($2$) algebra. If we consider cases with higher symmetry, such as SU($3$)~\cite{ODea2020}, we may obtain a different parent Hamiltonian~\cite{hashimotoSUN}.

We note that our parent Hamiltonian $\hat{\mathcal{P}}\hat{H}_0^2\hat{\mathcal{P}}$ is similar in form to the Hamiltonian used in the Shiraishi-Mori construction \cite{Shiraishi2017}. In their construction, the Hamiltonian is assumed to be
\begin{align}
\hat{H}=\sum_j\hat{P}_j^{\rm SM}\hat{h}_j^{\rm SM}\hat{P}_j^{\rm SM}+\hat{H}',\label{eq:Hamiltonian_SM}
\end{align}
where $\hat{h}_j^{\rm SM}$ is an arbitrary local Hamiltonian, $\hat{P}_j^{\rm SM}$ is a local projection operator satisfying $\hat{P}_j^{\rm SM}\cket{\psi}=0$ for any $j$ and $\cket{\psi}\in\mathcal{T}$, with $\mathcal{T}$ being a target Hilbert subspace. The term $\hat{H}'$ commutes with $\hat{P}_j^{\rm SM}$, i.e., $[\hat{H}', \hat{P}_j^{\rm SM}]=0$ for all $j$. However, our formalism differs from their construction because the projection operator $\hat{\mathcal{P}}$ is nonlocal. For example, in the case of the spin-1 XY model, the projection operator is given by $\hat{\mathcal{P}}=\prod_{j=1}^M(\hat{\tau}_j^z)^2$, which is nonlocal. Therefore, our formalism is distinct from the Shiraishi-Mori construction. 

Furthermore, we have identified a connection between QMBS and SUSY quantum mechanics. In our formalism, SUSY remains unbroken because the QMBS state is always a zero-energy eigenstate of the SUSY Hamiltonian. This raises an important question: What happens if SUSY is broken? This issue may be related to the robustness of QMBS states against perturbations. As discussed in Ref.~\cite{Gotta2023}, the QMBS states $\cket{S_n}$ evolve into the AQMBS states when a perturbation is introduced. In this case, the original QMBS states are no longer exact eigenstates of the perturbed Hamiltonian $\hat{H}+\hat{H}_{\rm pert}$, but their energy variance $\bra{S_n}\hat{H}^2_{\rm pert}\cket{S_n}-(\bra{S_n}\hat{H}_{\rm pert}\cket{S_n})^2$ vanishes in the thermodynamic limit, where we used the fact that $\cket{S_n}$ is an eigenstate of the unperturbed Hamiltonian $\hat{H}$. From the perspective of SUSY, this AQMBS state may be understood as a SUSY broken state.

\begin{acknowledgments}
The authors thank U. Miura, K. Totsuka, and L. Mazza for their helpful discussions. The authors also thank an anonymous referee for pointing out a flaw in Appendix A of an earlier version of this paper. This work was supported by JSPS KAKENHI Grants No.~JP22H05268 (M.K.), No.~JP25K00215 (M.K.), No.~JP20K03855 (Y.K.), No.~JP23K25783 (H.K.), No.~JP23K25790 (H.K.), JST ASPIRE No.~JPMJAP24C2 (M.K.), and MEXT KAKENHI Grant-in-Aid for Transformative Research Areas A “Extreme Universe” (KAKENHI Grant No.~JP21H05191). This research was also supported by Joint Research by the Institute for Molecular Science (IMS Program No. 23IMS1101) (Y.K.)
\end{acknowledgments}

%%%%%%%%%%%%%%%%%%%%%%%%%%%%%%%%%%%%%%%%%%%%%%%%%%%%%%%%%%%%%%
\section*{Data Availability}
%%%%%%%%%%%%%%%%%%%%%%%%%%%%%%%%%%%%%%%%%%%%%%%%%%%%%%%%%%%%%%

The data that support the findings of this article are
openly available~\cite{kunimi_2025_16891505}.

%%%%%%%%%%%%%%%%%%%%%%%%%%%%%%%%%%%%%%%%%%%%%%%%%%%%%%%%%%%%%%
\appendix
%%%%%%%%%%%%%%%%%%%%%%%%%%%%%%%%%%%%%%%%%%%%%%%%%%%%%%%%%%%%%%
%%%%%%%%%%%%%%%%%%%%%%%%%%%%%%%%%%%%%%%%%%%%%%%%%%%%%%%%%
%\clearpage
%\newpage
%\begin{widetext}
\section{Remark on locality of $\hat{\mathcal{P}}\hat{H}_0^2\hat{\mathcal{P}}$}\label{app:remark_locality}
%%%%%%%%%%%%%%%%%%%%%%%%%%%%%%%%%%%%%%%%%%%%%%%%%%%%%%%%%%%%%%
Here, we prove that the parent Hamiltonian $\hat{H}_{\rm p}=\hat{\mathcal{P}}\hat{H}_0^2\hat{\mathcal{P}}$ is local under the assumptions that $\hat{\mathcal{P}}\hat{h}_j\hat{\mathcal{P}}=0$ for any $j$, $\hat{h}_j$ is a local operator, and the subspace $\mathcal{H}_P$ has a tensor product structure. The subspace $\mathcal{H}_P$ can be written as $\mathcal{H}_P=(\mathcal{H}_P^{\rm L})^{\otimes M}$, where $\mathcal{H}_P^{\rm L}$ is a local Hilbert space.  To prove the locality of the parent Hamiltonian, we expand the parent Hamiltonian as
\begin{align}
\hat{H}_{\rm p}=\sum_{j,k}\hat{\mathcal{P}}\hat{h}_j\hat{h}_k\hat{\mathcal{P}}.\label{eq:expansion_parent_Hamiltonian}
\end{align}
Let $\cket{\bm{n}}$ and $\cket{\bm{m}}$ be direct product states in the subspace $\mathcal{H}_{P}$. The projection operator $\hat{\mathcal{P}}$ can be written as
\begin{align}
\hat{\mathcal{P}}&=\sum_{\bm{n}}\cket{\bm{n}}\bra{\bm{n}}.\label{eq:explict_expression_projection_operator}
\end{align}
Therefore, it is sufficient to consider the matrix element $\bra{\bm{n}}\hat{h}_j\hat{h}_k\cket{\bm{m}}$. 

Here, we define $S_j\equiv {\rm supp}(\hat{h}_j)$ and $S_k\equiv {\rm supp}(\hat{h}_k)$, where ${\rm supp}(\cdot)$ represents the support of the operator. Suppose that $S_j\cap S_k=\phi$, which means that the regions where the operators $\hat{h}_j$ and $\hat{h}_k$ act nontrivially do not overlap. Let $S$ be the complement of $S_j\cup S_k$. The set $S_j\cup S_k\cup S$ represents all lattice sites. From these settings, we can write the direct product state $\cket{\bm{n}}$ as $\cket{\bm{n}}=|\bm{n}_{S_j},\bm{n}_{S_k},\bm{n}_{S}\rangle$, where $|\bm{n}_{S_j}\rangle$, $|\bm{n}_{S_k}\rangle$, and $|\bm{n}_{S}\rangle$ represent the product states in $S_j$, $S_k$, and $S$, respectively.
The matrix element reduces to
\begin{align}
\bra{\bm{n}}\hat{h}_j\hat{h}_k\cket{\bm{m}}&=\langle\bm{n}_{S_j}|\hat{h}_j|\bm{m}_{S_j}\rangle\langle \bm{n}_{S_k}|\hat{h}_k|\bm{m}_{S_k}\rangle\delta_{\bm{n}_{S},\bm{m}_S}.\label{eq:rewrite_matrix_element_hj_hk}
\end{align}
From the assumptions that $\mathcal{H}_P$ has a tensor product structure and $\hat{\mathcal{P}}\hat{h}_j\hat{\mathcal{P}}=0$, we obtain $\langle\bm{n}_{S_j}|\hat{h}_j|\bm{m}_{S_j}\rangle=0$. Then, the matrix element $\langle \bm{n}|\hat{h}_j\hat{h}_k\cket{\bm{m}}$ must be zero when $S_j$ and $S_k$ do not have common elements. The nonvanishing terms of $\hat{H}_{\rm p}$ appear when $S_j$ and $S_k$ have common elements. Since $\hat{h}_j$ and $\hat{h}_k$ are local operators, the nonvanishing terms are also local operators. Therefore, we prove that $\hat{H}_{\rm p}$ is a local Hamiltonian under the appropriate assumptions.

Finally, we remark on the case where the Hilbert subspace $\mathcal{H}_P$ does not have a tensor product structure. In this case, the above property does not hold in general, and the locality of the parent Hamiltonian needs to be verified individually. As an example, we consider the Onsager scar model with periodic boundary conditions. In contrast to the case of open boundary conditions, the locality does not hold. For $M=10$ and $n=4$, we consider a state
\begin{align}\cket{\uparrow_1\uparrow_2\uparrow_3\uparrow_4\downarrow_5\downarrow_6\uparrow_7\uparrow_8\uparrow_9\uparrow_{10}}.\label{eq:example_Onsager_violate_locality}
\end{align}
Here, we apply the {\it XY} interaction terms $\hat{h}_4=\hat{S}_4^+\hat{S}_{5}^-+\hat{S}_4^-\hat{S}_{5}^+$ and $\hat{h}_6=\hat{S}_6^+\hat{S}_{7}^-+\hat{S}_6^-\hat{S}_{7}^+$ to the state (\ref{eq:example_Onsager_violate_locality}). The support of these operators does not have common elements. We obtain
\begin{align}
&\hat{h}_4\hat{h}_6\cket{\uparrow_1\uparrow_2\uparrow_3\uparrow_4\downarrow_5\downarrow_6\uparrow_7\uparrow_8\uparrow_9\uparrow_{10}}\notag \\
&\quad =\cket{\uparrow_1\uparrow_2\uparrow_3\downarrow_4\uparrow_5\uparrow_6\downarrow_7\uparrow_8\uparrow_9\uparrow_{10}}\in\mathcal{H}_P.\label{eq:apply_XY_term_Onsager_PBC}
\end{align}
We find that the state obtained after applying $\hat{h}_4\hat{h}_6$ is an element of $\mathcal{H}_P$ because it contains clusters consisting of an even number of up spins, i.e., $\uparrow_8\uparrow_9\uparrow_{10}\uparrow_1\uparrow_2\uparrow_3$ and $\uparrow_5\uparrow_6$. This result means that a non-tensor -product structure (nontrivial kinetic constraint) can lead to nonlocality in the parent Hamiltonian.

As other examples, we consider the DH and DWC models, whose Hilbert subspaces do not have a tensor product structure due to the kinetic constraint. In contrast to the Onsager scar model with periodic boundary conditions, we can show that the parent Hamiltonian of these models is local by direct calculations (see Appendix~\ref{app:derivation_of_parent_Hamiltonian_DH_model}).

%%%%%%%%%%%%%%%%%%%%%%%%%%%%%%%%%%%%%%%%%%%%%%%%%%%%%%%%%%%%%%
\section{Remark on the assumption of the off-diagonal elements of the parent Hamiltonian}\label{app:remark_off-diagonal_element_parent_Hamiltonian}
%%%%%%%%%%%%%%%%%%%%%%%%%%%%%%%%%%%%%%%%%%%%%%%%%%%%%%%%%%%%%%
Here, we discuss the off-diagonal elements of the parent Hamiltonian $\hat{\mathcal{P}}_n\hat{H}_0^2\hat{\mathcal{P}}_m$ for $n\not=m$. As mentioned in Sec.~\ref{subsec:Subspace}, when $[\hat{Q}^z, \hat{H}_0]=0$ holds, the off-diagonal element must vanish. Then, we consider the case $[\hat{Q}^z, \hat{H}_0]\not=0$. The DH model and the DWC model are examples of this case. 

Here, we present a sufficient condition for the vanishing of the off-diagonal elements of the parent Hamiltonian. We assume that the following condition is satisfied:
\begin{align}
[\hat{Q}^z, \hat{H}_0]\hat{\mathcal{P}}=\Delta\hat{H}_0\hat{\mathcal{P}},\label{eq:relation_Qz_and_H0}
\end{align}
where $\Delta$ is a nonzero real value. Under this assumption, for any $\cket{\psi}\in\mathcal{H}_{P_n}$, the state $\hat{H}_0\cket{\psi}\not=0$ belongs to $\mathcal{H}_Q$ and is an eigenstate of $\hat{Q}^z$ with an eigenvalue $Q^z_0+n+\Delta$. Using this relation, we see that the off-diagonal elements of the parent Hamiltonian vanish. Direct calculations show that both the DH and DWC models satisfy Eq.~(\ref{eq:relation_Qz_and_H0}) with $\Delta=1$.

%%%%%%%%%%%%%%%%%%%%%%%%%%%%%%%%%%%%%%%%%%%%%%%%%%%%%%%%%%%%%%
\section{Magnon excitation of the ferromagnetic Heisenberg model with open boundary conditions}\label{app:Magnon_excitation_ferro_Heisenberg_model}
%%%%%%%%%%%%%%%%%%%%%%%%%%%%%%%%%%%%%%%%%%%%%%%%%%%%%%%%%%%%%%
Here, we derive the excitation spectra of magnon excitations in the one-dimensional ferromagnetic Heisenberg model with open boundary conditions. The Hamiltonian is given by
\begin{align}
\hat{H}&=J\sum_{j=1}^{M-1}\left(\frac{1}{4}-\hat{\bm{S}}_j\cdot\hat{\bm{S}}_{j+1}\right),\label{eq:Hamiltonian_ferro_Heisenberg_model_app}
\end{align}
where $J>0$ is the strength of the exchange interaction. The ground states are given by
\begin{align}
\cket{{\rm GS}_n}&\equiv (\hat{S}_{\rm tot}^+)^n\cket{\downarrow_1,\downarrow_2,\ldots,\downarrow_M},\label{eq:definition_of_ground_state_ferro_Heisenberg_model}
\end{align}
where $\hat{S}_{\rm tot}^+\equiv \sum_{j=1}^M\hat{S}_j^+$ and $n=0,1,\ldots, M$ represents the number of up spins. The ground-state energy is zero: $\hat{H}\cket{{\rm GS}_n}=0$. 

The low-lying excitations of the ferromagnetic Heisenberg model are given by magnon excitations. To derive the excitation spectra, we define the operator
\begin{align}
\hat{S}^+_{\rm tot}(l)&\equiv \sum_{j=1}^Mf_j(l)\hat{S}_j^+,\label{eq:definition_of_Stot_k}\\
f_j(l)&\equiv \cos\left[\frac{\pi l(2j-1)}{2M}\right],\quad l=0,1,\ldots, M-1.\label{eq:definition_of_fj_k_for_Heisenberg_model}
\end{align}
Direct calculations show that
\begin{align}
[\hat{H}, \hat{S}_{\rm tot}^+(l)]\cket{{\rm GS}_0}&=J\left[1-\cos\left(\frac{\pi l}{M}\right)\right]\hat{S}_{\rm tot}^+(l)\cket{{\rm GS}_0}.\label{eq:commutator_for_magnon_excitation}
\end{align}
This result implies that the one-magnon excitation energy and eigenstate are given by
\begin{align}
\epsilon(l)&\equiv J\left[1-\cos\left(\frac{\pi l}{M}\right)\right].\label{eq:one-magnon_excitation_spectrum}\\
\cket{M_1(l)}&\equiv \hat{S}_{\rm tot}^+(l)\cket{{\rm GS}_0}.\label{eq:wave_function_one-magnon_excitation}
\end{align}
Using the commutation relation $[\hat{H}, \hat{S}_{\rm tot}^+]=0$ and $[\hat{S}_{\rm tot}^+(l), \hat{S}_{\rm tot}^+]=0$, we find that the low-lying excited state in the $n$-up spin sector is 
\begin{align}
\cket{M_n(l)}&=\hat{S}_{\rm tot}^+(l)\cket{{\rm GS}_{n-1}}\notag \\
&=\hat{S}_{\rm tot}^+(l)(\hat{S}_{\rm tot}^+)^{n-1}\cket{\downarrow_1,\downarrow_2,\ldots,\downarrow_M}.\label{eq:excitation_n_sector}
\end{align}
The excitation spectrum is identical to that of the one-magnon excitation due to the commutation relations. 

Here, we discuss the EE of the state $\cket{M_n(l)}$. To evaluate the entanglement entropy, we consider the matrix product state representation of the state. To this end, we consider the matrix product operator (MPO) representation of $\hat{S}_{\rm tot}^+(l)$, which is given by
\begin{widetext}
\begin{align}
\hat{S}_{\rm tot}^+(l)&=\sum_{\bm{n},\bm{n}'}\hat{W}_0(\hat{W}_1)^{n_1n_1'}(\hat{W}_2)^{n_2n_2'}\cdots (\hat{W}_M)^{n_Mn_M'}\hat{W}_{M+1}\cket{\bm{n}}\bra{\bm{n}'},\label{eq:MPO_rep_S^+_tot}\\
\hat{W}_i&=
\begin{bmatrix}
\hat{1}_i & f_i(l)\hat{S}_i^+\\
0 & \hat{1}_i
\end{bmatrix}
,\;\text{for } 1\le i\le M,\quad \hat{W}_0=[1\quad 0], \quad \hat{W}_{M+1}=[0\quad 1]^{\rm T},\label{eq:definition_of_MPO_for_Stot_k}
\end{align}
where $\cket{\bm{n}}=\cket{n_1,n_2,\ldots,n_M}$ with $n_i=\uparrow_i, \downarrow_i$ is the product basis, $\hat{W}_i$ is an operator-valued matrix, and $(\hat{W}_i)^{n_in_i'}=\bra{n_i}\hat{W}_i\cket{n_i'}$. The MPO representation of $(\hat{S}_{\rm tot})^{n-1}$ is given by
\begin{align}
 (\hat{S}_{\rm tot}^+)^{n-1}&=\sum_{\bm{n},\bm{n'}}\hat{S}_0(\hat{S}_1)^{n_1n_1'}(\hat{S}_2)^{n_2n_2'}\cdots (\hat{S}_M)^{n_Mn_M'}\hat{S}_{M+1}\cket{\bm{n}}\bra{\bm{n}'},\label{eq:definition_of_MPO_Stot_to_n-1}\\
\hat{S}_i&=\left.
\begin{bmatrix}
\hat{1}_i & \hat{S}_i^+ & 0 & & \\
0 & \hat{1}_i & \hat{S}_i^+ & 0 & \\
   &                    & \ddots &   &  & \\
   &                    &  & & \hat{1}_i & \hat{S}_i^+ \\
   &                    & &  & 0 & \hat{1}_i
\end{bmatrix}
\right\}n
,\quad \hat{S}_0=\underbrace{[1\quad 0\;\cdots\; 0]}_n,\quad \hat{S}_{M+1}=\underbrace{[0\quad 0\;\cdots\; 0\quad 1]^{\rm T}}_{n}.\label{eq:MPO_rep_Stot_to_n}
\end{align}
\end{widetext}
This can be obtained by the technique used in Sec.~IV B of Ref.~\cite{moudgalya2018entanglement}. Because the bond dimension of the product of MPOs is given by the product of the bond dimensions of the individual MPOs, the bond dimension of the state $\cket{M_n(l)}$ is at most $2n$. This implies that the entanglement entropy of the magnon excited state obeys a subvolume law scaling. We note that the entanglement entropy for $l=0$ has been calculated in Ref.~\cite{popkov2005logarithmic}.

%%%%%%%%%%%%%%%%%%%%%%%%%%%%%%%%%%%%%%%%%%%%%%%%%%%%%%%%%%%%%%
\section{Spin-1 XY model with long-range interaction}\label{app:spin-1_XY_with_long-range}
%%%%%%%%%%%%%%%%%%%%%%%%%%%%%%%%%%%%%%%%%%%%%%%%%%%%%%%%%%%%%%

Here, we discuss the spin-1 XY model with a long-range interaction. For simplicity, we impose the periodic boundary conditions in this appendix. The XY interaction term is given by
\begin{align}
\hat{H}_{XY}^{S=1}&=\frac{1}{2}\sum_{k=1,3,\ldots, M-1}J_k\sum_{j=1}^{M}(\hat{\tau}_j^x\hat{\tau}_{j+k}^x+\hat{\tau}_j^y\hat{\tau}_{j+k}^y),\label{eq:definition_of_XY_interaction_S=1_with_long_range}\\
J_k&\equiv \frac{J}{\min(k,M-k)^{\alpha}},\label{eq:definition_of_Jn_long-range_XY}
\end{align}
where we assume that $M$ is even and $\alpha$ is real. Here, $\min(k,M-k)$ represents the distance to the $k$th neighbor sites in the ring geometry. In this case, the parent Hamiltonian after the sublattice spin rotation becomes
\begin{align}
\hat{\tilde{H}}_{\rm p}&=\sum_{k=1,3,\ldots,M-1}\frac{J_k^2}{2}\sum_{j=1}^M\left(\frac{1}{4}-\hat{\bm{S}}_j\cdot\hat{\bm{S}}_{j+k}\right).\label{eq:long-range_Heisenberg_model}
\end{align}
The one-magnon excitation energy of Eq.~(\ref{eq:long-range_Heisenberg_model}) is given by
\begin{align}
\Delta E^2_1&=\sum_{k=1,3,\ldots,M-1}\frac{J_k^2}{2}\left[1-\cos\left(\frac{2\pi l k}{M}\right)\right].\label{eq:energy_variance_of_long_range_parent_Hamiltonian_XY_general_form}
\end{align}
where $l=0,1,2,\ldots, M-1$. Here, we evaluate Eq.~(\ref{eq:energy_variance_of_long_range_parent_Hamiltonian_XY_general_form}). First, we consider the case $\alpha=0$ (all-to-all interacting case). In this case, we can obtain the one-magnon excitation energy analytically:
\begin{align}
\Delta E^2_1=\frac{J^2}{4}M.\label{eq:energy_variance_all-to-all_case}
\end{align}
This result means that the energy variance diverges in the thermodynamic limit. For the power-law decaying interacting case ($\alpha>0$), we can show that the system is gapped for $\alpha\le 1/2$ and gapless for $\alpha>1/2$. We note that similar techniques have been used in Refs.~\cite{ma2024lieb,zhou2024validity}.

Here, we consider the case $l=1$, which corresponds to the lowest excited state. We rewrite Eq.~(\ref{eq:energy_variance_of_long_range_parent_Hamiltonian_XY_general_form}) as
\begin{align}
&\Delta E^2_1\notag \\
&=
\begin{cases}
\vspace{0.5em}\displaystyle{\sum_{k=1,3,\ldots,M/2-1}\frac{J^2}{k^{2\alpha}}\left[1-\cos\left(\frac{2\pi k}{M}\right)\right],\; M=4m, }\\
\vspace{0.5em}\displaystyle{\sum_{k=1,3,\ldots,M/2}\frac{J^2}{k^{2\alpha}}\left[1-\cos\left(\frac{2\pi k}{M}\right)\right]},\; M=4m+2, 
\end{cases}
\label{eq:rewrite_energy_variance_ring_geometry}
\end{align}
where $m$ is a positive integer. We first evaluate the upper bound of Eq.~(\ref{eq:rewrite_energy_variance_ring_geometry}). We use the following inequalities:
\begin{align}
1-\cos(x)&\le \frac{x^2}{2},\label{eq:inequality_for_cos_upper_bound}\\
2\sum_{k=1,3,\ldots,L-1}\frac{1}{k^{\beta}}&\le 2+\int^{L-1}_1dx\frac{1}{x^{\beta}},
\label{eq:inequality_for_power-law_decaying_part_upper_bound}
\end{align}
where $x$ is real, $\beta \ge 0$, and $L$ is a positive even integer. From these inequalities, we obtain
\begin{align}
\Delta E_1^2&\le 
\begin{cases}
\vspace{0.5em}\max\left[O(M^{-2}), O(M^{1-2\alpha})\right],\alpha\not=3/2,\\
O(M^{-2}\ln M),\quad \alpha=3/2.
\end{cases}
\label{eq:upper_bound_long-range_spin_XY}
\end{align}
The lower bound of $\Delta E_1^2$ can be obtained by the following inequalities:
\begin{align}
\frac{2}{\pi^2}x^2&\le1-\cos(x),\quad |x|\le \pi,\label{eq:inequality_cos_lower_bound}\\
\int^{L+1}_1dx\frac{1}{x^{\beta}}&\le 2\sum_{k=1,3,\ldots,L-1}\frac{1}{k^{\beta}}.\label{eq:inequality_for_power_law_decaying_part_lower_bound}
\end{align}
From these inequalities, we obtain the lower bound of $\Delta E_1^2$ for sufficiently large $M$:
\begin{align}
\Delta E_1^2\ge 
\begin{cases}
\vspace{0.5em}\max\left[O(M^{-2}), O(M^{1-2\alpha})\right],\alpha\not=3/2,\\
O(M^{-2}\ln M),\quad \alpha=3/2.
\end{cases}
\label{eq:lower_bound_energy_variance_spin1_XY}
\end{align}
Combining Eqs.~(\ref{eq:upper_bound_long-range_spin_XY}) and (\ref{eq:lower_bound_energy_variance_spin1_XY}), we can find that $\Delta E_1^2$ diverges  in the thermodynamic limit for $\alpha\le 1/2$ and converges to $0$ in the thermodynamic limit for $\alpha>1/2$. The one-magnon energy for $\alpha>1/2$ scales as
\begin{align}
\Delta E_1^2&=
\begin{cases}
O(M^{1-2\alpha}),\quad \text{for }1/2< \alpha<3/2,\\
O(M^{-2}\ln M),\quad \text{for }\alpha=3/2,\\
O(M^{-2}),\quad \text{for }3/2<\alpha.
\end{cases}
\label{eq:scaling_energy_gap_long-range_XY}
\end{align}

%%%%%%%%%%%%%%%%%%%%%%%%%%%%%%%%%%%%%%%%%%%%%%%%%%%%%%%%%%%%%%
\section{Derivation of the parent Hamiltonian of the DH model}\label{app:derivation_of_parent_Hamiltonian_DH_model}
%%%%%%%%%%%%%%%%%%%%%%%%%%%%%%%%%%%%%%%%%%%%%%%%%%%%%%%%%%%%%%
In this appendix, we derive the parent Hamiltonian of the DH model. To this end, we rewrite $\hat{H}_{\rm DM}$ in bracket notation:
\begin{align}
\hat{H}_{\rm DM}=\frac{D}{2}\sum_{j=1}^M\hat{L}_j,\label{eq:rewrite_bracket_form_DM_interaction_term}
\end{align}
where we introduced the operator $\hat{L}_j$ as{\small
\begin{align}
\hat{L}_j&\equiv \left(\cket{\uparrow_{j-1}\uparrow_j\downarrow_{j+1}}\bra{\uparrow_{j-1}\downarrow_j\downarrow_{j+1}}+\cket{\uparrow_{j-1}\downarrow_j\downarrow_{j+1}}\bra{\uparrow_{j-1}\uparrow_j\downarrow_{j+1}}\right.\notag \\
&\left.-\cket{\downarrow_{j-1}\uparrow_j\uparrow_{j+1}}\bra{\downarrow_{j-1}\downarrow_j\uparrow_{j+1}}-\cket{\downarrow_{j-1}\downarrow_j\uparrow_{j+1}}\bra{\downarrow_{j-1}\uparrow_j\uparrow_{j+1}}\right).\label{eq:definition_of_Lj}
\end{align}
}As discussed in Sec.~\ref{subsec:DH_model}, $\cket{\uparrow_0}=0$ and $\cket{\uparrow_{M+1}}=0$. Using this operator, we calculate the parent Hamiltonian:
\begin{align}
\hat{H}_{\rm p}&=\hat{\mathcal{P}}\hat{H}_{\rm DM}^2\hat{\mathcal{P}}\notag \\
&=\frac{D^2}{4}\sum_{j=1}^M\hat{\mathcal{P}}\left[\hat{L}_j^2+(\hat{L}_j\hat{L}_{j+1}+\hat{L}_{j+1}\hat{L}_j)\right]\hat{\mathcal{P}}\notag \\
&\equiv \hat{H}_{{\rm p}1}+\hat{H}_{{\rm p}2},\label{eq:parent_Hamiltonian_calculation1_DH_model}
\end{align}
where we used the properties that $\hat{\mathcal{P}}\hat{L}_j\hat{L}_k\hat{\mathcal{P}}$ for $|j-k|\ge 2$ vanish because this term always breaks the constraint that there is no $\uparrow_j\uparrow_{j+1}$ configuration in $\mathcal{H}_{P}$. The first term of Eq.~(\ref{eq:parent_Hamiltonian_calculation1_DH_model}) becomes
\begin{align}
\hat{H}_{{\rm p}1}&=\frac{D^2}{4}\sum_{j=1}^M\hat{\mathcal{P}}(\cket{\uparrow_{j-1}\downarrow_{j}\downarrow_{j+1}}\bra{\uparrow_{j-1}\downarrow_{j}\downarrow_{j+1}}\notag \\
&\hspace{5.0em}+\cket{\downarrow_{j-1}\downarrow_{j}\uparrow_{j+1}}\bra{\downarrow_{j-1}\downarrow_{j}\uparrow_{j+1}})\hat{\mathcal{P}}\notag \\
&=\frac{D^2}{4}\sum_{j=1}^M\hat{\mathcal{P}}(\hat{P}_{j-1}'\hat{P}_j\hat{P}_{j+1}+\hat{P}_{j-1}\hat{P}_j\hat{P}_{j+1}')\hat{\mathcal{P}}\notag \\
&=\frac{D^2}{4}\sum_{j=1}^{M-1}\hat{\mathcal{P}}(\hat{P}_j'\hat{P}_{j+1}\hat{P}_{j+2}+\hat{P}_{j-1}\hat{P}_j\hat{P}_{j+1}')\hat{\mathcal{P}}\notag \\
&=\frac{D^2}{2}\sum_{j=1}^{M-1}\hat{\mathcal{P}}\hat{P}_{j-1}\left(\frac{1}{4}-\hat{S}_j^z\hat{S}_{j+1}^z\right)\hat{P}_{j+2}\hat{\mathcal{P}},\label{eq:calculation_first_term_parent_Hamiltonian_DH_model}
\end{align}
where we defined $\hat{P}_j'\equiv 1/2+\hat{S}_j^z=\cket{\uparrow_j}\bra{\uparrow_j}$ and $\hat{P}_0'=\hat{P}_{M+1}'=0$. The second term of Eq.~(\ref{eq:parent_Hamiltonian_calculation1_DH_model}) becomes
\begin{align}
&\hat{H}_{{\rm p}2}\notag \\
&=\frac{D^2}{4}\sum_{j=1}^{M-1}\hat{\mathcal{P}}(\cket{\uparrow_{j-1}\downarrow_j\downarrow_{j+1}}\bra{\uparrow_{j-1}\uparrow_{j+1}\downarrow_j}\otimes\hat{1}_{j+2}\notag \\
&\hspace{5.0em}-\cket{\downarrow_{j-1}\downarrow_j\uparrow_{j+1}}\bra{\downarrow_{j-1}\uparrow_j\uparrow_{j+1}}\otimes\hat{1}_{j+2})\notag \\
&\hspace{5.0em}\times (\hat{1}_{j-1}\otimes\cket{\uparrow_{j}\uparrow_{j+1}\downarrow_{j+2}}\bra{\uparrow_{j}\downarrow_{j+1}\downarrow_{j+2}}\notag \\
&\hspace{5.0em}-\hat{1}_{j-1}\otimes\cket{\downarrow_{j}\uparrow_{j+1}\uparrow_{j+2}}\bra{\downarrow_{j}\downarrow_{j+1}\uparrow_{j+2}})\hat{\mathcal{P}}\notag \\
&\quad +H.c.\notag \\
&=\frac{D^2}{4}\hat{\mathcal{P}}\sum_{j=1}^{M-1}(-\cket{\downarrow_{j-1}\downarrow_{j}\uparrow_{j+1}\downarrow_{j+2}}\bra{\downarrow_{j-1}\uparrow_{j}\downarrow_{j+1}\downarrow_{j+2}}\notag \\
&\hspace{6em}-\cket{\downarrow_{j-1}\uparrow_{j}\downarrow_{j+1}\downarrow_{j+2}}\bra{\downarrow_{j-1}\downarrow_{j}\uparrow_{j+1}\downarrow_{j+2}})\hat{\mathcal{P}}\notag \\
&=-\frac{D^2}{2}\sum_{j=1}^{M-1}\hat{\mathcal{P}}\hat{P}_{j-1}(\hat{S}_j^x\hat{S}_{j+1}^x+\hat{S}_j^y\hat{S}_{j+1}^y)\hat{P}_{j+2}\hat{\mathcal{P}},\label{eq:calculation_second_term_parent_Hamiltonian_DH_model}
\end{align}
where $\hat{1}_j=\hat{P}_j+\hat{P}_j'$ represents the identity operator at site $j$. Therefore, adding Eqs.~(\ref{eq:calculation_first_term_parent_Hamiltonian_DH_model}) and (\ref{eq:calculation_second_term_parent_Hamiltonian_DH_model}), we obtain the parent Hamiltonian (\ref{eq:parent_Hamiltonian_for_DH_model}). We note that the parent Hamiltonian of the DWC model can be obtained in a similar manner.

%%%%%%%%%%%%%%%%%%%%%%%%%%%%%%%%%%%%%%%%%%%%%%%%%%%%%%%%%%%%%%
\section{Perfect revival originating from the AQMBS state in the DH model}\label{app:perfect_revival_DH}
%%%%%%%%%%%%%%%%%%%%%%%%%%%%%%%%%%%%%%%%%%%%%%%%%%%%%%%%%%%%%%
Here, we show that the DH model exhibits perfect revivals originating from the AQMBS. Before discussing perfect revivals, we point out that the Hamiltonian of the DH model commutes with the soliton number operator \cite{Kodama2023,Kunimi2024}:
\begin{align}
\hat{N}_{\rm sol}\equiv \sum_{j=0}^M\left(\frac{1}{4}-\hat{S}_j^z\hat{S}_{j+1}^z\right),\label{eq:definition_of_soliton_number_operator_DH}
\end{align}
where $\hat{S}_0^{z}=\hat{S}_{M+1}^z=-1/2$. The soliton number operator takes the values $\hat{N}_{\rm sol}=0,1,\ldots, M/2$. In the subspace $\mathcal{H}_{P_n}$, the value of the soliton number operator is equal to $n$.

The perfect revival occurs in the sector with $\hat{N}_{\rm sol}=M/2$, whose dimension is $M+1$. We can write down all the states in this sector as
\begin{subequations}   
\begin{align}
\cket{1}&\equiv \cket{\downarrow_1,\uparrow_2,\downarrow_3,\uparrow_4,\ldots, \downarrow_{M-1},\uparrow_M}\in \mathcal{H}_P,\label{eq:state1_DH_revival}\\
\cket{2}&\equiv \cket{\uparrow_1,\uparrow_2,\downarrow_3,\uparrow_4,\ldots, \downarrow_{M-1},\uparrow_M}\in \mathcal{H}_Q,\label{eq:state2_DH_revival}\\
\cket{3}&\equiv \cket{\uparrow_1,\downarrow_2,\downarrow_3,\uparrow_4,\ldots, \downarrow_{M-1},\uparrow_M}\in \mathcal{H}_P,\label{eq:state3_DH_revival}\\
\cket{4}&\equiv \cket{\uparrow_1,\downarrow_2,\uparrow_3,\uparrow_4,\ldots, \downarrow_{M-1},\uparrow_M}\in \mathcal{H}_Q,\label{eq:state4_DH_revival}\\
&\;\;\vdots\notag \\
\cket{M}&\equiv \cket{\uparrow_1,\downarrow_2,\uparrow_3,\downarrow_4,\ldots, \uparrow_{M-1},\uparrow_M}\in \mathcal{H}_Q,\label{eq:stateM_DH_revival}\\
\cket{M+1}&\equiv \cket{\uparrow_1,\downarrow_2,\uparrow_3,\downarrow_4,\ldots, \uparrow_{M-1},\downarrow_M}\in \mathcal{H}_P.\label{eq:stateM+1_DH_revival}
\end{align}
\end{subequations}
Using the above basis, we can write down the matrix representation of $\hat{H}$ as{\small
\begin{align}
&H_{\rm DH}\equiv \notag \\
&
\begin{bmatrix}
0      & -D/2 & 0       &          &              &  & &  \\
-D/2 & -h     & D/2  &          &              &  & &  \\
0        & D/2 & 0      &- D/2 &              &  &  &  \\
          & 0       & -D/2&- h     & D/2      &  &  &  \\
          &          &         &          & \ddots &  &  &  \\
          &          &         &          &              & -D/2 & -h  & D/2 \\          
          &          &         &          &              & 0 & D/2 & 0 \\         
\end{bmatrix}
.\label{eq:matrix_representation_DH_model_revival}
\end{align}
}This matrix is identical to the matrix representation of the Rice-Mele model~\cite{Rice1982}. This matrix can be diagonalized analytically. For simplicity, we assume $D, h>0$. The eigenvalue equation is given by
\begin{widetext}
\begin{align}
H_{\rm DH}\bm{z}_{\pm}^{(l)}&=E_{\pm}^{(l)}\bm{z}_{\pm}^{(l)},\;l=1,2,\ldots, M/2,\quad \text{for }E_{\pm}^{(l)}\not=0,\label{eq:eiganvalue_equation_for_DH_revival}
\end{align}
\begin{subequations}   
\begin{align}   
E_{\pm}^{(l)}&\equiv \pm\sqrt{\epsilon_l^2+\frac{h^2}{4}}-\frac{h}{2}\equiv E_{\pm}^{'(l)}-\frac{h}{2},\label{eq:definition_of_Epm_l_for_DH_revival}\\
\epsilon_l&\equiv D\cos\left(\frac{\pi l}{M+2}\right),\label{eq:definition_of_epsilon_l_for_DH_revival}\\
\bm{z}_{\pm}^{(l)}&\equiv U\bm{x}_{\pm}^{(l)},\label{eq:definition_of_z_for_DH_revival}\\
U&\equiv {\rm diag}[+1,-1,-1,+1,+1,-1,\ldots],\label{eq:definition_of_U_for_DH_revival}\\
\bm{x}_+^{(l)}&\equiv [+(a^{(l)}+b^{(l)})v_1^{(l)}, (a^{(l)}-b^{(l)})v_2^{(l)},\ldots +(a^{(l)}+b^{(l)})v_{M+1}^{(l)}]^{\rm T},\label{eq:definition_of_x+_for_DH_revival}\\
\bm{x}_-^{(l)}&\equiv [-(a^{(l)}-b^{(l)})v_1^{(l)}, (a^{(l)}+b^{(l)})v_2^{(l)},\ldots -(a^{(l)}-b^{(l)})v_{M+1}^{(l)}]^{\rm T},\label{eq:definition_of_x-_for_DH_revival}\\
a^{(l)}&\equiv \sqrt{\frac{E_{+}^{'(l)}+\epsilon_l}{2E_{+}^{'(l)}}},\quad b^{(l)}\equiv \sqrt{\frac{E_{+}^{'(l)}-\epsilon_l}{2E_{+}^{'(l)}}},\label{eq:definition_of_a_and_b_for_DH_revival}\\
v_j^{(l)}&\equiv \sqrt{\frac{2}{M+2}}\sin\left(\frac{\pi l j}{M+2}\right),\label{eq:definition_of_vj_for_DH_revival}\\
H_{\rm DH}\bm{z}_0&=\bm{0},\label{eq:eigenvalue_equation_zero_for_DH_revival}\\
\bm{z}_0&\equiv \sqrt{\frac{2}{M+2}}[1,0,1,0,1,\ldots,0,1]^{\rm T}.\label{eq:definition_of_z0_for_DH_revival}
\end{align}
\end{subequations}
\end{widetext}
We note that the zero-energy eigenstate $\bm{z}_0$ corresponds to the QMBS state for $n=M/2$. From these results, we can construct the states that have nonzero amplitude only in $\mathcal{H}_P$:
\begin{align}
\bm{A}^{(l)}&\equiv \frac{1}{\sqrt{2}}\left\{[a^{(l)}+b^{(l)}]\bm{z}_+^{(l)}-[a^{(l)}-b^{(l)}]\bm{z}_-^{(l)}\right\}\notag \\
&=\sqrt{2}[v_1^{(l)}, 0, -v_3^{(l)}, 0, v_5^{(l)},0,\ldots,]^{\rm T}.\label{eq:localized_state_in_H_P}
\end{align}
The bracket form is given by
\begin{align}
&|\bm{A}^{(l)}\rangle\notag \\
&=\frac{2}{\sqrt{M+2}}\sum_{s=1}^{M/2+1}(-1)^{s+1}\sin\left[\frac{\pi l (2s-1)}{M+2}\right]\cket{2s-1}.\label{eq:bracket_form_localized_Hp_state}
\end{align}

Then, we consider the AQMBS state for the $n=M/2$ case. As shown in Sec.~\ref{subsec:DH_model}, the AQMBS states of the DH model can be obtained by the low-lying excited states of the ferromagnetic Heisenberg model with $M-n+1$ sites after mapping. For $n=M/2$, we obtain the AQMBS states as (see Appendix \ref{app:Magnon_excitation_ferro_Heisenberg_model})
\begin{align}
\cket{AS_{M/2}(l')}&=\hat{\tilde{S}}_{\rm tot}^+(l')(\hat{\tilde{S}}_{\rm tot}^+)^{M/2-1}\cket{\downarrow_1,\downarrow_2,\ldots,\downarrow_{M/2+1}},\label{eq:AQMBS_for_ferro_Heisenberg_after_mapping}\\
\hat{\tilde{S}}_{\rm tot}^+&\equiv \sum_{j=1}^{M/2+1}\hat{\tilde{S}}_j^+,\quad \hat{\tilde{S}}_{\rm tot}^-(l')\equiv \sum_{j=1}^{M/2+1}\tilde{f}_j(l')\hat{\tilde{S}}_j^+,\label{eq:definition_of_rasing_operators_after_mapping}\\
\tilde{f}_j(l')&\equiv \cos\left[\frac{\pi l'(2j-1)}{M+2}\right],\quad l'=0,1,\ldots, M/2.\label{eq:definition_of_tilde_f_k}
\end{align}
The above results are written by the variables after mapping. We need to express Eq.~(\ref{eq:AQMBS_for_ferro_Heisenberg_after_mapping}) in the original basis defined by Eqs.~(\ref{eq:state1_DH_revival})--(\ref{eq:stateM+1_DH_revival}). Because the AQMBS states belong to $\mathcal{H}_P$, Eq.~(\ref{eq:AQMBS_for_ferro_Heisenberg_after_mapping}) has a nonzero overlap with the states $\cket{j}$ for odd $j$. Taking into account the mapping rule, the AQMBS state (\ref{eq:AQMBS_for_ferro_Heisenberg_after_mapping}) can be written as{\small
\begin{align}
&\cket{AS_{M/2}(l')}\notag \\
&=\frac{-2}{\sqrt{M+2}}\sum_{s=1}^{M/2+1}\cos\left[\frac{\pi l'(2s-1)}{M+2}\right]\cket{2s-1}\notag \\
&=\frac{2}{\sqrt{M+2}}\hspace{-0.5em}\sum_{s=1}^{M/2+1}\hspace{-0.5em}(-1)^{s}\sin\left[\frac{\pi(l'+M/2+1)(2s-1)}{M+2}\right]\hspace{-0.2em}\cket{2s-1}.\label{eq:AQMBS_state_DH_model_original_basis}
\end{align}
}Comparing Eq.~(\ref{eq:AQMBS_state_DH_model_original_basis}) with Eq.~(\ref{eq:localized_state_in_H_P}), we find that $|\bm{A}^{(l)}\rangle$ is proportional to the AQMBS state when we set $l=l'+M/2+1$. Therefore, we can analytically calculate the fidelity $|\bra{AS_{M/2}(l')}e^{-i\hat{H}t/\hbar}\cket{AS_{M/2}(l')}|^2$:{\small
\begin{align}
&|\bra{AS_{M/2}(l')}e^{-i\hat{H}t/\hbar}\cket{AS_{M/2}(l')}|^2\notag \\
&=\frac{1}{2}\left\{1+\frac{h^2/4}{\epsilon_l^2+h^2/4}+\frac{\epsilon_l^2}{\epsilon_l^2+h^2/4}\cos\left[2\sqrt{\epsilon_l^2+h^2/4}t/\hbar\right]\right\}.\label{eq:Fidelty_DH_model_perfect_revial}
\end{align}
}This result implies that the fidelity becomes $1$ at the integer multiple of $T\equiv \pi \hbar/\sqrt{\epsilon_l^2+h^2/4}$. Therefore, we prove that the perfect revival occurs in the DH model starting from the AQMBS state for $n=M/2$.

%%%%%%%%%%%%%%%%%%%%%%%%%%%%%%%%%%%%%%%%%%%%%%%%%%%%%%%%%%%%%%
\section{Derivation of the parent Hamiltonian of the Onsager scar model}\label{app:derivation_of_parent_Hamiltonian_Onsager_scar_model}
%%%%%%%%%%%%%%%%%%%%%%%%%%%%%%%%%%%%%%%%%%%%%%%%%%%%%%%%%%%%%%
In this appendix, we derive the parent Hamiltonian of the Onsager scar model. To this end, we rewrite $\hat{H}_{XY}^{S=1/2}$ in bracket notation:
\begin{align}
\hat{H}_{XY}^{S=1/2}&=-J\sum_{j=1}^{M-1}\hat{L}_j,\label{eq:bracket_form_S=1/2_XY}\\
\hat{L}_j&\equiv \cket{\uparrow_j\downarrow_{j+1}}\bra{\downarrow_j\uparrow_{j+1}}+\cket{\downarrow_j\uparrow_{j+1}}\bra{\uparrow_j\downarrow_{j+1}}.\label{eq:definition_of_L_Onsager_scar}
\end{align}
The parent Hamiltonian becomes
\begin{align}
\hat{H}_{\rm p}&=J^2\hat{\mathcal{P}}\sum_{j=1}^{M-1}\hat{L}_j^2\hat{\mathcal{P}}+J^2\hat{\mathcal{P}}\sum_{j,k,j\not=k}\hat{L}_j\hat{L}_k\hat{\mathcal{P}}\notag \\
&\equiv \hat{H}_{{\rm p}1}+\hat{H}_{{\rm p}2}.\label{eq:calculation_parent_Hamiltonian_Onsager_scar}
\end{align}
The first term of Eq.~(\ref{eq:calculation_parent_Hamiltonian_Onsager_scar}) becomes
\begin{align}
\hat{H}_{{\rm p}1}&=2J^2\sum_{j=1}^{M-1}\hat{\mathcal{P}}\left(\frac{1}{4}-\hat{S}_j^z\hat{S}_{j+1}^z\right)\hat{\mathcal{P}}.\label{eq:calculation_first_term_parent_Hamiltonian_Onsager_scar}
\end{align}
The second term of Eq.~(\ref{eq:calculation_parent_Hamiltonian_Onsager_scar}) can be written as
\begin{align}
\hat{H}_{{\rm p}2}&=J^2\hat{\mathcal{P}}\left[\sum_{j=1}^{M-2}(\hat{L}_j\hat{L}_{j+1}+\hat{L}_{j+1}\hat{L}_j)\right.\notag \\
&\left.\hspace{5.0em}+2\sum_{j=1}^{M-3}\hat{L}_j\hat{L}_{j+2}+\cdots\right]\hat{\mathcal{P}},\label{eq:rewrite_Hp2_for_Onsager_scar_model}
\end{align}
where we used $[\hat{L}_j, \hat{L}_k]=0$ for $|j-k|\ge 2$. The first term of Eq.~(\ref{eq:rewrite_Hp2_for_Onsager_scar_model}) reduces to
\begin{align}
&\hat{\mathcal{P}}\sum_{j=1}^{M-2}(\hat{L}_j\hat{L}_{j+1}+\hat{L}_{j+1}\hat{L}_j)\hat{\mathcal{P}}\notag \\
&=\hat{\mathcal{P}}\sum_{j=1}^{M-2}( \cket{\uparrow_j\downarrow_{j+1}\downarrow_{j+2}}\bra{\downarrow_j\downarrow_{j+1}\uparrow_{j+2}}\notag \\
&\hspace{3.5em}+\cket{\downarrow_j\downarrow_{j+1}\uparrow_{j+2}}\bra{\uparrow_j\downarrow_{j+1}\downarrow_{j+2}}\notag \\
&\hspace{3.5em}+\cket{\downarrow_j\uparrow_{j+1}\uparrow_{j+2}}\bra{\uparrow_j\uparrow_{j+1}\downarrow_{j+2}}\notag \\
&\hspace{3.5em}+\cket{\uparrow_j\uparrow_{j+1}\downarrow_{j+2}}\bra{\downarrow_{j}\uparrow_{j+1}\uparrow_{j+2}} )\hat{\mathcal{P}}\notag \\
&=\hat{\mathcal{P}}\sum_{j=1}^{M-2}(\cket{\downarrow_j\uparrow_{j+1}\uparrow_{j+2}}\bra{\uparrow_j\uparrow_{j+1}\downarrow_{j+2}}\notag \\
&\hspace{3.5em}+\cket{\uparrow_j\uparrow_{j+1}\downarrow_{j+2}}\bra{\downarrow_{j}\uparrow_{j+1}\uparrow_{j+2}})\hat{\mathcal{P}}\notag \\
&=2\hat{\mathcal{P}}\sum_{j=1}^{M-2}(\hat{S}_j^x\hat{P}'_{j+1}\hat{S}_{j+2}^x+\hat{S}_j^y\hat{P}'_{j+1}\hat{S}_{j+2}^y)\hat{\mathcal{P}}.\label{eq:calculation_Lj_Lj+1_Onsager_scar}
\end{align}
Here, we used the fact that $\hat{\mathcal{P}}\cket{\uparrow_j\downarrow_{j+1}\downarrow_{j+2}}\bra{\downarrow_j\downarrow_{j+1}\uparrow_{j+2}}\hat{\mathcal{P}}=0$ and $\hat{\mathcal{P}}\cket{\downarrow_j\downarrow_{j+1}\uparrow_{j+2}}\bra{\uparrow_j\downarrow_{j+1}\downarrow_{j+2}}\hat{\mathcal{P}}=0$ because these terms create a cluster of an odd number of up spins. For the same reason,  the terms $\hat{\mathcal{P}}\hat{L}_j\hat{L}_{j+k}\hat{\mathcal{P}}\;(k\ge 2)$ are zero. Therefore, we obtain the parent Hamiltonian of the Onsager scar model (\ref{eq:parent_Hamiltonian_Onsager_scar}).

%%%%%%%%%%%%%%%%%%%%%%%%%%%%%%%%%%%%%%%%%%%%%%%%%%%%%%%%%%%%%%
\section{Derivation of the parent Hamiltonian of the nonmaximal spin scar model}\label{app:derivation_of_parent_Hamiltonian_nonmaximal_spin_scar_model}
%%%%%%%%%%%%%%%%%%%%%%%%%%%%%%%%%%%%%%%%%%%%%%%%%%%%%%%%%%%%%%
In this appendix, we derive the parent Hamiltonian of the nonmaximal spin scar model. To this end, we rewrite $\hat{H}_{{\rm DM}0}$ in bracket notation:
\begin{align}
\hat{H}_{{\rm DM}0}&=iD_z\sum_{j=2}^{M-1}\hat{L}_j,\label{eq:bracket_form_HDM0_nonmaximal_spin_scar}
\end{align}
where we defined 
\begin{align}
\hat{L}_j&\equiv \cket{+_{j-1}0_j-_{j+1}}\bra{0_{j-1}0_j0_{j+1}}\notag \\
&+\cket{0_{j-1}0_j0_{j+1}}\bra{-_{j-1}0_{j}+_{j+1}}\notag \\
&+\cket{+_{j-1}0_j0_{j+1}}\bra{0_{j-1}0_j+_{j+1}}\notag \\
&+\cket{0_{j-1}0_j-_{j+1}}\bra{-_{j-1}0_j0_{j+1}}\notag \\
&-\cket{0_{j-1}0_j0_{j+1}}\bra{+_{j-1}0_j-_{j+1}}\notag \\
&-\cket{-_{j-1}0_j+_{j+1}}\bra{0_{j-1}0_j0_{j+1}}\notag \\
&-\cket{0_{j-1}0_j+_{j+1}}\bra{+_{j-1}0_j0_{j+1}}\notag \\
&-\cket{-_{j-1}0_j0_{j+1}}\bra{0_{j-1}0_j-_{j+1}}.\label{eq:definition_of_Lj_nonmaximal_spin_scar}
\end{align}
The parent Hamiltonian becomes
\begin{align}
\hat{H}_{\rm p}&=-D^2_z\hat{\mathcal{P}}\sum_{j=2}^{M-1}\hat{L}_j^2\hat{\mathcal{P}}-D^2_z\sum_{j,k,j\not=k}\hat{\mathcal{P}}\hat{L}_j\hat{L}_{k}\hat{\mathcal{P}}\notag \\
&\equiv \hat{H}_{{\rm p}1}+\hat{H}_{{\rm p}2}.\label{eq:calculation_parent_Hamiltonian_nomaximal_spin_scar}
\end{align}
The first term of Eq.~(\ref{eq:calculation_parent_Hamiltonian_nomaximal_spin_scar}) becomes
\begin{align}
\hat{H}_{{\rm p}1}&=-D^2_z\sum_{j=2}^{M-1}\hat{\mathcal{P}}(\cket{+_{j-1}0_j-_{j+1}}\bra{-_{j-1}0_j+_{j+1}}\notag \\
&\hspace{5.5em}+\cket{-_{j-1}0_j+_{j+1}}\bra{+_{j-1}0_j-_{j+1}}\notag \\
&\hspace{5.5em}-\cket{+_{j-1}0_j-_{j+1}}\bra{+_{j-1}0_j-_{j+1}}\notag \\
&\hspace{5.5em}-\cket{-_{j-1}0_j+_{j+1}}\bra{-_{j-1}0_j+_{j+1}})\hat{\mathcal{P}}\notag \\
&=D_z^2\hat{\mathcal{P}}\sum_{j=2}^{M-1}[1-(\hat{\tau}_j^z)^2]\notag \\
&\quad \times\left[-\frac{1}{4}(\hat{\tau}_{j-1}^+)^2(\hat{\tau}_{j+1}^-)^2-\frac{1}{4}(\hat{\tau}_{j-1}^-)^2(\hat{\tau}_{j+1}^+)^2\right.\notag \\
&\quad \left. \quad \quad +\frac{1}{2}(\hat{\tau}_{j-1}^z)^2(\hat{\tau}_{j+1}^z)^2-\frac{1}{2}\hat{\tau}_{j-1}^z\hat{\tau}_{j+1}^z\right]\hat{\mathcal{P}}.\label{eq:calculation_first_term_parent_Hamiltonian_nonmaximal_spin_scar}
\end{align}
The other terms in Eq.~(\ref{eq:calculation_parent_Hamiltonian_nomaximal_spin_scar}) become zero because of the projection operator. Therefore, we obtain the parent Hamiltonian of the nonmaximal spin scar model (\ref{eq:parent_Hamiltonian_nonmaximal_spin}).

%%%%%%%%%%%%%%%%%%%%%%%%%%%%%%%%%%%%%%%%%%%%%%%%%%%%%%%%%%%%%%
\section{Derivation of the expression of the AQMBS state for the nonmaximal spin scar model}\label{app:derivation_of_AQMBS_nonmaximal_spin_scar_model}
%%%%%%%%%%%%%%%%%%%%%%%%%%%%%%%%%%%%%%%%%%%%%%%%%%%%%%%%%%%%%%
Here, we discuss the expression of $\cket{AS_n}$ for the nonmaximal spin scar model. As shown in Sec.~\ref{subsec:Nonmaximal_spins}, the parent Hamiltonian of the nonmaximal spin scar model is given by the ferromagnetic Heisenberg model on the two independent ladders. A magnon excited state in the even-parity sector is given by
\begin{align}
\cket{AS_n}&=\frac{1}{\sqrt{2}}(\cket{AS_n^{\rm odd}}+\cket{AS_n^{\rm even}}).\label{eq:re_expression_AS_n_for_nonmaximal_spin_scar_app}
\end{align}
The expressions of the states $\cket{AS_n^{\rm odd}}$ and $\cket{AS_n^{\rm even}}$ can be obtained using Eq.~(\ref{eq:excitation_n_sector}) and the mapping rule (\ref{eq:mapping_upper_ladder}) and (\ref{eq:mapping_lower_ladder}). 

We have derived the matrix product presentation of the magnon excited state of the ferromagnetic Heisenberg chain (see Appendix~\ref{app:Magnon_excitation_ferro_Heisenberg_model}). For convenience, we define the matrix product representation of the state $\cket{M_n(l)}$ for the site number $M/2$ as
\begin{align}
\cket{M_n(l)}&=\sum_{n_1,n_2,\ldots,n_{M/2}}u\mathcal{M}^{n_1}\mathcal{M}^{n_2}\cdots\mathcal{M}^{n_{M/2}}v,\label{eq:definition_of_MPS_representation_n_magnon_for_nms}\\
u&\equiv \underbrace{
\begin{bmatrix}
1 & 0 & \cdots & 0
\end{bmatrix}
}_{2n},\quad v\equiv \underbrace{
\begin{bmatrix}
0 & 0 & \cdots & 0 & 1
\end{bmatrix}^{\rm T}}_{2n},\label{eq:definition_of_boundary_vector_Mn}
\end{align}
where $\mathcal{M}^{n_i}$ is the MPS whose bond dimension is $2n$ and $n_i=\uparrow_i \text{or }\downarrow_i$. Since the state $\cket{M_n(l)}$ is a spin-1/2 state, we need to extend the definition of the MPS from spin-1/2 to spin-1. To this end, we assign $\mathcal{M}^{n_i=0}=0$ and $\mathcal{M}^{n_i=+_i, -_i}=\mathcal{M}^{n_i=\uparrow_i, \downarrow_i}$. Although we do not write down the explicit expression of $\mathcal{M}^{n_i}$, it can be obtained using Eqs.~(\ref{eq:definition_of_MPO_for_Stot_k}) and (\ref{eq:MPO_rep_Stot_to_n}). 
From this expression and the mapping rules, the wave function before mapping can be obtained by inserting $0$s at the odd bonds or even bonds. This can be easily achieved by the MPS representation by inserting an MPS to $\cket{M_n(l)}$. We obtain
\begin{align}
\cket{AS_n^{\rm odd}}&\propto \sum_{\bm{n}}u\mathcal{Z}^{n_1}\tilde{\mathcal{M}}^{n_2}\mathcal{Z}^{n_3}\tilde{\mathcal{M}}^{n_4}\cdots \mathcal{Z}^{n_{M-1}}\tilde{\mathcal{M}}^{n_M}v,\\
\cket{AS_n^{\rm even}}&\propto \sum_{\bm{n}}u\tilde{\mathcal{M}}^{n_1}\mathcal{Z}^{n_2}\tilde{\mathcal{M}}^{n_3}\mathcal{Z}^{n_4}\cdots \tilde{\mathcal{M}}^{n_{M-1}}\mathcal{Z}^{n_{M}}v,\\
\tilde{\mathcal{M}}^{n_{2j-1}}&=\tilde{\mathcal{M}}^{n_{2j}}\equiv \mathcal{M}^{n_j},\label{eq:definition_of_tilde_M_nms}\\
\mathcal{Z}^{n_i}&\equiv \delta_{n_i,0}{\rm diag}\underbrace{(\cket{0_i}, \cket{0_i},\ldots, \cket{0_i})}_{2n}.\label{eq:definition_of_Zn_MPS}
\end{align}
From these expressions, we find that the bond dimensions of these states are still $2n$. Therefore, we prove that the bond dimension of the state $\cket{AS_n}$ is at most $4n$.

%%%%%%%%%%%%%%%%%%%%%%%%%%%%%%%%%%%%%%%%%%%%%%%%%%%%%%%%%%%%%%

%\end{widetext}
%\clearpage
%\bibliographystyle{apsrev4-2}
\bibliography{references}

\begin{thebibliography}{113}%
\makeatletter
\providecommand \@ifxundefined [1]{%
 \@ifx{#1\undefined}
}%
\providecommand \@ifnum [1]{%
 \ifnum #1\expandafter \@firstoftwo
 \else \expandafter \@secondoftwo
 \fi
}%
\providecommand \@ifx [1]{%
 \ifx #1\expandafter \@firstoftwo
 \else \expandafter \@secondoftwo
 \fi
}%
\providecommand \natexlab [1]{#1}%
\providecommand \enquote  [1]{``#1''}%
\providecommand \bibnamefont  [1]{#1}%
\providecommand \bibfnamefont [1]{#1}%
\providecommand \citenamefont [1]{#1}%
\providecommand \href@noop [0]{\@secondoftwo}%
\providecommand \href [0]{\begingroup \@sanitize@url \@href}%
\providecommand \@href[1]{\@@startlink{#1}\@@href}%
\providecommand \@@href[1]{\endgroup#1\@@endlink}%
\providecommand \@sanitize@url [0]{\catcode `\\12\catcode `\$12\catcode
  `\&12\catcode `\#12\catcode `\^12\catcode `\_12\catcode `\%12\relax}%
\providecommand \@@startlink[1]{}%
\providecommand \@@endlink[0]{}%
\providecommand \url  [0]{\begingroup\@sanitize@url \@url }%
\providecommand \@url [1]{\endgroup\@href {#1}{\urlprefix }}%
\providecommand \urlprefix  [0]{URL }%
\providecommand \Eprint [0]{\href }%
\providecommand \doibase [0]{https://doi.org/}%
\providecommand \selectlanguage [0]{\@gobble}%
\providecommand \bibinfo  [0]{\@secondoftwo}%
\providecommand \bibfield  [0]{\@secondoftwo}%
\providecommand \translation [1]{[#1]}%
\providecommand \BibitemOpen [0]{}%
\providecommand \bibitemStop [0]{}%
\providecommand \bibitemNoStop [0]{.\EOS\space}%
\providecommand \EOS [0]{\spacefactor3000\relax}%
\providecommand \BibitemShut  [1]{\csname bibitem#1\endcsname}%
\let\auto@bib@innerbib\@empty
%</preamble>
\bibitem [{\citenamefont {Deutsch}(1991)}]{Deutsch1991}%
  \BibitemOpen
  \bibfield  {author} {\bibinfo {author} {\bibfnamefont {J.~M.}\ \bibnamefont
  {Deutsch}},\ }\bibfield  {title} {\bibinfo {title} {Quantum statistical
  mechanics in a closed system},\ }\href@noop {} {\bibfield  {journal}
  {\bibinfo  {journal} {Phys. Rev. A}\ }\textbf {\bibinfo {volume} {43}},\
  \bibinfo {pages} {2046} (\bibinfo {year} {1991})}\BibitemShut {NoStop}%
\bibitem [{\citenamefont {Srednicki}(1994)}]{Srednicki1994}%
  \BibitemOpen
  \bibfield  {author} {\bibinfo {author} {\bibfnamefont {M.}~\bibnamefont
  {Srednicki}},\ }\bibfield  {title} {\bibinfo {title} {Chaos and quantum
  thermalization},\ }\href@noop {} {\bibfield  {journal} {\bibinfo  {journal}
  {Phys. Rev. E}\ }\textbf {\bibinfo {volume} {50}},\ \bibinfo {pages} {888}
  (\bibinfo {year} {1994})}\BibitemShut {NoStop}%
\bibitem [{\citenamefont {Rigol}\ \emph {et~al.}(2008)\citenamefont {Rigol},
  \citenamefont {Dunjko},\ and\ \citenamefont {Olshanii}}]{Rigol2008}%
  \BibitemOpen
  \bibfield  {author} {\bibinfo {author} {\bibfnamefont {M.}~\bibnamefont
  {Rigol}}, \bibinfo {author} {\bibfnamefont {V.}~\bibnamefont {Dunjko}},\ and\
  \bibinfo {author} {\bibfnamefont {M.}~\bibnamefont {Olshanii}},\ }\bibfield
  {title} {\bibinfo {title} {Thermalization and its mechanism for generic
  isolated quantum systems},\ }\href@noop {} {\bibfield  {journal} {\bibinfo
  {journal} {Nature}\ }\textbf {\bibinfo {volume} {452}},\ \bibinfo {pages}
  {854} (\bibinfo {year} {2008})}\BibitemShut {NoStop}%
\bibitem [{\citenamefont {D'Alessio}\ \emph {et~al.}(2016)\citenamefont
  {D'Alessio}, \citenamefont {Kafri}, \citenamefont {Polkovnikov},\ and\
  \citenamefont {Rigol}}]{DAlessio2016}%
  \BibitemOpen
  \bibfield  {author} {\bibinfo {author} {\bibfnamefont {L.}~\bibnamefont
  {D'Alessio}}, \bibinfo {author} {\bibfnamefont {Y.}~\bibnamefont {Kafri}},
  \bibinfo {author} {\bibfnamefont {A.}~\bibnamefont {Polkovnikov}},\ and\
  \bibinfo {author} {\bibfnamefont {M.}~\bibnamefont {Rigol}},\ }\bibfield
  {title} {\bibinfo {title} {From quantum chaos and eigenstate thermalization
  to statistical mechanics and thermodynamics},\ }\href@noop {} {\bibfield
  {journal} {\bibinfo  {journal} {Adv. Phys.}\ }\textbf {\bibinfo {volume}
  {65}},\ \bibinfo {pages} {239} (\bibinfo {year} {2016})}\BibitemShut
  {NoStop}%
\bibitem [{\citenamefont {Mori}\ \emph {et~al.}(2018)\citenamefont {Mori},
  \citenamefont {Ikeda}, \citenamefont {Kaminishi},\ and\ \citenamefont
  {Ueda}}]{Mori2018}%
  \BibitemOpen
  \bibfield  {author} {\bibinfo {author} {\bibfnamefont {T.}~\bibnamefont
  {Mori}}, \bibinfo {author} {\bibfnamefont {T.~N.}\ \bibnamefont {Ikeda}},
  \bibinfo {author} {\bibfnamefont {E.}~\bibnamefont {Kaminishi}},\ and\
  \bibinfo {author} {\bibfnamefont {M.}~\bibnamefont {Ueda}},\ }\bibfield
  {title} {\bibinfo {title} {Thermalization and prethermalization in isolated
  quantum systems: a theoretical overview},\ }\href@noop {} {\bibfield
  {journal} {\bibinfo  {journal} {J. Phys. B: At. Mol. Opt. Phys.}\ }\textbf
  {\bibinfo {volume} {51}},\ \bibinfo {pages} {112001} (\bibinfo {year}
  {2018})}\BibitemShut {NoStop}%
\bibitem [{\citenamefont {Gornyi}\ \emph {et~al.}(2005)\citenamefont {Gornyi},
  \citenamefont {Mirlin},\ and\ \citenamefont {Polyakov}}]{Gornyi2005}%
  \BibitemOpen
  \bibfield  {author} {\bibinfo {author} {\bibfnamefont {I.~V.}\ \bibnamefont
  {Gornyi}}, \bibinfo {author} {\bibfnamefont {A.~D.}\ \bibnamefont {Mirlin}},\
  and\ \bibinfo {author} {\bibfnamefont {D.~G.}\ \bibnamefont {Polyakov}},\
  }\bibfield  {title} {\bibinfo {title} {Interacting electrons in disordered
  wires: Anderson localization and {low-T} transport},\ }\href@noop {}
  {\bibfield  {journal} {\bibinfo  {journal} {Phys. Rev. Lett.}\ }\textbf
  {\bibinfo {volume} {95}},\ \bibinfo {pages} {206603} (\bibinfo {year}
  {2005})}\BibitemShut {NoStop}%
\bibitem [{\citenamefont {Basko}\ \emph {et~al.}(2007)\citenamefont {Basko},
  \citenamefont {Aleiner},\ and\ \citenamefont {Altshuler}}]{Basko2007}%
  \BibitemOpen
  \bibfield  {author} {\bibinfo {author} {\bibfnamefont {D.~M.}\ \bibnamefont
  {Basko}}, \bibinfo {author} {\bibfnamefont {I.~L.}\ \bibnamefont {Aleiner}},\
  and\ \bibinfo {author} {\bibfnamefont {B.~L.}\ \bibnamefont {Altshuler}},\
  }\bibfield  {title} {\bibinfo {title} {Possible experimental manifestations
  of the many-body localization},\ }\href@noop {} {\bibfield  {journal}
  {\bibinfo  {journal} {Phys. Rev. B}\ }\textbf {\bibinfo {volume} {76}},\
  \bibinfo {pages} {052203} (\bibinfo {year} {2007})}\BibitemShut {NoStop}%
\bibitem [{\citenamefont {Huse}\ \emph {et~al.}(2014)\citenamefont {Huse},
  \citenamefont {Nandkishore},\ and\ \citenamefont {Oganesyan}}]{Huse2014}%
  \BibitemOpen
  \bibfield  {author} {\bibinfo {author} {\bibfnamefont {D.~A.}\ \bibnamefont
  {Huse}}, \bibinfo {author} {\bibfnamefont {R.}~\bibnamefont {Nandkishore}},\
  and\ \bibinfo {author} {\bibfnamefont {V.}~\bibnamefont {Oganesyan}},\
  }\bibfield  {title} {\bibinfo {title} {Phenomenology of fully
  many-body-localized systems},\ }\href@noop {} {\bibfield  {journal} {\bibinfo
   {journal} {Phys. Rev. B}\ }\textbf {\bibinfo {volume} {90}},\ \bibinfo
  {pages} {174202} (\bibinfo {year} {2014})}\BibitemShut {NoStop}%
\bibitem [{\citenamefont {Bernien}\ \emph {et~al.}(2017)\citenamefont
  {Bernien}, \citenamefont {Schwartz}, \citenamefont {Keesling}, \citenamefont
  {Levine}, \citenamefont {Omran}, \citenamefont {Pichler}, \citenamefont
  {Choi}, \citenamefont {Zibrov}, \citenamefont {Endres}, \citenamefont
  {Greiner}, \citenamefont {Vuleti\'c},\ and\ \citenamefont
  {Lukin}}]{Bernien2017}%
  \BibitemOpen
  \bibfield  {author} {\bibinfo {author} {\bibfnamefont {H.}~\bibnamefont
  {Bernien}}, \bibinfo {author} {\bibfnamefont {S.}~\bibnamefont {Schwartz}},
  \bibinfo {author} {\bibfnamefont {A.}~\bibnamefont {Keesling}}, \bibinfo
  {author} {\bibfnamefont {H.}~\bibnamefont {Levine}}, \bibinfo {author}
  {\bibfnamefont {A.}~\bibnamefont {Omran}}, \bibinfo {author} {\bibfnamefont
  {H.}~\bibnamefont {Pichler}}, \bibinfo {author} {\bibfnamefont
  {S.}~\bibnamefont {Choi}}, \bibinfo {author} {\bibfnamefont {A.~S.}\
  \bibnamefont {Zibrov}}, \bibinfo {author} {\bibfnamefont {M.}~\bibnamefont
  {Endres}}, \bibinfo {author} {\bibfnamefont {M.}~\bibnamefont {Greiner}},
  \bibinfo {author} {\bibfnamefont {V.}~\bibnamefont {Vuleti\'c}},\ and\
  \bibinfo {author} {\bibfnamefont {M.~D.}\ \bibnamefont {Lukin}},\ }\bibfield
  {title} {\bibinfo {title} {Probing many-body dynamics on a 51-atom quantum
  simulator},\ }\href@noop {} {\bibfield  {journal} {\bibinfo  {journal}
  {Nature}\ }\textbf {\bibinfo {volume} {551}},\ \bibinfo {pages} {579}
  (\bibinfo {year} {2017})}\BibitemShut {NoStop}%
\bibitem [{\citenamefont {Turner}\ \emph
  {et~al.}(2018{\natexlab{a}})\citenamefont {Turner}, \citenamefont
  {Michailidis}, \citenamefont {Abanin}, \citenamefont {Serbyn},\ and\
  \citenamefont {Papi{\'c}}}]{Turner2018}%
  \BibitemOpen
  \bibfield  {author} {\bibinfo {author} {\bibfnamefont {C.~J.}\ \bibnamefont
  {Turner}}, \bibinfo {author} {\bibfnamefont {A.~A.}\ \bibnamefont
  {Michailidis}}, \bibinfo {author} {\bibfnamefont {D.~A.}\ \bibnamefont
  {Abanin}}, \bibinfo {author} {\bibfnamefont {M.}~\bibnamefont {Serbyn}},\
  and\ \bibinfo {author} {\bibfnamefont {Z.}~\bibnamefont {Papi{\'c}}},\
  }\bibfield  {title} {\bibinfo {title} {Weak ergodicity breaking from quantum
  many-body scars},\ }\href@noop {} {\bibfield  {journal} {\bibinfo  {journal}
  {Nat. Phys.}\ }\textbf {\bibinfo {volume} {14}},\ \bibinfo {pages} {745}
  (\bibinfo {year} {2018}{\natexlab{a}})}\BibitemShut {NoStop}%
\bibitem [{\citenamefont {Turner}\ \emph
  {et~al.}(2018{\natexlab{b}})\citenamefont {Turner}, \citenamefont
  {Michailidis}, \citenamefont {Abanin}, \citenamefont {Serbyn},\ and\
  \citenamefont {Papi{\'c}}}]{Turner2018prb}%
  \BibitemOpen
  \bibfield  {author} {\bibinfo {author} {\bibfnamefont {C.}~\bibnamefont
  {Turner}}, \bibinfo {author} {\bibfnamefont {A.}~\bibnamefont {Michailidis}},
  \bibinfo {author} {\bibfnamefont {D.}~\bibnamefont {Abanin}}, \bibinfo
  {author} {\bibfnamefont {M.}~\bibnamefont {Serbyn}},\ and\ \bibinfo {author}
  {\bibfnamefont {Z.}~\bibnamefont {Papi{\'c}}},\ }\bibfield  {title} {\bibinfo
  {title} {Quantum scarred eigenstates in a {Rydberg} atom chain: Entanglement,
  breakdown of thermalization, and stability to perturbations},\ }\href@noop {}
  {\bibfield  {journal} {\bibinfo  {journal} {Phys. Rev. B}\ }\textbf {\bibinfo
  {volume} {98}},\ \bibinfo {pages} {155134} (\bibinfo {year}
  {2018}{\natexlab{b}})}\BibitemShut {NoStop}%
\bibitem [{\citenamefont {Sala}\ \emph {et~al.}(2020)\citenamefont {Sala},
  \citenamefont {Rakovszky}, \citenamefont {Verresen}, \citenamefont {Knap},\
  and\ \citenamefont {Pollmann}}]{Sala2020}%
  \BibitemOpen
  \bibfield  {author} {\bibinfo {author} {\bibfnamefont {P.}~\bibnamefont
  {Sala}}, \bibinfo {author} {\bibfnamefont {T.}~\bibnamefont {Rakovszky}},
  \bibinfo {author} {\bibfnamefont {R.}~\bibnamefont {Verresen}}, \bibinfo
  {author} {\bibfnamefont {M.}~\bibnamefont {Knap}},\ and\ \bibinfo {author}
  {\bibfnamefont {F.}~\bibnamefont {Pollmann}},\ }\bibfield  {title} {\bibinfo
  {title} {Ergodicity breaking arising from {Hilbert} space fragmentation in
  dipole-conserving {Hamiltonians}},\ }\href@noop {} {\bibfield  {journal}
  {\bibinfo  {journal} {Physs Rev. X}\ }\textbf {\bibinfo {volume} {10}},\
  \bibinfo {pages} {011047} (\bibinfo {year} {2020})}\BibitemShut {NoStop}%
\bibitem [{\citenamefont {Khemani}\ \emph {et~al.}(2020)\citenamefont
  {Khemani}, \citenamefont {Hermele},\ and\ \citenamefont
  {Nandkishore}}]{Khemani2020}%
  \BibitemOpen
  \bibfield  {author} {\bibinfo {author} {\bibfnamefont {V.}~\bibnamefont
  {Khemani}}, \bibinfo {author} {\bibfnamefont {M.}~\bibnamefont {Hermele}},\
  and\ \bibinfo {author} {\bibfnamefont {R.}~\bibnamefont {Nandkishore}},\
  }\bibfield  {title} {\bibinfo {title} {Localization from {Hilbert} space
  shattering: From theory to physical realizations},\ }\href@noop {} {\bibfield
   {journal} {\bibinfo  {journal} {Phys. Rev. B}\ }\textbf {\bibinfo {volume}
  {101}},\ \bibinfo {pages} {174204} (\bibinfo {year} {2020})}\BibitemShut
  {NoStop}%
\bibitem [{\citenamefont {Serbyn}\ \emph {et~al.}(2021)\citenamefont {Serbyn},
  \citenamefont {Abanin},\ and\ \citenamefont {Papi{\'c}}}]{Serbyn2021}%
  \BibitemOpen
  \bibfield  {author} {\bibinfo {author} {\bibfnamefont {M.}~\bibnamefont
  {Serbyn}}, \bibinfo {author} {\bibfnamefont {D.~A.}\ \bibnamefont {Abanin}},\
  and\ \bibinfo {author} {\bibfnamefont {Z.}~\bibnamefont {Papi{\'c}}},\
  }\bibfield  {title} {\bibinfo {title} {Quantum many-body scars and weak
  breaking of ergodicity},\ }\href@noop {} {\bibfield  {journal} {\bibinfo
  {journal} {Nat. Phys.}\ }\textbf {\bibinfo {volume} {17}},\ \bibinfo {pages}
  {675} (\bibinfo {year} {2021})}\BibitemShut {NoStop}%
\bibitem [{\citenamefont {Papi{\'c}}(2022)}]{Papic2022}%
  \BibitemOpen
  \bibfield  {author} {\bibinfo {author} {\bibfnamefont {Z.}~\bibnamefont
  {Papi{\'c}}},\ }\bibfield  {title} {\bibinfo {title} {Weak ergodicity
  breaking through the lens of quantum entanglement},\ }in\ \href@noop {}
  {\emph {\bibinfo {booktitle} {Entanglement in Spin Chains: From Theory to
  Quantum Technology Applications}}},\ \bibinfo {editor} {edited by\ \bibinfo
  {editor} {\bibfnamefont {S.~B.}\ \bibnamefont {A.~Bayat}}\ and\ \bibinfo
  {editor} {\bibfnamefont {H.}~\bibnamefont {Johannesson}}}\ (\bibinfo
  {publisher} {Springer},\ \bibinfo {address} {Cham},\ \bibinfo {year} {2022})\
  pp.\ \bibinfo {pages} {341--395}\BibitemShut {NoStop}%
\bibitem [{\citenamefont {Moudgalya}\ \emph {et~al.}(2022)\citenamefont
  {Moudgalya}, \citenamefont {Bernevig},\ and\ \citenamefont
  {Regnault}}]{Moudgalya2022quantum}%
  \BibitemOpen
  \bibfield  {author} {\bibinfo {author} {\bibfnamefont {S.}~\bibnamefont
  {Moudgalya}}, \bibinfo {author} {\bibfnamefont {B.~A.}\ \bibnamefont
  {Bernevig}},\ and\ \bibinfo {author} {\bibfnamefont {N.}~\bibnamefont
  {Regnault}},\ }\bibfield  {title} {\bibinfo {title} {Quantum many-body scars
  and {Hilbert} space fragmentation: a review of exact results},\ }\href@noop
  {} {\bibfield  {journal} {\bibinfo  {journal} {Rep. Prog. Phys.}\ }\textbf
  {\bibinfo {volume} {85}},\ \bibinfo {pages} {086501} (\bibinfo {year}
  {2022})}\BibitemShut {NoStop}%
\bibitem [{\citenamefont {Chandran}\ \emph {et~al.}(2023)\citenamefont
  {Chandran}, \citenamefont {Iadecola}, \citenamefont {Khemani},\ and\
  \citenamefont {Moessner}}]{Chandran2023quantum}%
  \BibitemOpen
  \bibfield  {author} {\bibinfo {author} {\bibfnamefont {A.}~\bibnamefont
  {Chandran}}, \bibinfo {author} {\bibfnamefont {T.}~\bibnamefont {Iadecola}},
  \bibinfo {author} {\bibfnamefont {V.}~\bibnamefont {Khemani}},\ and\ \bibinfo
  {author} {\bibfnamefont {R.}~\bibnamefont {Moessner}},\ }\bibfield  {title}
  {\bibinfo {title} {Quantum many-body scars: A quasiparticle perspective},\
  }\href@noop {} {\bibfield  {journal} {\bibinfo  {journal} {Annu. Rev.
  Condens. Matter Phys.}\ }\textbf {\bibinfo {volume} {14}},\ \bibinfo {pages}
  {443} (\bibinfo {year} {2023})}\BibitemShut {NoStop}%
\bibitem [{\citenamefont {Choi}\ \emph {et~al.}(2019)\citenamefont {Choi},
  \citenamefont {Turner}, \citenamefont {Pichler}, \citenamefont {Ho},
  \citenamefont {Michailidis}, \citenamefont {Papi{\'c}}, \citenamefont
  {Serbyn}, \citenamefont {Lukin},\ and\ \citenamefont {Abanin}}]{Choi2019}%
  \BibitemOpen
  \bibfield  {author} {\bibinfo {author} {\bibfnamefont {S.}~\bibnamefont
  {Choi}}, \bibinfo {author} {\bibfnamefont {C.~J.}\ \bibnamefont {Turner}},
  \bibinfo {author} {\bibfnamefont {H.}~\bibnamefont {Pichler}}, \bibinfo
  {author} {\bibfnamefont {W.~W.}\ \bibnamefont {Ho}}, \bibinfo {author}
  {\bibfnamefont {A.~A.}\ \bibnamefont {Michailidis}}, \bibinfo {author}
  {\bibfnamefont {Z.}~\bibnamefont {Papi{\'c}}}, \bibinfo {author}
  {\bibfnamefont {M.}~\bibnamefont {Serbyn}}, \bibinfo {author} {\bibfnamefont
  {M.~D.}\ \bibnamefont {Lukin}},\ and\ \bibinfo {author} {\bibfnamefont
  {D.~A.}\ \bibnamefont {Abanin}},\ }\bibfield  {title} {\bibinfo {title}
  {Emergent {SU(2)} dynamics and perfect quantum many-body scars},\ }\href@noop
  {} {\bibfield  {journal} {\bibinfo  {journal} {Phys. Rev. Lett.}\ }\textbf
  {\bibinfo {volume} {122}},\ \bibinfo {pages} {220603} (\bibinfo {year}
  {2019})}\BibitemShut {NoStop}%
\bibitem [{\citenamefont {Ho}\ \emph {et~al.}(2019)\citenamefont {Ho},
  \citenamefont {Choi}, \citenamefont {Pichler},\ and\ \citenamefont
  {Lukin}}]{Ho2019}%
  \BibitemOpen
  \bibfield  {author} {\bibinfo {author} {\bibfnamefont {W.~W.}\ \bibnamefont
  {Ho}}, \bibinfo {author} {\bibfnamefont {S.}~\bibnamefont {Choi}}, \bibinfo
  {author} {\bibfnamefont {H.}~\bibnamefont {Pichler}},\ and\ \bibinfo {author}
  {\bibfnamefont {M.~D.}\ \bibnamefont {Lukin}},\ }\bibfield  {title} {\bibinfo
  {title} {Periodic orbits, entanglement, and quantum many-body scars in
  constrained models: Matrix product state approach},\ }\href@noop {}
  {\bibfield  {journal} {\bibinfo  {journal} {Phys. Rev. Lett.}\ }\textbf
  {\bibinfo {volume} {122}},\ \bibinfo {pages} {040603} (\bibinfo {year}
  {2019})}\BibitemShut {NoStop}%
\bibitem [{\citenamefont {Lin}\ and\ \citenamefont
  {Motrunich}(2019)}]{Lin2019}%
  \BibitemOpen
  \bibfield  {author} {\bibinfo {author} {\bibfnamefont {C.-J.}\ \bibnamefont
  {Lin}}\ and\ \bibinfo {author} {\bibfnamefont {O.~I.}\ \bibnamefont
  {Motrunich}},\ }\bibfield  {title} {\bibinfo {title} {Exact quantum many-body
  scar states in the {Rydberg-blockaded} atom chain},\ }\href@noop {}
  {\bibfield  {journal} {\bibinfo  {journal} {Phys. Rev. Lett.}\ }\textbf
  {\bibinfo {volume} {122}},\ \bibinfo {pages} {173401} (\bibinfo {year}
  {2019})}\BibitemShut {NoStop}%
\bibitem [{\citenamefont {Schecter}\ and\ \citenamefont
  {Iadecola}(2019)}]{Schecter2019}%
  \BibitemOpen
  \bibfield  {author} {\bibinfo {author} {\bibfnamefont {M.}~\bibnamefont
  {Schecter}}\ and\ \bibinfo {author} {\bibfnamefont {T.}~\bibnamefont
  {Iadecola}},\ }\bibfield  {title} {\bibinfo {title} {Weak ergodicity breaking
  and quantum many-body scars in spin-1 {XY} magnets},\ }\href@noop {}
  {\bibfield  {journal} {\bibinfo  {journal} {Phys. Rev. Lett.}\ }\textbf
  {\bibinfo {volume} {123}},\ \bibinfo {pages} {147201} (\bibinfo {year}
  {2019})}\BibitemShut {NoStop}%
\bibitem [{\citenamefont {Shiraishi}(2019)}]{Shiraishi2019}%
  \BibitemOpen
  \bibfield  {author} {\bibinfo {author} {\bibfnamefont {N.}~\bibnamefont
  {Shiraishi}},\ }\bibfield  {title} {\bibinfo {title} {Connection between
  quantum-many-body scars and the {Affleck--Kennedy--Lieb--Tasaki} model from
  the viewpoint of embedded {Hamiltonians}},\ }\href@noop {} {\bibfield
  {journal} {\bibinfo  {journal} {J. Stat. Mech.}\ }\textbf {\bibinfo {volume}
  {2019}},\ \bibinfo {pages} {083103} (\bibinfo {year} {2019})}\BibitemShut
  {NoStop}%
\bibitem [{\citenamefont {Ok}\ \emph {et~al.}(2019)\citenamefont {Ok},
  \citenamefont {Choo}, \citenamefont {Mudry}, \citenamefont {Castelnovo},
  \citenamefont {Chamon},\ and\ \citenamefont {Neupert}}]{Ok2019}%
  \BibitemOpen
  \bibfield  {author} {\bibinfo {author} {\bibfnamefont {S.}~\bibnamefont
  {Ok}}, \bibinfo {author} {\bibfnamefont {K.}~\bibnamefont {Choo}}, \bibinfo
  {author} {\bibfnamefont {C.}~\bibnamefont {Mudry}}, \bibinfo {author}
  {\bibfnamefont {C.}~\bibnamefont {Castelnovo}}, \bibinfo {author}
  {\bibfnamefont {C.}~\bibnamefont {Chamon}},\ and\ \bibinfo {author}
  {\bibfnamefont {T.}~\bibnamefont {Neupert}},\ }\bibfield  {title} {\bibinfo
  {title} {Topological many-body scar states in dimensions one, two, and
  three},\ }\href@noop {} {\bibfield  {journal} {\bibinfo  {journal} {Phys.
  Rev. Res.}\ }\textbf {\bibinfo {volume} {1}},\ \bibinfo {pages} {033144}
  (\bibinfo {year} {2019})}\BibitemShut {NoStop}%
\bibitem [{\citenamefont {Iadecola}\ and\ \citenamefont
  {Schecter}(2020)}]{Iadecola2020}%
  \BibitemOpen
  \bibfield  {author} {\bibinfo {author} {\bibfnamefont {T.}~\bibnamefont
  {Iadecola}}\ and\ \bibinfo {author} {\bibfnamefont {M.}~\bibnamefont
  {Schecter}},\ }\bibfield  {title} {\bibinfo {title} {Quantum many-body scar
  states with emergent kinetic constraints and finite-entanglement revivals},\
  }\href@noop {} {\bibfield  {journal} {\bibinfo  {journal} {Phys. Rev. B}\
  }\textbf {\bibinfo {volume} {101}},\ \bibinfo {pages} {024306} (\bibinfo
  {year} {2020})}\BibitemShut {NoStop}%
\bibitem [{\citenamefont {Bull}\ \emph {et~al.}(2020)\citenamefont {Bull},
  \citenamefont {Desaules},\ and\ \citenamefont {Papi{\'c}}}]{Bull2020}%
  \BibitemOpen
  \bibfield  {author} {\bibinfo {author} {\bibfnamefont {K.}~\bibnamefont
  {Bull}}, \bibinfo {author} {\bibfnamefont {J.-Y.}\ \bibnamefont {Desaules}},\
  and\ \bibinfo {author} {\bibfnamefont {Z.}~\bibnamefont {Papi{\'c}}},\
  }\bibfield  {title} {\bibinfo {title} {Quantum scars as embeddings of weakly
  broken {Lie} algebra representations},\ }\href@noop {} {\bibfield  {journal}
  {\bibinfo  {journal} {Phys. Rev. B}\ }\textbf {\bibinfo {volume} {101}},\
  \bibinfo {pages} {165139} (\bibinfo {year} {2020})}\BibitemShut {NoStop}%
\bibitem [{\citenamefont {Chattopadhyay}\ \emph {et~al.}(2020)\citenamefont
  {Chattopadhyay}, \citenamefont {Pichler}, \citenamefont {Lukin},\ and\
  \citenamefont {Ho}}]{Chattopadhyay2020}%
  \BibitemOpen
  \bibfield  {author} {\bibinfo {author} {\bibfnamefont {S.}~\bibnamefont
  {Chattopadhyay}}, \bibinfo {author} {\bibfnamefont {H.}~\bibnamefont
  {Pichler}}, \bibinfo {author} {\bibfnamefont {M.~D.}\ \bibnamefont {Lukin}},\
  and\ \bibinfo {author} {\bibfnamefont {W.~W.}\ \bibnamefont {Ho}},\
  }\bibfield  {title} {\bibinfo {title} {Quantum many-body scars from virtual
  entangled pairs},\ }\href@noop {} {\bibfield  {journal} {\bibinfo  {journal}
  {Phys. Rev. B}\ }\textbf {\bibinfo {volume} {101}},\ \bibinfo {pages}
  {174308} (\bibinfo {year} {2020})}\BibitemShut {NoStop}%
\bibitem [{\citenamefont {Lee}\ \emph {et~al.}(2020)\citenamefont {Lee},
  \citenamefont {Melendrez}, \citenamefont {Pal},\ and\ \citenamefont
  {Changlani}}]{Lee2020}%
  \BibitemOpen
  \bibfield  {author} {\bibinfo {author} {\bibfnamefont {K.}~\bibnamefont
  {Lee}}, \bibinfo {author} {\bibfnamefont {R.}~\bibnamefont {Melendrez}},
  \bibinfo {author} {\bibfnamefont {A.}~\bibnamefont {Pal}},\ and\ \bibinfo
  {author} {\bibfnamefont {H.~J.}\ \bibnamefont {Changlani}},\ }\bibfield
  {title} {\bibinfo {title} {Exact three-colored quantum scars from geometric
  frustration},\ }\href@noop {} {\bibfield  {journal} {\bibinfo  {journal}
  {Phys. Rev. B}\ }\textbf {\bibinfo {volume} {101}},\ \bibinfo {pages}
  {241111} (\bibinfo {year} {2020})}\BibitemShut {NoStop}%
\bibitem [{\citenamefont {Alhambra}\ \emph {et~al.}(2020)\citenamefont
  {Alhambra}, \citenamefont {Anshu},\ and\ \citenamefont
  {Wilming}}]{Alhambra2020}%
  \BibitemOpen
  \bibfield  {author} {\bibinfo {author} {\bibfnamefont {{\'A}.~M.}\
  \bibnamefont {Alhambra}}, \bibinfo {author} {\bibfnamefont {A.}~\bibnamefont
  {Anshu}},\ and\ \bibinfo {author} {\bibfnamefont {H.}~\bibnamefont
  {Wilming}},\ }\bibfield  {title} {\bibinfo {title} {Revivals imply quantum
  many-body scars},\ }\href@noop {} {\bibfield  {journal} {\bibinfo  {journal}
  {Phys. Rev. B}\ }\textbf {\bibinfo {volume} {101}},\ \bibinfo {pages}
  {205107} (\bibinfo {year} {2020})}\BibitemShut {NoStop}%
\bibitem [{\citenamefont {Lin}\ \emph {et~al.}(2020)\citenamefont {Lin},
  \citenamefont {Calvera},\ and\ \citenamefont {Hsieh}}]{Lin2020}%
  \BibitemOpen
  \bibfield  {author} {\bibinfo {author} {\bibfnamefont {C.-J.}\ \bibnamefont
  {Lin}}, \bibinfo {author} {\bibfnamefont {V.}~\bibnamefont {Calvera}},\ and\
  \bibinfo {author} {\bibfnamefont {T.~H.}\ \bibnamefont {Hsieh}},\ }\bibfield
  {title} {\bibinfo {title} {Quantum many-body scar states in two-dimensional
  {Rydberg} atom arrays},\ }\href@noop {} {\bibfield  {journal} {\bibinfo
  {journal} {Phys. Rev. B}\ }\textbf {\bibinfo {volume} {101}},\ \bibinfo
  {pages} {220304} (\bibinfo {year} {2020})}\BibitemShut {NoStop}%
\bibitem [{\citenamefont {Mark}\ and\ \citenamefont
  {Motrunich}(2020)}]{Mark2020eta}%
  \BibitemOpen
  \bibfield  {author} {\bibinfo {author} {\bibfnamefont {D.~K.}\ \bibnamefont
  {Mark}}\ and\ \bibinfo {author} {\bibfnamefont {O.~I.}\ \bibnamefont
  {Motrunich}},\ }\bibfield  {title} {\bibinfo {title} {$\eta$-pairing states
  as true scars in an extended {Hubbard} model},\ }\href@noop {} {\bibfield
  {journal} {\bibinfo  {journal} {Phys. Rev. B}\ }\textbf {\bibinfo {volume}
  {102}},\ \bibinfo {pages} {075132} (\bibinfo {year} {2020})}\BibitemShut
  {NoStop}%
\bibitem [{\citenamefont {Mark}\ \emph {et~al.}(2020)\citenamefont {Mark},
  \citenamefont {Lin},\ and\ \citenamefont {Motrunich}}]{Mark2020unified}%
  \BibitemOpen
  \bibfield  {author} {\bibinfo {author} {\bibfnamefont {D.~K.}\ \bibnamefont
  {Mark}}, \bibinfo {author} {\bibfnamefont {C.-J.}\ \bibnamefont {Lin}},\ and\
  \bibinfo {author} {\bibfnamefont {O.~I.}\ \bibnamefont {Motrunich}},\
  }\bibfield  {title} {\bibinfo {title} {Unified structure for exact towers of
  scar states in the {Affleck-Kennedy-Lieb-Tasaki} and other models},\
  }\href@noop {} {\bibfield  {journal} {\bibinfo  {journal} {Phys. Rev. B}\
  }\textbf {\bibinfo {volume} {101}},\ \bibinfo {pages} {195131} (\bibinfo
  {year} {2020})}\BibitemShut {NoStop}%
\bibitem [{\citenamefont {Moudgalya}\ \emph
  {et~al.}(2020{\natexlab{a}})\citenamefont {Moudgalya}, \citenamefont
  {Regnault},\ and\ \citenamefont {Bernevig}}]{Moudgalya2020eta}%
  \BibitemOpen
  \bibfield  {author} {\bibinfo {author} {\bibfnamefont {S.}~\bibnamefont
  {Moudgalya}}, \bibinfo {author} {\bibfnamefont {N.}~\bibnamefont
  {Regnault}},\ and\ \bibinfo {author} {\bibfnamefont {B.~A.}\ \bibnamefont
  {Bernevig}},\ }\bibfield  {title} {\bibinfo {title} {$\eta$-pairing in
  hubbard models: From spectrum generating algebras to quantum many-body
  scars},\ }\href@noop {} {\bibfield  {journal} {\bibinfo  {journal} {Phys.
  Rev. B}\ }\textbf {\bibinfo {volume} {102}},\ \bibinfo {pages} {085140}
  (\bibinfo {year} {2020}{\natexlab{a}})}\BibitemShut {NoStop}%
\bibitem [{\citenamefont {Moudgalya}\ \emph
  {et~al.}(2020{\natexlab{b}})\citenamefont {Moudgalya}, \citenamefont
  {O'Brien}, \citenamefont {Bernevig}, \citenamefont {Fendley},\ and\
  \citenamefont {Regnault}}]{Moudgalya2020large}%
  \BibitemOpen
  \bibfield  {author} {\bibinfo {author} {\bibfnamefont {S.}~\bibnamefont
  {Moudgalya}}, \bibinfo {author} {\bibfnamefont {E.}~\bibnamefont {O'Brien}},
  \bibinfo {author} {\bibfnamefont {B.~A.}\ \bibnamefont {Bernevig}}, \bibinfo
  {author} {\bibfnamefont {P.}~\bibnamefont {Fendley}},\ and\ \bibinfo {author}
  {\bibfnamefont {N.}~\bibnamefont {Regnault}},\ }\bibfield  {title} {\bibinfo
  {title} {Large classes of quantum scarred hamiltonians from matrix product
  states},\ }\href@noop {} {\bibfield  {journal} {\bibinfo  {journal} {Phys.
  Rev. B}\ }\textbf {\bibinfo {volume} {102}},\ \bibinfo {pages} {085120}
  (\bibinfo {year} {2020}{\natexlab{b}})}\BibitemShut {NoStop}%
\bibitem [{\citenamefont {Michailidis}\ \emph {et~al.}(2020)\citenamefont
  {Michailidis}, \citenamefont {Turner}, \citenamefont {Papi{\'c}},
  \citenamefont {Abanin},\ and\ \citenamefont {Serbyn}}]{Michailidis2020}%
  \BibitemOpen
  \bibfield  {author} {\bibinfo {author} {\bibfnamefont {A.}~\bibnamefont
  {Michailidis}}, \bibinfo {author} {\bibfnamefont {C.}~\bibnamefont {Turner}},
  \bibinfo {author} {\bibfnamefont {Z.}~\bibnamefont {Papi{\'c}}}, \bibinfo
  {author} {\bibfnamefont {D.}~\bibnamefont {Abanin}},\ and\ \bibinfo {author}
  {\bibfnamefont {M.}~\bibnamefont {Serbyn}},\ }\bibfield  {title} {\bibinfo
  {title} {Stabilizing two-dimensional quantum scars by deformation and
  synchronization},\ }\href@noop {} {\bibfield  {journal} {\bibinfo  {journal}
  {Phys. Rev. Res.}\ }\textbf {\bibinfo {volume} {2}},\ \bibinfo {pages}
  {022065} (\bibinfo {year} {2020})}\BibitemShut {NoStop}%
\bibitem [{\citenamefont {O'Dea}\ \emph {et~al.}(2020)\citenamefont {O'Dea},
  \citenamefont {Burnell}, \citenamefont {Chandran},\ and\ \citenamefont
  {Khemani}}]{ODea2020}%
  \BibitemOpen
  \bibfield  {author} {\bibinfo {author} {\bibfnamefont {N.}~\bibnamefont
  {O'Dea}}, \bibinfo {author} {\bibfnamefont {F.}~\bibnamefont {Burnell}},
  \bibinfo {author} {\bibfnamefont {A.}~\bibnamefont {Chandran}},\ and\
  \bibinfo {author} {\bibfnamefont {V.}~\bibnamefont {Khemani}},\ }\bibfield
  {title} {\bibinfo {title} {From tunnels to towers: Quantum scars from {Lie}
  algebras and q-deformed {Lie} algebras},\ }\href@noop {} {\bibfield
  {journal} {\bibinfo  {journal} {Phys. Rev. Res.}\ }\textbf {\bibinfo {volume}
  {2}},\ \bibinfo {pages} {043305} (\bibinfo {year} {2020})}\BibitemShut
  {NoStop}%
\bibitem [{\citenamefont {Shibata}\ \emph {et~al.}(2020)\citenamefont
  {Shibata}, \citenamefont {Yoshioka},\ and\ \citenamefont
  {Katsura}}]{Shibata2020}%
  \BibitemOpen
  \bibfield  {author} {\bibinfo {author} {\bibfnamefont {N.}~\bibnamefont
  {Shibata}}, \bibinfo {author} {\bibfnamefont {N.}~\bibnamefont {Yoshioka}},\
  and\ \bibinfo {author} {\bibfnamefont {H.}~\bibnamefont {Katsura}},\
  }\bibfield  {title} {\bibinfo {title} {Onsager's scars in disordered spin
  chains},\ }\href@noop {} {\bibfield  {journal} {\bibinfo  {journal} {Phys.
  Rev. Lett.}\ }\textbf {\bibinfo {volume} {124}},\ \bibinfo {pages} {180604}
  (\bibinfo {year} {2020})}\BibitemShut {NoStop}%
\bibitem [{\citenamefont {Pakrouski}\ \emph {et~al.}(2020)\citenamefont
  {Pakrouski}, \citenamefont {Pallegar}, \citenamefont {Popov},\ and\
  \citenamefont {Klebanov}}]{Pakrouski2020}%
  \BibitemOpen
  \bibfield  {author} {\bibinfo {author} {\bibfnamefont {K.}~\bibnamefont
  {Pakrouski}}, \bibinfo {author} {\bibfnamefont {P.~N.}\ \bibnamefont
  {Pallegar}}, \bibinfo {author} {\bibfnamefont {F.~K.}\ \bibnamefont
  {Popov}},\ and\ \bibinfo {author} {\bibfnamefont {I.~R.}\ \bibnamefont
  {Klebanov}},\ }\bibfield  {title} {\bibinfo {title} {Many-body scars as a
  group invariant sector of {Hilbert} space},\ }\href@noop {} {\bibfield
  {journal} {\bibinfo  {journal} {Phys. Rev. Lett.}\ }\textbf {\bibinfo
  {volume} {125}},\ \bibinfo {pages} {230602} (\bibinfo {year}
  {2020})}\BibitemShut {NoStop}%
\bibitem [{\citenamefont {Kuno}\ \emph {et~al.}(2020)\citenamefont {Kuno},
  \citenamefont {Mizoguchi},\ and\ \citenamefont {Hatsugai}}]{Kuno2020}%
  \BibitemOpen
  \bibfield  {author} {\bibinfo {author} {\bibfnamefont {Y.}~\bibnamefont
  {Kuno}}, \bibinfo {author} {\bibfnamefont {T.}~\bibnamefont {Mizoguchi}},\
  and\ \bibinfo {author} {\bibfnamefont {Y.}~\bibnamefont {Hatsugai}},\
  }\bibfield  {title} {\bibinfo {title} {Flat band quantum scar},\ }\href@noop
  {} {\bibfield  {journal} {\bibinfo  {journal} {Phy. Rev. B}\ }\textbf
  {\bibinfo {volume} {102}},\ \bibinfo {pages} {241115} (\bibinfo {year}
  {2020})}\BibitemShut {NoStop}%
\bibitem [{\citenamefont {Mizuta}\ \emph {et~al.}(2020)\citenamefont {Mizuta},
  \citenamefont {Takasan},\ and\ \citenamefont {Kawakami}}]{Mizuta2020}%
  \BibitemOpen
  \bibfield  {author} {\bibinfo {author} {\bibfnamefont {K.}~\bibnamefont
  {Mizuta}}, \bibinfo {author} {\bibfnamefont {K.}~\bibnamefont {Takasan}},\
  and\ \bibinfo {author} {\bibfnamefont {N.}~\bibnamefont {Kawakami}},\
  }\bibfield  {title} {\bibinfo {title} {Exact floquet quantum many-body scars
  under {Rydberg} blockade},\ }\href@noop {} {\bibfield  {journal} {\bibinfo
  {journal} {Phys. Rev. Res.}\ }\textbf {\bibinfo {volume} {2}},\ \bibinfo
  {pages} {033284} (\bibinfo {year} {2020})}\BibitemShut {NoStop}%
\bibitem [{\citenamefont {Hudomal}\ \emph {et~al.}(2020)\citenamefont
  {Hudomal}, \citenamefont {Vasi{\'c}}, \citenamefont {Regnault},\ and\
  \citenamefont {Papi{\'c}}}]{Hudomal2020}%
  \BibitemOpen
  \bibfield  {author} {\bibinfo {author} {\bibfnamefont {A.}~\bibnamefont
  {Hudomal}}, \bibinfo {author} {\bibfnamefont {I.}~\bibnamefont {Vasi{\'c}}},
  \bibinfo {author} {\bibfnamefont {N.}~\bibnamefont {Regnault}},\ and\
  \bibinfo {author} {\bibfnamefont {Z.}~\bibnamefont {Papi{\'c}}},\ }\bibfield
  {title} {\bibinfo {title} {Quantum scars of bosons with correlated hopping},\
  }\href@noop {} {\bibfield  {journal} {\bibinfo  {journal} {Commun. Phys.}\
  }\textbf {\bibinfo {volume} {3}},\ \bibinfo {pages} {99} (\bibinfo {year}
  {2020})}\BibitemShut {NoStop}%
\bibitem [{\citenamefont {Surace}\ \emph {et~al.}(2020)\citenamefont {Surace},
  \citenamefont {Giudici},\ and\ \citenamefont {Dalmonte}}]{Surace2020}%
  \BibitemOpen
  \bibfield  {author} {\bibinfo {author} {\bibfnamefont {F.~M.}\ \bibnamefont
  {Surace}}, \bibinfo {author} {\bibfnamefont {G.}~\bibnamefont {Giudici}},\
  and\ \bibinfo {author} {\bibfnamefont {M.}~\bibnamefont {Dalmonte}},\
  }\bibfield  {title} {\bibinfo {title} {Weak-ergodicity-breaking via lattice
  supersymmetry},\ }\href@noop {} {\bibfield  {journal} {\bibinfo  {journal}
  {Quantum}\ }\textbf {\bibinfo {volume} {4}},\ \bibinfo {pages} {339}
  (\bibinfo {year} {2020})}\BibitemShut {NoStop}%
\bibitem [{\citenamefont {Sugiura}\ \emph {et~al.}(2021)\citenamefont
  {Sugiura}, \citenamefont {Kuwahara},\ and\ \citenamefont
  {Saito}}]{Sugiura2021}%
  \BibitemOpen
  \bibfield  {author} {\bibinfo {author} {\bibfnamefont {S.}~\bibnamefont
  {Sugiura}}, \bibinfo {author} {\bibfnamefont {T.}~\bibnamefont {Kuwahara}},\
  and\ \bibinfo {author} {\bibfnamefont {K.}~\bibnamefont {Saito}},\ }\bibfield
   {title} {\bibinfo {title} {Many-body scar state intrinsic to periodically
  driven system},\ }\href@noop {} {\bibfield  {journal} {\bibinfo  {journal}
  {Phys. Rev. Res.}\ }\textbf {\bibinfo {volume} {3}},\ \bibinfo {pages}
  {L012010} (\bibinfo {year} {2021})}\BibitemShut {NoStop}%
\bibitem [{\citenamefont {Pakrouski}\ \emph {et~al.}(2021)\citenamefont
  {Pakrouski}, \citenamefont {Pallegar}, \citenamefont {Popov},\ and\
  \citenamefont {Klebanov}}]{Pakrouski2021}%
  \BibitemOpen
  \bibfield  {author} {\bibinfo {author} {\bibfnamefont {K.}~\bibnamefont
  {Pakrouski}}, \bibinfo {author} {\bibfnamefont {P.~N.}\ \bibnamefont
  {Pallegar}}, \bibinfo {author} {\bibfnamefont {F.~K.}\ \bibnamefont
  {Popov}},\ and\ \bibinfo {author} {\bibfnamefont {I.~R.}\ \bibnamefont
  {Klebanov}},\ }\bibfield  {title} {\bibinfo {title} {Group theoretic approach
  to many-body scar states in fermionic lattice models},\ }\href@noop {}
  {\bibfield  {journal} {\bibinfo  {journal} {Phys. Rev. Res.}\ }\textbf
  {\bibinfo {volume} {3}},\ \bibinfo {pages} {043156} (\bibinfo {year}
  {2021})}\BibitemShut {NoStop}%
\bibitem [{\citenamefont {Mondragon-Shem}\ \emph {et~al.}(2021)\citenamefont
  {Mondragon-Shem}, \citenamefont {Vavilov},\ and\ \citenamefont
  {Martin}}]{Mondragon2021}%
  \BibitemOpen
  \bibfield  {author} {\bibinfo {author} {\bibfnamefont {I.}~\bibnamefont
  {Mondragon-Shem}}, \bibinfo {author} {\bibfnamefont {M.~G.}\ \bibnamefont
  {Vavilov}},\ and\ \bibinfo {author} {\bibfnamefont {I.}~\bibnamefont
  {Martin}},\ }\bibfield  {title} {\bibinfo {title} {Fate of quantum many-body
  scars in the presence of disorder},\ }\href@noop {} {\bibfield  {journal}
  {\bibinfo  {journal} {PRX Quantum}\ }\textbf {\bibinfo {volume} {2}},\
  \bibinfo {pages} {030349} (\bibinfo {year} {2021})}\BibitemShut {NoStop}%
\bibitem [{\citenamefont {Chertkov}\ and\ \citenamefont
  {Clark}(2021)}]{chertkov2021motif}%
  \BibitemOpen
  \bibfield  {author} {\bibinfo {author} {\bibfnamefont {E.}~\bibnamefont
  {Chertkov}}\ and\ \bibinfo {author} {\bibfnamefont {B.~K.}\ \bibnamefont
  {Clark}},\ }\bibfield  {title} {\bibinfo {title} {Motif magnetism and quantum
  many-body scars},\ }\href@noop {} {\bibfield  {journal} {\bibinfo  {journal}
  {Phys. Rev. B}\ }\textbf {\bibinfo {volume} {104}},\ \bibinfo {pages}
  {104410} (\bibinfo {year} {2021})}\BibitemShut {NoStop}%
\bibitem [{\citenamefont {Ren}\ \emph {et~al.}(2021)\citenamefont {Ren},
  \citenamefont {Liang},\ and\ \citenamefont {Fang}}]{Ren2021quasisymmetry}%
  \BibitemOpen
  \bibfield  {author} {\bibinfo {author} {\bibfnamefont {J.}~\bibnamefont
  {Ren}}, \bibinfo {author} {\bibfnamefont {C.}~\bibnamefont {Liang}},\ and\
  \bibinfo {author} {\bibfnamefont {C.}~\bibnamefont {Fang}},\ }\bibfield
  {title} {\bibinfo {title} {Quasisymmetry groups and many-body scar
  dynamics},\ }\href@noop {} {\bibfield  {journal} {\bibinfo  {journal} {Phys.
  Rev. Lett.}\ }\textbf {\bibinfo {volume} {126}},\ \bibinfo {pages} {120604}
  (\bibinfo {year} {2021})}\BibitemShut {NoStop}%
\bibitem [{\citenamefont {Ren}\ \emph {et~al.}(2022)\citenamefont {Ren},
  \citenamefont {Liang},\ and\ \citenamefont {Fang}}]{Ren2022deformed}%
  \BibitemOpen
  \bibfield  {author} {\bibinfo {author} {\bibfnamefont {J.}~\bibnamefont
  {Ren}}, \bibinfo {author} {\bibfnamefont {C.}~\bibnamefont {Liang}},\ and\
  \bibinfo {author} {\bibfnamefont {C.}~\bibnamefont {Fang}},\ }\bibfield
  {title} {\bibinfo {title} {Deformed symmetry structures and quantum many-body
  scar subspaces},\ }\href@noop {} {\bibfield  {journal} {\bibinfo  {journal}
  {Phys. Rev. Res.}\ }\textbf {\bibinfo {volume} {4}},\ \bibinfo {pages}
  {013155} (\bibinfo {year} {2022})}\BibitemShut {NoStop}%
\bibitem [{\citenamefont {Langlett}\ \emph {et~al.}(2022)\citenamefont
  {Langlett}, \citenamefont {Yang}, \citenamefont {Wildeboer}, \citenamefont
  {Gorshkov}, \citenamefont {Iadecola},\ and\ \citenamefont
  {Xu}}]{Langlett2022}%
  \BibitemOpen
  \bibfield  {author} {\bibinfo {author} {\bibfnamefont {C.~M.}\ \bibnamefont
  {Langlett}}, \bibinfo {author} {\bibfnamefont {Z.-C.}\ \bibnamefont {Yang}},
  \bibinfo {author} {\bibfnamefont {J.}~\bibnamefont {Wildeboer}}, \bibinfo
  {author} {\bibfnamefont {A.~V.}\ \bibnamefont {Gorshkov}}, \bibinfo {author}
  {\bibfnamefont {T.}~\bibnamefont {Iadecola}},\ and\ \bibinfo {author}
  {\bibfnamefont {S.}~\bibnamefont {Xu}},\ }\bibfield  {title} {\bibinfo
  {title} {Rainbow scars: From area to volume law},\ }\href@noop {} {\bibfield
  {journal} {\bibinfo  {journal} {Phys. Rev. B}\ }\textbf {\bibinfo {volume}
  {105}},\ \bibinfo {pages} {L060301} (\bibinfo {year} {2022})}\BibitemShut
  {NoStop}%
\bibitem [{\citenamefont {Tamura}\ and\ \citenamefont
  {Katsura}(2022)}]{Tamura2022}%
  \BibitemOpen
  \bibfield  {author} {\bibinfo {author} {\bibfnamefont {K.}~\bibnamefont
  {Tamura}}\ and\ \bibinfo {author} {\bibfnamefont {H.}~\bibnamefont
  {Katsura}},\ }\bibfield  {title} {\bibinfo {title} {Quantum many-body scars
  of spinless fermions with density-assisted hopping in higher dimensions},\
  }\href@noop {} {\bibfield  {journal} {\bibinfo  {journal} {Phys. Rev. B}\
  }\textbf {\bibinfo {volume} {106}},\ \bibinfo {pages} {144306} (\bibinfo
  {year} {2022})}\BibitemShut {NoStop}%
\bibitem [{\citenamefont {Desaules}\ \emph {et~al.}(2022)\citenamefont
  {Desaules}, \citenamefont {Pietracaprina}, \citenamefont {Papi{\'c}},
  \citenamefont {Goold},\ and\ \citenamefont {Pappalardi}}]{Desaules2022}%
  \BibitemOpen
  \bibfield  {author} {\bibinfo {author} {\bibfnamefont {J.-Y.}\ \bibnamefont
  {Desaules}}, \bibinfo {author} {\bibfnamefont {F.}~\bibnamefont
  {Pietracaprina}}, \bibinfo {author} {\bibfnamefont {Z.}~\bibnamefont
  {Papi{\'c}}}, \bibinfo {author} {\bibfnamefont {J.}~\bibnamefont {Goold}},\
  and\ \bibinfo {author} {\bibfnamefont {S.}~\bibnamefont {Pappalardi}},\
  }\bibfield  {title} {\bibinfo {title} {Extensive multipartite entanglement
  from su (2) quantum many-body scars},\ }\href@noop {} {\bibfield  {journal}
  {\bibinfo  {journal} {Phys. Rev. Lett.}\ }\textbf {\bibinfo {volume} {129}},\
  \bibinfo {pages} {020601} (\bibinfo {year} {2022})}\BibitemShut {NoStop}%
\bibitem [{\citenamefont {Gotta}\ \emph {et~al.}(2022)\citenamefont {Gotta},
  \citenamefont {Mazza}, \citenamefont {Simon},\ and\ \citenamefont
  {Roux}}]{Gotta2022}%
  \BibitemOpen
  \bibfield  {author} {\bibinfo {author} {\bibfnamefont {L.}~\bibnamefont
  {Gotta}}, \bibinfo {author} {\bibfnamefont {L.}~\bibnamefont {Mazza}},
  \bibinfo {author} {\bibfnamefont {P.}~\bibnamefont {Simon}},\ and\ \bibinfo
  {author} {\bibfnamefont {G.}~\bibnamefont {Roux}},\ }\bibfield  {title}
  {\bibinfo {title} {Exact many-body scars based on pairs or multimers in a
  chain of spinless fermions},\ }\href@noop {} {\bibfield  {journal} {\bibinfo
  {journal} {Phys. Rev. B}\ }\textbf {\bibinfo {volume} {106}},\ \bibinfo
  {pages} {235147} (\bibinfo {year} {2022})}\BibitemShut {NoStop}%
\bibitem [{\citenamefont {Tang}\ \emph {et~al.}(2022)\citenamefont {Tang},
  \citenamefont {O'Dea},\ and\ \citenamefont {Chandran}}]{Tang2022}%
  \BibitemOpen
  \bibfield  {author} {\bibinfo {author} {\bibfnamefont {L.-H.}\ \bibnamefont
  {Tang}}, \bibinfo {author} {\bibfnamefont {N.}~\bibnamefont {O'Dea}},\ and\
  \bibinfo {author} {\bibfnamefont {A.}~\bibnamefont {Chandran}},\ }\bibfield
  {title} {\bibinfo {title} {Multimagnon quantum many-body scars from tensor
  operators},\ }\href@noop {} {\bibfield  {journal} {\bibinfo  {journal} {Phys.
  Rev. Res.}\ }\textbf {\bibinfo {volume} {4}},\ \bibinfo {pages} {043006}
  (\bibinfo {year} {2022})}\BibitemShut {NoStop}%
\bibitem [{\citenamefont {Wildeboer}\ \emph {et~al.}(2022)\citenamefont
  {Wildeboer}, \citenamefont {Langlett}, \citenamefont {Yang}, \citenamefont
  {Gorshkov}, \citenamefont {Iadecola},\ and\ \citenamefont
  {Xu}}]{wildeboer2022quantum}%
  \BibitemOpen
  \bibfield  {author} {\bibinfo {author} {\bibfnamefont {J.}~\bibnamefont
  {Wildeboer}}, \bibinfo {author} {\bibfnamefont {C.~M.}\ \bibnamefont
  {Langlett}}, \bibinfo {author} {\bibfnamefont {Z.-C.}\ \bibnamefont {Yang}},
  \bibinfo {author} {\bibfnamefont {A.~V.}\ \bibnamefont {Gorshkov}}, \bibinfo
  {author} {\bibfnamefont {T.}~\bibnamefont {Iadecola}},\ and\ \bibinfo
  {author} {\bibfnamefont {S.}~\bibnamefont {Xu}},\ }\bibfield  {title}
  {\bibinfo {title} {Quantum many-body scars from {Einstein-Podolsky-Rosen}
  states in bilayer systems},\ }\href@noop {} {\bibfield  {journal} {\bibinfo
  {journal} {Phys. Rev. B}\ }\textbf {\bibinfo {volume} {106}},\ \bibinfo
  {pages} {205142} (\bibinfo {year} {2022})}\BibitemShut {NoStop}%
\bibitem [{\citenamefont {Omiya}\ and\ \citenamefont
  {M{\"u}ller}(2023{\natexlab{a}})}]{Omiya2023}%
  \BibitemOpen
  \bibfield  {author} {\bibinfo {author} {\bibfnamefont {K.}~\bibnamefont
  {Omiya}}\ and\ \bibinfo {author} {\bibfnamefont {M.}~\bibnamefont
  {M{\"u}ller}},\ }\bibfield  {title} {\bibinfo {title} {Fractionalization
  paves the way to local projector embeddings of quantum many-body scars},\
  }\href@noop {} {\bibfield  {journal} {\bibinfo  {journal} {Phys. Rev. B}\
  }\textbf {\bibinfo {volume} {108}},\ \bibinfo {pages} {054412} (\bibinfo
  {year} {2023}{\natexlab{a}})}\BibitemShut {NoStop}%
\bibitem [{\citenamefont {Omiya}\ and\ \citenamefont
  {M{\"u}ller}(2023{\natexlab{b}})}]{Omiya2023quantum}%
  \BibitemOpen
  \bibfield  {author} {\bibinfo {author} {\bibfnamefont {K.}~\bibnamefont
  {Omiya}}\ and\ \bibinfo {author} {\bibfnamefont {M.}~\bibnamefont
  {M{\"u}ller}},\ }\bibfield  {title} {\bibinfo {title} {Quantum many-body
  scars in bipartite {Rydberg} arrays originating from hidden projector
  embedding},\ }\href@noop {} {\bibfield  {journal} {\bibinfo  {journal} {Phys.
  Rev. A}\ }\textbf {\bibinfo {volume} {107}},\ \bibinfo {pages} {023318}
  (\bibinfo {year} {2023}{\natexlab{b}})}\BibitemShut {NoStop}%
\bibitem [{\citenamefont {Desaules}\ \emph
  {et~al.}(2023{\natexlab{a}})\citenamefont {Desaules}, \citenamefont
  {Banerjee}, \citenamefont {Hudomal}, \citenamefont {Papi{\'c}}, \citenamefont
  {Sen},\ and\ \citenamefont {Halimeh}}]{Desaules2023}%
  \BibitemOpen
  \bibfield  {author} {\bibinfo {author} {\bibfnamefont {J.-Y.}\ \bibnamefont
  {Desaules}}, \bibinfo {author} {\bibfnamefont {D.}~\bibnamefont {Banerjee}},
  \bibinfo {author} {\bibfnamefont {A.}~\bibnamefont {Hudomal}}, \bibinfo
  {author} {\bibfnamefont {Z.}~\bibnamefont {Papi{\'c}}}, \bibinfo {author}
  {\bibfnamefont {A.}~\bibnamefont {Sen}},\ and\ \bibinfo {author}
  {\bibfnamefont {J.~C.}\ \bibnamefont {Halimeh}},\ }\bibfield  {title}
  {\bibinfo {title} {Weak ergodicity breaking in the {Schwinger} model},\
  }\href@noop {} {\bibfield  {journal} {\bibinfo  {journal} {Phys. Rev. B}\
  }\textbf {\bibinfo {volume} {107}},\ \bibinfo {pages} {L201105} (\bibinfo
  {year} {2023}{\natexlab{a}})}\BibitemShut {NoStop}%
\bibitem [{\citenamefont {Desaules}\ \emph
  {et~al.}(2023{\natexlab{b}})\citenamefont {Desaules}, \citenamefont
  {Hudomal}, \citenamefont {Banerjee}, \citenamefont {Sen}, \citenamefont
  {Papi{\'c}},\ and\ \citenamefont {Halimeh}}]{Desaules2023prominent}%
  \BibitemOpen
  \bibfield  {author} {\bibinfo {author} {\bibfnamefont {J.-Y.}\ \bibnamefont
  {Desaules}}, \bibinfo {author} {\bibfnamefont {A.}~\bibnamefont {Hudomal}},
  \bibinfo {author} {\bibfnamefont {D.}~\bibnamefont {Banerjee}}, \bibinfo
  {author} {\bibfnamefont {A.}~\bibnamefont {Sen}}, \bibinfo {author}
  {\bibfnamefont {Z.}~\bibnamefont {Papi{\'c}}},\ and\ \bibinfo {author}
  {\bibfnamefont {J.~C.}\ \bibnamefont {Halimeh}},\ }\bibfield  {title}
  {\bibinfo {title} {Prominent quantum many-body scars in a truncated
  {Schwinger} model},\ }\href@noop {} {\bibfield  {journal} {\bibinfo
  {journal} {Phys. Rev. B}\ }\textbf {\bibinfo {volume} {107}},\ \bibinfo
  {pages} {205112} (\bibinfo {year} {2023}{\natexlab{b}})}\BibitemShut
  {NoStop}%
\bibitem [{\citenamefont {Sanada}\ \emph {et~al.}(2023)\citenamefont {Sanada},
  \citenamefont {Miao},\ and\ \citenamefont {Katsura}}]{Sanada2023}%
  \BibitemOpen
  \bibfield  {author} {\bibinfo {author} {\bibfnamefont {K.}~\bibnamefont
  {Sanada}}, \bibinfo {author} {\bibfnamefont {Y.}~\bibnamefont {Miao}},\ and\
  \bibinfo {author} {\bibfnamefont {H.}~\bibnamefont {Katsura}},\ }\bibfield
  {title} {\bibinfo {title} {Quantum many-body scars in spin models with
  multibody interactions},\ }\href@noop {} {\bibfield  {journal} {\bibinfo
  {journal} {Phys. Rev. B}\ }\textbf {\bibinfo {volume} {108}},\ \bibinfo
  {pages} {155102} (\bibinfo {year} {2023})}\BibitemShut {NoStop}%
\bibitem [{\citenamefont {Iversen}\ and\ \citenamefont
  {Nielsen}(2023)}]{Iversen2023}%
  \BibitemOpen
  \bibfield  {author} {\bibinfo {author} {\bibfnamefont {M.}~\bibnamefont
  {Iversen}}\ and\ \bibinfo {author} {\bibfnamefont {A.~E.}\ \bibnamefont
  {Nielsen}},\ }\bibfield  {title} {\bibinfo {title} {Tower of quantum scars in
  a partially many-body localized system},\ }\href@noop {} {\bibfield
  {journal} {\bibinfo  {journal} {Phys. Rev. B}\ }\textbf {\bibinfo {volume}
  {107}},\ \bibinfo {pages} {205140} (\bibinfo {year} {2023})}\BibitemShut
  {NoStop}%
\bibitem [{\citenamefont {Halimeh}\ \emph {et~al.}(2023)\citenamefont
  {Halimeh}, \citenamefont {Barbiero}, \citenamefont {Hauke}, \citenamefont
  {Grusdt},\ and\ \citenamefont {Bohrdt}}]{Lalimeh2023}%
  \BibitemOpen
  \bibfield  {author} {\bibinfo {author} {\bibfnamefont {J.~C.}\ \bibnamefont
  {Halimeh}}, \bibinfo {author} {\bibfnamefont {L.}~\bibnamefont {Barbiero}},
  \bibinfo {author} {\bibfnamefont {P.}~\bibnamefont {Hauke}}, \bibinfo
  {author} {\bibfnamefont {F.}~\bibnamefont {Grusdt}},\ and\ \bibinfo {author}
  {\bibfnamefont {A.}~\bibnamefont {Bohrdt}},\ }\bibfield  {title} {\bibinfo
  {title} {Robust quantum many-body scars in lattice gauge theories},\
  }\href@noop {} {\bibfield  {journal} {\bibinfo  {journal} {Quantum}\ }\textbf
  {\bibinfo {volume} {7}},\ \bibinfo {pages} {1004} (\bibinfo {year}
  {2023})}\BibitemShut {NoStop}%
\bibitem [{\citenamefont {Kaneko}\ \emph {et~al.}(2024)\citenamefont {Kaneko},
  \citenamefont {Kunimi},\ and\ \citenamefont {Danshita}}]{Kaneko2024}%
  \BibitemOpen
  \bibfield  {author} {\bibinfo {author} {\bibfnamefont {R.}~\bibnamefont
  {Kaneko}}, \bibinfo {author} {\bibfnamefont {M.}~\bibnamefont {Kunimi}},\
  and\ \bibinfo {author} {\bibfnamefont {I.}~\bibnamefont {Danshita}},\
  }\bibfield  {title} {\bibinfo {title} {Quantum many-body scars in the
  {Bose-Hubbard} model with a three-body constraint},\ }\href@noop {}
  {\bibfield  {journal} {\bibinfo  {journal} {Phys. Rev. A}\ }\textbf {\bibinfo
  {volume} {109}},\ \bibinfo {pages} {L011301} (\bibinfo {year}
  {2024})}\BibitemShut {NoStop}%
\bibitem [{\citenamefont {Wang}\ \emph {et~al.}(2024)\citenamefont {Wang},
  \citenamefont {Yuan}, \citenamefont {Zhang}, \citenamefont {Wang},
  \citenamefont {Deng},\ and\ \citenamefont {Duan}}]{Wang2024}%
  \BibitemOpen
  \bibfield  {author} {\bibinfo {author} {\bibfnamefont {H.-R.}\ \bibnamefont
  {Wang}}, \bibinfo {author} {\bibfnamefont {D.}~\bibnamefont {Yuan}}, \bibinfo
  {author} {\bibfnamefont {S.-Y.}\ \bibnamefont {Zhang}}, \bibinfo {author}
  {\bibfnamefont {Z.}~\bibnamefont {Wang}}, \bibinfo {author} {\bibfnamefont
  {D.-L.}\ \bibnamefont {Deng}},\ and\ \bibinfo {author} {\bibfnamefont
  {L.-M.}\ \bibnamefont {Duan}},\ }\bibfield  {title} {\bibinfo {title}
  {Embedding quantum many-body scars into decoherence-free subspaces},\
  }\href@noop {} {\bibfield  {journal} {\bibinfo  {journal} {Phys. Rev. Lett.}\
  }\textbf {\bibinfo {volume} {132}},\ \bibinfo {pages} {150401} (\bibinfo
  {year} {2024})}\BibitemShut {NoStop}%
\bibitem [{\citenamefont {Kunimi}\ \emph {et~al.}(2024)\citenamefont {Kunimi},
  \citenamefont {Tomita}, \citenamefont {Katsura},\ and\ \citenamefont
  {Kato}}]{Kunimi2024}%
  \BibitemOpen
  \bibfield  {author} {\bibinfo {author} {\bibfnamefont {M.}~\bibnamefont
  {Kunimi}}, \bibinfo {author} {\bibfnamefont {T.}~\bibnamefont {Tomita}},
  \bibinfo {author} {\bibfnamefont {H.}~\bibnamefont {Katsura}},\ and\ \bibinfo
  {author} {\bibfnamefont {Y.}~\bibnamefont {Kato}},\ }\bibfield  {title}
  {\bibinfo {title} {Proposal for simulating quantum spin models with the
  {Dzyaloshinskii-Moriya} interaction using {Rydberg} atoms and the
  construction of asymptotic quantum many-body scar states},\ }\href@noop {}
  {\bibfield  {journal} {\bibinfo  {journal} {Phys. Rev. A}\ }\textbf {\bibinfo
  {volume} {110}},\ \bibinfo {pages} {043312} (\bibinfo {year}
  {2024})}\BibitemShut {NoStop}%
\bibitem [{\citenamefont {Matsui}(2024)}]{Matsui2024}%
  \BibitemOpen
  \bibfield  {author} {\bibinfo {author} {\bibfnamefont {C.}~\bibnamefont
  {Matsui}},\ }\bibfield  {title} {\bibinfo {title} {Exactly solvable subspaces
  of nonintegrable spin chains with boundaries and quasiparticle
  interactions},\ }\href@noop {} {\bibfield  {journal} {\bibinfo  {journal}
  {Phys. Rev. B}\ }\textbf {\bibinfo {volume} {109}},\ \bibinfo {pages}
  {104307} (\bibinfo {year} {2024})}\BibitemShut {NoStop}%
\bibitem [{\citenamefont {Teretenkov}\ and\ \citenamefont
  {Lychkovskiy}(2024)}]{teretenkov2024duality}%
  \BibitemOpen
  \bibfield  {author} {\bibinfo {author} {\bibfnamefont {A.}~\bibnamefont
  {Teretenkov}}\ and\ \bibinfo {author} {\bibfnamefont {O.}~\bibnamefont
  {Lychkovskiy}},\ }\bibfield  {title} {\bibinfo {title} {Duality between open
  systems and closed bilayer systems: {Thermofield} double states as quantum
  many-body scars},\ }\href@noop {} {\bibfield  {journal} {\bibinfo  {journal}
  {Phys. Rev. B}\ }\textbf {\bibinfo {volume} {110}},\ \bibinfo {pages}
  {L241105} (\bibinfo {year} {2024})}\BibitemShut {NoStop}%
\bibitem [{\citenamefont {Nakagawa}\ \emph {et~al.}(2024)\citenamefont
  {Nakagawa}, \citenamefont {Katsura},\ and\ \citenamefont
  {Ueda}}]{Nakagawa2024}%
  \BibitemOpen
  \bibfield  {author} {\bibinfo {author} {\bibfnamefont {M.}~\bibnamefont
  {Nakagawa}}, \bibinfo {author} {\bibfnamefont {H.}~\bibnamefont {Katsura}},\
  and\ \bibinfo {author} {\bibfnamefont {M.}~\bibnamefont {Ueda}},\ }\bibfield
  {title} {\bibinfo {title} {Exact eigenstates of multicomponent {Hubbard}
  models: {SU(${\mathit N}$)} magnetic $\eta$ pairing, weak ergodicity
  breaking, and partial integrability},\ }\href@noop {} {\bibfield  {journal}
  {\bibinfo  {journal} {Phys. Rev. Res.}\ }\textbf {\bibinfo {volume} {6}},\
  \bibinfo {pages} {043259} (\bibinfo {year} {2024})}\BibitemShut {NoStop}%
\bibitem [{\citenamefont {Moudgalya}\ and\ \citenamefont
  {Motrunich}(2024{\natexlab{a}})}]{Moudgalya2024exhaustive}%
  \BibitemOpen
  \bibfield  {author} {\bibinfo {author} {\bibfnamefont {S.}~\bibnamefont
  {Moudgalya}}\ and\ \bibinfo {author} {\bibfnamefont {O.~I.}\ \bibnamefont
  {Motrunich}},\ }\bibfield  {title} {\bibinfo {title} {Exhaustive
  characterization of quantum many-body scars using commutant algebras},\
  }\href@noop {} {\bibfield  {journal} {\bibinfo  {journal} {Phys. Rev. X}\
  }\textbf {\bibinfo {volume} {14}},\ \bibinfo {pages} {041069} (\bibinfo
  {year} {2024}{\natexlab{a}})}\BibitemShut {NoStop}%
\bibitem [{\citenamefont {Moudgalya}\ and\ \citenamefont
  {Motrunich}(2024{\natexlab{b}})}]{Moudgalya2024symmetries}%
  \BibitemOpen
  \bibfield  {author} {\bibinfo {author} {\bibfnamefont {S.}~\bibnamefont
  {Moudgalya}}\ and\ \bibinfo {author} {\bibfnamefont {O.~I.}\ \bibnamefont
  {Motrunich}},\ }\bibfield  {title} {\bibinfo {title} {Symmetries as ground
  states of local superoperators: Hydrodynamic implications},\ }\href@noop {}
  {\bibfield  {journal} {\bibinfo  {journal} {PRX Quantum}\ }\textbf {\bibinfo
  {volume} {5}},\ \bibinfo {pages} {040330} (\bibinfo {year}
  {2024}{\natexlab{b}})}\BibitemShut {NoStop}%
\bibitem [{\citenamefont {Sanada}\ \emph {et~al.}(2024)\citenamefont {Sanada},
  \citenamefont {Miao},\ and\ \citenamefont {Katsura}}]{Sanada2024a}%
  \BibitemOpen
  \bibfield  {author} {\bibinfo {author} {\bibfnamefont {K.}~\bibnamefont
  {Sanada}}, \bibinfo {author} {\bibfnamefont {Y.}~\bibnamefont {Miao}},\ and\
  \bibinfo {author} {\bibfnamefont {H.}~\bibnamefont {Katsura}},\ }\bibfield
  {title} {\bibinfo {title} {Towers of quantum many-body scars from integrable
  boundary states},\ }\href@noop {} {\bibfield  {journal} {\bibinfo  {journal}
  {arXiv:2411.01270}\ } (\bibinfo {year} {2024})}\BibitemShut {NoStop}%
\bibitem [{\citenamefont {Imai}\ and\ \citenamefont {Tsuji}(2025)}]{Imai2025}%
  \BibitemOpen
  \bibfield  {author} {\bibinfo {author} {\bibfnamefont {S.}~\bibnamefont
  {Imai}}\ and\ \bibinfo {author} {\bibfnamefont {N.}~\bibnamefont {Tsuji}},\
  }\bibfield  {title} {\bibinfo {title} {Quantum many-body scars with
  unconventional superconducting pairing symmetries via multibody
  interactions},\ }\href@noop {} {\bibfield  {journal} {\bibinfo  {journal}
  {Phys. Rev. Res.}\ }\textbf {\bibinfo {volume} {7}},\ \bibinfo {pages}
  {013064} (\bibinfo {year} {2025})}\BibitemShut {NoStop}%
\bibitem [{\citenamefont {Calaj\'o}\ \emph {et~al.}(2025)\citenamefont
  {Calaj\'o}, \citenamefont {Cataldi}, \citenamefont {Rigobello}, \citenamefont
  {Wanisch}, \citenamefont {Magnifico}, \citenamefont {Silvi}, \citenamefont
  {Montangero},\ and\ \citenamefont {Halimeh}}]{Calajo2025}%
  \BibitemOpen
  \bibfield  {author} {\bibinfo {author} {\bibfnamefont {G.}~\bibnamefont
  {Calaj\'o}}, \bibinfo {author} {\bibfnamefont {G.}~\bibnamefont {Cataldi}},
  \bibinfo {author} {\bibfnamefont {M.}~\bibnamefont {Rigobello}}, \bibinfo
  {author} {\bibfnamefont {D.}~\bibnamefont {Wanisch}}, \bibinfo {author}
  {\bibfnamefont {G.}~\bibnamefont {Magnifico}}, \bibinfo {author}
  {\bibfnamefont {P.}~\bibnamefont {Silvi}}, \bibinfo {author} {\bibfnamefont
  {S.}~\bibnamefont {Montangero}},\ and\ \bibinfo {author} {\bibfnamefont
  {J.~C.}\ \bibnamefont {Halimeh}},\ }\bibfield  {title} {\bibinfo {title}
  {Quantum many-body scarring in a {non-Abelian} lattice gauge theory},\ }\href
  {https://doi.org/10.1103/PhysRevResearch.7.013322} {\bibfield  {journal}
  {\bibinfo  {journal} {Phys. Rev. Res.}\ }\textbf {\bibinfo {volume} {7}},\
  \bibinfo {pages} {013322} (\bibinfo {year} {2025})}\BibitemShut {NoStop}%
\bibitem [{\citenamefont {Bluvstein}\ \emph {et~al.}(2021)\citenamefont
  {Bluvstein}, \citenamefont {Omran}, \citenamefont {Levine}, \citenamefont
  {Keesling}, \citenamefont {Semeghini}, \citenamefont {Ebadi}, \citenamefont
  {Wang}, \citenamefont {Michailidis}, \citenamefont {Maskara}, \citenamefont
  {Ho}, \citenamefont {Choi}, \citenamefont {Serbyn}, \citenamefont {Greiner},
  \citenamefont {Vuleti\'c},\ and\ \citenamefont
  {Lukin}}]{Bluvstein2021controlling}%
  \BibitemOpen
  \bibfield  {author} {\bibinfo {author} {\bibfnamefont {D.}~\bibnamefont
  {Bluvstein}}, \bibinfo {author} {\bibfnamefont {A.}~\bibnamefont {Omran}},
  \bibinfo {author} {\bibfnamefont {H.}~\bibnamefont {Levine}}, \bibinfo
  {author} {\bibfnamefont {A.}~\bibnamefont {Keesling}}, \bibinfo {author}
  {\bibfnamefont {G.}~\bibnamefont {Semeghini}}, \bibinfo {author}
  {\bibfnamefont {S.}~\bibnamefont {Ebadi}}, \bibinfo {author} {\bibfnamefont
  {T.~T.}\ \bibnamefont {Wang}}, \bibinfo {author} {\bibfnamefont {A.~A.}\
  \bibnamefont {Michailidis}}, \bibinfo {author} {\bibfnamefont
  {N.}~\bibnamefont {Maskara}}, \bibinfo {author} {\bibfnamefont {W.~W.}\
  \bibnamefont {Ho}}, \bibinfo {author} {\bibfnamefont {S.}~\bibnamefont
  {Choi}}, \bibinfo {author} {\bibfnamefont {M.}~\bibnamefont {Serbyn}},
  \bibinfo {author} {\bibfnamefont {M.}~\bibnamefont {Greiner}}, \bibinfo
  {author} {\bibfnamefont {V.}~\bibnamefont {Vuleti\'c}},\ and\ \bibinfo
  {author} {\bibfnamefont {M.~D.}\ \bibnamefont {Lukin}},\ }\bibfield  {title}
  {\bibinfo {title} {Controlling quantum many-body dynamics in driven {Rydberg}
  atom arrays},\ }\href@noop {} {\bibfield  {journal} {\bibinfo  {journal}
  {Science}\ }\textbf {\bibinfo {volume} {371}},\ \bibinfo {pages} {1355}
  (\bibinfo {year} {2021})}\BibitemShut {NoStop}%
\bibitem [{\citenamefont {Zhang}\ \emph {et~al.}(2023)\citenamefont {Zhang},
  \citenamefont {Dong}, \citenamefont {Gao}, \citenamefont {Zhao},
  \citenamefont {Hao}, \citenamefont {Desaules}, \citenamefont {Guo},
  \citenamefont {Chen}, \citenamefont {Deng}, \citenamefont {Liu},
  \citenamefont {Ren}, \citenamefont {Yao}, \citenamefont {Zhang},
  \citenamefont {Xu}, \citenamefont {Wang}, \citenamefont {Jin}, \citenamefont
  {Zhu}, \citenamefont {Zhang}, \citenamefont {Hekang}, \citenamefont {Song},
  \citenamefont {Wang}, \citenamefont {Liu}, \citenamefont {Papi\'c},
  \citenamefont {Ying}, \citenamefont {Wang},\ and\ \citenamefont
  {Lai}}]{Zhang2023}%
  \BibitemOpen
  \bibfield  {author} {\bibinfo {author} {\bibfnamefont {P.}~\bibnamefont
  {Zhang}}, \bibinfo {author} {\bibfnamefont {H.}~\bibnamefont {Dong}},
  \bibinfo {author} {\bibfnamefont {Y.}~\bibnamefont {Gao}}, \bibinfo {author}
  {\bibfnamefont {L.}~\bibnamefont {Zhao}}, \bibinfo {author} {\bibfnamefont
  {J.}~\bibnamefont {Hao}}, \bibinfo {author} {\bibfnamefont {J.-Y.}\
  \bibnamefont {Desaules}}, \bibinfo {author} {\bibfnamefont {Q.}~\bibnamefont
  {Guo}}, \bibinfo {author} {\bibfnamefont {J.}~\bibnamefont {Chen}}, \bibinfo
  {author} {\bibfnamefont {J.}~\bibnamefont {Deng}}, \bibinfo {author}
  {\bibfnamefont {B.}~\bibnamefont {Liu}}, \bibinfo {author} {\bibfnamefont
  {W.}~\bibnamefont {Ren}}, \bibinfo {author} {\bibfnamefont {Y.}~\bibnamefont
  {Yao}}, \bibinfo {author} {\bibfnamefont {X.}~\bibnamefont {Zhang}}, \bibinfo
  {author} {\bibfnamefont {S.}~\bibnamefont {Xu}}, \bibinfo {author}
  {\bibfnamefont {K.}~\bibnamefont {Wang}}, \bibinfo {author} {\bibfnamefont
  {F.}~\bibnamefont {Jin}}, \bibinfo {author} {\bibfnamefont {X.}~\bibnamefont
  {Zhu}}, \bibinfo {author} {\bibfnamefont {B.}~\bibnamefont {Zhang}}, \bibinfo
  {author} {\bibfnamefont {L.}~\bibnamefont {Hekang}}, \bibinfo {author}
  {\bibfnamefont {C.}~\bibnamefont {Song}}, \bibinfo {author} {\bibfnamefont
  {Z.}~\bibnamefont {Wang}}, \bibinfo {author} {\bibfnamefont {F.}~\bibnamefont
  {Liu}}, \bibinfo {author} {\bibfnamefont {Z.}~\bibnamefont {Papi\'c}},
  \bibinfo {author} {\bibfnamefont {L.}~\bibnamefont {Ying}}, \bibinfo {author}
  {\bibfnamefont {H.}~\bibnamefont {Wang}},\ and\ \bibinfo {author}
  {\bibfnamefont {Y.-C.}\ \bibnamefont {Lai}},\ }\bibfield  {title} {\bibinfo
  {title} {Many-body {Hilbert} space scarring on a superconducting processor},\
  }\href@noop {} {\bibfield  {journal} {\bibinfo  {journal} {Nature Physics}\
  }\textbf {\bibinfo {volume} {19}},\ \bibinfo {pages} {120} (\bibinfo {year}
  {2023})}\BibitemShut {NoStop}%
\bibitem [{\citenamefont {Su}\ \emph {et~al.}(2023)\citenamefont {Su},
  \citenamefont {Sun}, \citenamefont {Hudomal}, \citenamefont {Desaules},
  \citenamefont {Zhou}, \citenamefont {Yang}, \citenamefont {Halimeh},
  \citenamefont {Yuan}, \citenamefont {Papi{\'c}},\ and\ \citenamefont
  {Pan}}]{Su2023}%
  \BibitemOpen
  \bibfield  {author} {\bibinfo {author} {\bibfnamefont {G.-X.}\ \bibnamefont
  {Su}}, \bibinfo {author} {\bibfnamefont {H.}~\bibnamefont {Sun}}, \bibinfo
  {author} {\bibfnamefont {A.}~\bibnamefont {Hudomal}}, \bibinfo {author}
  {\bibfnamefont {J.-Y.}\ \bibnamefont {Desaules}}, \bibinfo {author}
  {\bibfnamefont {Z.-Y.}\ \bibnamefont {Zhou}}, \bibinfo {author}
  {\bibfnamefont {B.}~\bibnamefont {Yang}}, \bibinfo {author} {\bibfnamefont
  {J.~C.}\ \bibnamefont {Halimeh}}, \bibinfo {author} {\bibfnamefont {Z.-S.}\
  \bibnamefont {Yuan}}, \bibinfo {author} {\bibfnamefont {Z.}~\bibnamefont
  {Papi{\'c}}},\ and\ \bibinfo {author} {\bibfnamefont {J.-W.}\ \bibnamefont
  {Pan}},\ }\bibfield  {title} {\bibinfo {title} {Observation of many-body
  scarring in a {Bose-Hubbard} quantum simulator},\ }\href@noop {} {\bibfield
  {journal} {\bibinfo  {journal} {Phys. Rev. Res.}\ }\textbf {\bibinfo {volume}
  {5}},\ \bibinfo {pages} {023010} (\bibinfo {year} {2023})}\BibitemShut
  {NoStop}%
\bibitem [{\citenamefont {Zhao}\ \emph {et~al.}(2025)\citenamefont {Zhao},
  \citenamefont {Datla}, \citenamefont {Tian}, \citenamefont {Aliyu},\ and\
  \citenamefont {Loh}}]{Zhao2025}%
  \BibitemOpen
  \bibfield  {author} {\bibinfo {author} {\bibfnamefont {L.}~\bibnamefont
  {Zhao}}, \bibinfo {author} {\bibfnamefont {P.~R.}\ \bibnamefont {Datla}},
  \bibinfo {author} {\bibfnamefont {W.}~\bibnamefont {Tian}}, \bibinfo {author}
  {\bibfnamefont {M.~M.}\ \bibnamefont {Aliyu}},\ and\ \bibinfo {author}
  {\bibfnamefont {H.}~\bibnamefont {Loh}},\ }\bibfield  {title} {\bibinfo
  {title} {Observation of quantum thermalization restricted to {Hilbert} space
  fragments and $\mathbb{Z}_{2k}$ scars},\ }\href@noop {} {\bibfield  {journal}
  {\bibinfo  {journal} {Phys. Rev. X}\ }\textbf {\bibinfo {volume} {15}},\
  \bibinfo {pages} {011035} (\bibinfo {year} {2025})}\BibitemShut {NoStop}%
\bibitem [{\citenamefont {Gotta}\ \emph {et~al.}(2023)\citenamefont {Gotta},
  \citenamefont {Moudgalya},\ and\ \citenamefont {Mazza}}]{Gotta2023}%
  \BibitemOpen
  \bibfield  {author} {\bibinfo {author} {\bibfnamefont {L.}~\bibnamefont
  {Gotta}}, \bibinfo {author} {\bibfnamefont {S.}~\bibnamefont {Moudgalya}},\
  and\ \bibinfo {author} {\bibfnamefont {L.}~\bibnamefont {Mazza}},\ }\bibfield
   {title} {\bibinfo {title} {Asymptotic quantum many-body scars},\ }\href@noop
  {} {\bibfield  {journal} {\bibinfo  {journal} {Phys. Rev. Lett.}\ }\textbf
  {\bibinfo {volume} {131}},\ \bibinfo {pages} {190401} (\bibinfo {year}
  {2023})}\BibitemShut {NoStop}%
\bibitem [{\citenamefont {Mandelstam}\ and\ \citenamefont
  {Tamm}(1945)}]{Mandelstam1945}%
  \BibitemOpen
  \bibfield  {author} {\bibinfo {author} {\bibfnamefont {L.}~\bibnamefont
  {Mandelstam}}\ and\ \bibinfo {author} {\bibfnamefont {I.}~\bibnamefont
  {Tamm}},\ }\bibfield  {title} {\bibinfo {title} {The uncertainty relation
  between energy and time in nonrelativistic quantum mechanics},\ }\href@noop
  {} {\bibfield  {journal} {\bibinfo  {journal} {J. Phys. USSR}\ }\textbf
  {\bibinfo {volume} {9}},\ \bibinfo {pages} {249} (\bibinfo {year}
  {1945})}\BibitemShut {NoStop}%
\bibitem [{\citenamefont {Gong}\ and\ \citenamefont
  {Hamazaki}(2022)}]{Gong2022}%
  \BibitemOpen
  \bibfield  {author} {\bibinfo {author} {\bibfnamefont {Z.}~\bibnamefont
  {Gong}}\ and\ \bibinfo {author} {\bibfnamefont {R.}~\bibnamefont
  {Hamazaki}},\ }\bibfield  {title} {\bibinfo {title} {Bounds in nonequilibrium
  quantum dynamics},\ }\href@noop {} {\bibfield  {journal} {\bibinfo  {journal}
  {Int. J. Mod. Phys. B}\ }\textbf {\bibinfo {volume} {36}},\ \bibinfo {pages}
  {2230007} (\bibinfo {year} {2022})}\BibitemShut {NoStop}%
\bibitem [{\citenamefont {Ren}\ \emph {et~al.}(2024)\citenamefont {Ren},
  \citenamefont {Wang},\ and\ \citenamefont {Fang}}]{Ren2024}%
  \BibitemOpen
  \bibfield  {author} {\bibinfo {author} {\bibfnamefont {J.}~\bibnamefont
  {Ren}}, \bibinfo {author} {\bibfnamefont {Y.-P.}\ \bibnamefont {Wang}},\ and\
  \bibinfo {author} {\bibfnamefont {C.}~\bibnamefont {Fang}},\ }\bibfield
  {title} {\bibinfo {title} {{Quasi-Nambu-Goldstone} modes in many-body scar
  models},\ }\href@noop {} {\bibfield  {journal} {\bibinfo  {journal} {Phys.
  Rev. B}\ }\textbf {\bibinfo {volume} {110}},\ \bibinfo {pages} {245101}
  (\bibinfo {year} {2024})}\BibitemShut {NoStop}%
\bibitem [{\citenamefont {Wei}\ and\ \citenamefont
  {Zhang}(2025)}]{wei2025spectroscopic}%
  \BibitemOpen
  \bibfield  {author} {\bibinfo {author} {\bibfnamefont {W.}~\bibnamefont
  {Wei}}\ and\ \bibinfo {author} {\bibfnamefont {L.}~\bibnamefont {Zhang}},\
  }\bibfield  {title} {\bibinfo {title} {Spectroscopic features of quantum
  many-body scar states},\ }\href@noop {} {\bibfield  {journal} {\bibinfo
  {journal} {Chin. Phys. Lett.}\ }\textbf {\bibinfo {volume} {42}},\ \bibinfo
  {pages} {020502} (\bibinfo {year} {2025})}\BibitemShut {NoStop}%
\bibitem [{\citenamefont {Kodama}\ \emph {et~al.}(2023)\citenamefont {Kodama},
  \citenamefont {Tanaka},\ and\ \citenamefont {Kato}}]{Kodama2023}%
  \BibitemOpen
  \bibfield  {author} {\bibinfo {author} {\bibfnamefont {S.}~\bibnamefont
  {Kodama}}, \bibinfo {author} {\bibfnamefont {A.}~\bibnamefont {Tanaka}},\
  and\ \bibinfo {author} {\bibfnamefont {Y.}~\bibnamefont {Kato}},\ }\bibfield
  {title} {\bibinfo {title} {Spin parity effects in a monoaxial chiral
  ferromagnetic chain},\ }\href@noop {} {\bibfield  {journal} {\bibinfo
  {journal} {Phys. Rev. B}\ }\textbf {\bibinfo {volume} {107}},\ \bibinfo
  {pages} {024403} (\bibinfo {year} {2023})}\BibitemShut {NoStop}%
\bibitem [{\citenamefont {Witten}(1981)}]{Witten1981}%
  \BibitemOpen
  \bibfield  {author} {\bibinfo {author} {\bibfnamefont {E.}~\bibnamefont
  {Witten}},\ }\bibfield  {title} {\bibinfo {title} {Dynamical breaking of
  supersymmetry},\ }\href@noop {} {\bibfield  {journal} {\bibinfo  {journal}
  {Nucl. Phys. B}\ }\textbf {\bibinfo {volume} {188}},\ \bibinfo {pages} {513}
  (\bibinfo {year} {1981})}\BibitemShut {NoStop}%
\bibitem [{\citenamefont {Witten}(1982)}]{Witten1982}%
  \BibitemOpen
  \bibfield  {author} {\bibinfo {author} {\bibfnamefont {E.}~\bibnamefont
  {Witten}},\ }\bibfield  {title} {\bibinfo {title} {Constraints on
  supersymmetry breaking},\ }\href@noop {} {\bibfield  {journal} {\bibinfo
  {journal} {Nucl. Phys. B}\ }\textbf {\bibinfo {volume} {202}},\ \bibinfo
  {pages} {253} (\bibinfo {year} {1982})}\BibitemShut {NoStop}%
\bibitem [{\citenamefont {Fendley}\ \emph
  {et~al.}(2003{\natexlab{a}})\citenamefont {Fendley}, \citenamefont
  {Schoutens},\ and\ \citenamefont {de~Boer}}]{Fendley2003lattice}%
  \BibitemOpen
  \bibfield  {author} {\bibinfo {author} {\bibfnamefont {P.}~\bibnamefont
  {Fendley}}, \bibinfo {author} {\bibfnamefont {K.}~\bibnamefont {Schoutens}},\
  and\ \bibinfo {author} {\bibfnamefont {J.}~\bibnamefont {de~Boer}},\
  }\bibfield  {title} {\bibinfo {title} {Lattice models with $\mathcal{N}=2$
  supersymmetry},\ }\href@noop {} {\bibfield  {journal} {\bibinfo  {journal}
  {Phys. Rev. Lett.}\ }\textbf {\bibinfo {volume} {90}},\ \bibinfo {pages}
  {120402} (\bibinfo {year} {2003}{\natexlab{a}})}\BibitemShut {NoStop}%
\bibitem [{\citenamefont {Fendley}\ \emph
  {et~al.}(2003{\natexlab{b}})\citenamefont {Fendley}, \citenamefont
  {Nienhuis},\ and\ \citenamefont {Schoutens}}]{Fendley2003lattice_fermion}%
  \BibitemOpen
  \bibfield  {author} {\bibinfo {author} {\bibfnamefont {P.}~\bibnamefont
  {Fendley}}, \bibinfo {author} {\bibfnamefont {B.}~\bibnamefont {Nienhuis}},\
  and\ \bibinfo {author} {\bibfnamefont {K.}~\bibnamefont {Schoutens}},\
  }\bibfield  {title} {\bibinfo {title} {Lattice fermion models with
  supersymmetry},\ }\href@noop {} {\bibfield  {journal} {\bibinfo  {journal}
  {J. Phys. A: Math. Gen.}\ }\textbf {\bibinfo {volume} {36}},\ \bibinfo
  {pages} {12399} (\bibinfo {year} {2003}{\natexlab{b}})}\BibitemShut {NoStop}%
\bibitem [{Note1()}]{Note1}%
  \BibitemOpen
  \bibinfo {note} {This relation can be obtained by the assumption $\protect
  \hat {\protect \mathcal {P}}\protect \hat {h}_j\protect \hat {\protect
  \mathcal {P}}=0$. From the definition of $\protect \hat {\protect \mathcal
  {P}}$, we have $\protect \hat {\protect \mathcal {P}}=\DOTSB \sum@ \slimits@
  _{\left |\protect \bm {n}\right \rangle \in \protect \mathcal {H}_P}\left
  |\protect \bm {n}\right \rangle \left \langle \protect \bm {n}\right |$,
  where $\left |\protect \bm {n}\right \rangle $ is a direct product state.
  Then, we obtain $\protect \hat {\protect \mathcal {P}}\protect \hat
  {h}_j\protect \hat {\protect \mathcal {P}}=\DOTSB \sum@ \slimits@ _{\left
  |\protect \bm {n}\right \rangle ,\left |\protect \bm {m}\right \rangle \in
  \protect \mathcal {H}_P}\left \langle \protect \bm {n}\right |\protect \hat
  {h}_j\left |\protect \bm {m}\right \rangle \left |\protect \bm {n}\right
  \rangle \left \langle \protect \bm {m}\right |=0$. For this equation to hold,
  it is necessary that $\left \langle \protect \bm {n}\right |\protect \hat
  {h}_j\left |\protect \bm {m}\right \rangle =0$ for any $\left |\protect \bm
  {n}\right \rangle $ and $\left |\protect \bm {m}\right \rangle \in \protect
  \mathcal {H}_P$. As a special case, we obtain $\protect \hat {\protect
  \mathcal {P}}_n\protect \hat {h}_j\protect \hat {\protect \mathcal
  {P}}_n=0$.}\BibitemShut {Stop}%
\bibitem [{Note2()}]{Note2}%
  \BibitemOpen
  \bibinfo {note} {The relation $\protect \hat {\protect \mathcal
  {P}}_n(\protect \hat {H}_{\protect \rm SG}+\protect \hat {H}_{\protect \rm
  sym})\protect \hat {\protect \mathcal {Q}}=0$ can be proven as follows: Since
  $\protect \hat {Q}^z$ is the generator of the onsite symmetry, we can show
  that $\protect \hat {Q}^z\left |\protect \bm {m}\right \rangle =Q^z_n\left
  |\protect \bm {m}\right \rangle $ holds for any product state $\left
  |\protect \bm {m}\right \rangle $ in $\protect \mathcal {H}_{P_n}$, where
  $Q^z_n$ is an eigenvalue of $\protect \hat {Q}^z$ in $\protect \mathcal
  {H}_{P_n}$. Furthermore, we also obtain $\protect \hat {H}_{\protect \rm
  sym}\left |\protect \bm {m}\right \rangle =\protect \mathcal {E}_0\left
  |\protect \bm {m}\right \rangle $, where $\protect \mathcal {E}_0$ is a real
  number. From these relations, $\protect \hat {Q}^z$ and $\protect \hat
  {H}_{\protect \rm sym}$ commute with $\protect \hat {\protect \mathcal
  {P}}_n$. Therefore, we obtain $\protect \hat {\protect \mathcal
  {P}}_n(\protect \hat {H}_{\protect \rm SG}+\protect \hat {H}_{\protect \rm
  sym})\protect \hat {\protect \mathcal {Q}}=(\protect \hat {H}_{\protect \rm
  SG}+\protect \hat {H}_{\protect \rm sym})\protect \hat {\protect \mathcal
  {P}}_n\protect \hat {\protect \mathcal {Q}}=0$.}\BibitemShut {Stop}%
\bibitem [{\citenamefont {Moudgalya}\ \emph {et~al.}(2018)\citenamefont
  {Moudgalya}, \citenamefont {Regnault},\ and\ \citenamefont
  {Bernevig}}]{moudgalya2018entanglement}%
  \BibitemOpen
  \bibfield  {author} {\bibinfo {author} {\bibfnamefont {S.}~\bibnamefont
  {Moudgalya}}, \bibinfo {author} {\bibfnamefont {N.}~\bibnamefont
  {Regnault}},\ and\ \bibinfo {author} {\bibfnamefont {B.~A.}\ \bibnamefont
  {Bernevig}},\ }\bibfield  {title} {\bibinfo {title} {Entanglement of exact
  excited states of {Affleck-Kennedy-Lieb-Tasaki models}: {Exact} results,
  many-body scars, and violation of the strong eigenstate thermalization
  hypothesis},\ }\href@noop {} {\bibfield  {journal} {\bibinfo  {journal}
  {Phys. Rev. B}\ }\textbf {\bibinfo {volume} {98}},\ \bibinfo {pages} {235156}
  (\bibinfo {year} {2018})}\BibitemShut {NoStop}%
\bibitem [{\citenamefont {Lieb}\ \emph {et~al.}(1961)\citenamefont {Lieb},
  \citenamefont {Schultz},\ and\ \citenamefont {Mattis}}]{lieb1961two}%
  \BibitemOpen
  \bibfield  {author} {\bibinfo {author} {\bibfnamefont {E.}~\bibnamefont
  {Lieb}}, \bibinfo {author} {\bibfnamefont {T.}~\bibnamefont {Schultz}},\ and\
  \bibinfo {author} {\bibfnamefont {D.}~\bibnamefont {Mattis}},\ }\bibfield
  {title} {\bibinfo {title} {Two soluble models of an antiferromagnetic
  chain},\ }\href@noop {} {\bibfield  {journal} {\bibinfo  {journal} {Ann.
  Phys.}\ }\textbf {\bibinfo {volume} {16}},\ \bibinfo {pages} {407} (\bibinfo
  {year} {1961})}\BibitemShut {NoStop}%
\bibitem [{\citenamefont {Tasaki}(2020)}]{tasaki2020physics}%
  \BibitemOpen
  \bibfield  {author} {\bibinfo {author} {\bibfnamefont {H.}~\bibnamefont
  {Tasaki}},\ }\href@noop {} {\emph {\bibinfo {title} {Physics and
  {Mathematics} of {Quantum} {Many-Body} {Systems}}}}\ (\bibinfo  {publisher}
  {Springer},\ \bibinfo {address} {New York},\ \bibinfo {year}
  {2020})\BibitemShut {NoStop}%
\bibitem [{Note3()}]{Note3}%
  \BibitemOpen
  \bibinfo {note} {We prove that every positive energy level is at least
  twofold degenerate. Let $|\psi \rangle $ be a simultaneous eigenstate of
  $\protect \hat {H}_{\protect \rm SUSY}$ and $(-1)^{\protect \hat {F}}$ with
  eigenvalues $E>0$ and $p=\pm 1$, respectively. If $\protect \hat {{\protect
  \tilde Q}}|\psi \rangle =0$, then $|\phi \rangle = \protect \hat {{\protect
  \tilde Q}}^\dagger |\psi \rangle $ is an eigenstate of $\protect \hat
  {H}_{\protect \rm SUSY}$ with the same eigenvalue $E$. This can be seen as
  follows: We first note that $|\phi \rangle \protect \ne 0$, since otherwise
  we would have $\protect \hat {{\protect \tilde Q}}|\psi \rangle =\protect
  \hat {{\protect \tilde Q}}^\dagger |\psi \rangle =0$, which implies $E=0$,
  contradicting our assumption that $E>0$. We then see that $|\psi \rangle $
  and $|\phi \rangle $ must be orthogonal, since they have opposite fermionic
  parity. Thus, it follows from Eq. \protect \eqref
  {eq:commutation_relation_SUSY_Hamiltonian_and_supercharge} that $|\psi
  \rangle $ and $|\phi \rangle $ form a degenerate pair. If, on the other hand,
  $\protect \hat {{\protect \tilde Q}}|\psi \rangle \protect \ne 0$, then we
  define $|\chi \rangle = \protect \hat {{\protect \tilde Q}} |\psi \rangle $,
  which is an eigenstate of $\protect \hat {H}_{\protect \rm SUSY}$ with
  eigenvalue $E$. By the same reasoning as before, $|\chi \rangle $ must be
  nonzero and orthogonal to $|\psi \rangle $. Thus, in both cases, we obtain
  the desired result.}\BibitemShut {Stop}%
\bibitem [{\citenamefont {Kitazawa}\ \emph {et~al.}(2003)\citenamefont
  {Kitazawa}, \citenamefont {Hijii},\ and\ \citenamefont
  {Nomura}}]{Kitazawa2003}%
  \BibitemOpen
  \bibfield  {author} {\bibinfo {author} {\bibfnamefont {A.}~\bibnamefont
  {Kitazawa}}, \bibinfo {author} {\bibfnamefont {K.}~\bibnamefont {Hijii}},\
  and\ \bibinfo {author} {\bibfnamefont {K.}~\bibnamefont {Nomura}},\
  }\bibfield  {title} {\bibinfo {title} {An {SU (2)} symmetry of the
  one-dimensional spin-1 {XY} model},\ }\href@noop {} {\bibfield  {journal}
  {\bibinfo  {journal} {J. Phys. A}\ }\textbf {\bibinfo {volume} {36}},\
  \bibinfo {pages} {L351} (\bibinfo {year} {2003})}\BibitemShut {NoStop}%
\bibitem [{\citenamefont {Yang}(1989)}]{Yang1989}%
  \BibitemOpen
  \bibfield  {author} {\bibinfo {author} {\bibfnamefont {C.~N.}\ \bibnamefont
  {Yang}},\ }\bibfield  {title} {\bibinfo {title} {$\eta$ pairing and
  off-diagonal long-range order in a {Hubbard} model},\ }\href@noop {}
  {\bibfield  {journal} {\bibinfo  {journal} {Phys, Rev. Lett.}\ }\textbf
  {\bibinfo {volume} {63}},\ \bibinfo {pages} {2144} (\bibinfo {year}
  {1989})}\BibitemShut {NoStop}%
\bibitem [{\citenamefont {Hirsch}(1989)}]{Hirsch1989}%
  \BibitemOpen
  \bibfield  {author} {\bibinfo {author} {\bibfnamefont {J.}~\bibnamefont
  {Hirsch}},\ }\bibfield  {title} {\bibinfo {title} {Bond-charge repulsion and
  hole superconductivity},\ }\href@noop {} {\bibfield  {journal} {\bibinfo
  {journal} {Physica C}\ }\textbf {\bibinfo {volume} {158}},\ \bibinfo {pages}
  {326} (\bibinfo {year} {1989})}\BibitemShut {NoStop}%
\bibitem [{\citenamefont {Arrachea}\ and\ \citenamefont
  {Aligia}(1994)}]{Arrachea1994}%
  \BibitemOpen
  \bibfield  {author} {\bibinfo {author} {\bibfnamefont {L.}~\bibnamefont
  {Arrachea}}\ and\ \bibinfo {author} {\bibfnamefont {A.}~\bibnamefont
  {Aligia}},\ }\bibfield  {title} {\bibinfo {title} {Exact solution of a
  {Hubbard} chain with bond-charge interaction},\ }\href@noop {} {\bibfield
  {journal} {\bibinfo  {journal} {Phys. Rev. Lett.}\ }\textbf {\bibinfo
  {volume} {73}},\ \bibinfo {pages} {2240} (\bibinfo {year}
  {1994})}\BibitemShut {NoStop}%
\bibitem [{\citenamefont {Schadschneider}(1995)}]{Schadschneider1995}%
  \BibitemOpen
  \bibfield  {author} {\bibinfo {author} {\bibfnamefont {A.}~\bibnamefont
  {Schadschneider}},\ }\bibfield  {title} {\bibinfo {title} {Superconductivity
  in an exactly solvable {Hubbard} model with bond-charge interaction},\
  }\href@noop {} {\bibfield  {journal} {\bibinfo  {journal} {Phys. Rev. B}\
  }\textbf {\bibinfo {volume} {51}},\ \bibinfo {pages} {10386} (\bibinfo {year}
  {1995})}\BibitemShut {NoStop}%
\bibitem [{\citenamefont {G{\'o}mez-Santos}(1993)}]{Gomez1993}%
  \BibitemOpen
  \bibfield  {author} {\bibinfo {author} {\bibfnamefont {G.}~\bibnamefont
  {G{\'o}mez-Santos}},\ }\bibfield  {title} {\bibinfo {title} {Generalized
  hard-core fermions in one dimension: An exactly solvable {Luttinger}
  liquid},\ }\href@noop {} {\bibfield  {journal} {\bibinfo  {journal} {Phys.
  Rev. Lett.}\ }\textbf {\bibinfo {volume} {70}},\ \bibinfo {pages} {3780}
  (\bibinfo {year} {1993})}\BibitemShut {NoStop}%
\bibitem [{\citenamefont {Cheong}\ and\ \citenamefont
  {Henley}(2009)}]{Cheong2009}%
  \BibitemOpen
  \bibfield  {author} {\bibinfo {author} {\bibfnamefont {S.-A.}\ \bibnamefont
  {Cheong}}\ and\ \bibinfo {author} {\bibfnamefont {C.~L.}\ \bibnamefont
  {Henley}},\ }\bibfield  {title} {\bibinfo {title} {Exact ground states and
  correlation functions of chain and ladder models of interacting hardcore
  bosons or spinless fermions},\ }\href@noop {} {\bibfield  {journal} {\bibinfo
   {journal} {Phys. Rev. B}\ }\textbf {\bibinfo {volume} {80}},\ \bibinfo
  {pages} {165124} (\bibinfo {year} {2009})}\BibitemShut {NoStop}%
\bibitem [{\citenamefont {Onsager}(1944)}]{Onsager1944}%
  \BibitemOpen
  \bibfield  {author} {\bibinfo {author} {\bibfnamefont {L.}~\bibnamefont
  {Onsager}},\ }\bibfield  {title} {\bibinfo {title} {Crystal statistics. {I}.
  {A} two-dimensional model with an order-disorder transition},\ }\href@noop {}
  {\bibfield  {journal} {\bibinfo  {journal} {Phys. Rev.}\ }\textbf {\bibinfo
  {volume} {65}},\ \bibinfo {pages} {117} (\bibinfo {year} {1944})}\BibitemShut
  {NoStop}%
\bibitem [{\citenamefont {Vernier}\ \emph {et~al.}(2019)\citenamefont
  {Vernier}, \citenamefont {O'Brien},\ and\ \citenamefont
  {Fendley}}]{vernier2019onsager}%
  \BibitemOpen
  \bibfield  {author} {\bibinfo {author} {\bibfnamefont {E.}~\bibnamefont
  {Vernier}}, \bibinfo {author} {\bibfnamefont {E.}~\bibnamefont {O'Brien}},\
  and\ \bibinfo {author} {\bibfnamefont {P.}~\bibnamefont {Fendley}},\
  }\bibfield  {title} {\bibinfo {title} {{Onsager symmetries in
  $U(1)$-invariant clock models}},\ }\href@noop {} {\bibfield  {journal}
  {\bibinfo  {journal} {J. Stat. Mech.}\ }\textbf {\bibinfo {volume} {2019}},\
  \bibinfo {pages} {043107} (\bibinfo {year} {2019})}\BibitemShut {NoStop}%
\bibitem [{\citenamefont {Bariev}(1991)}]{Bariev1991}%
  \BibitemOpen
  \bibfield  {author} {\bibinfo {author} {\bibfnamefont {R.}~\bibnamefont
  {Bariev}},\ }\bibfield  {title} {\bibinfo {title} {Integrable spin chain with
  two-and three-particle interactions},\ }\href@noop {} {\bibfield  {journal}
  {\bibinfo  {journal} {J. Phys. A: Math. Gen.}\ }\textbf {\bibinfo {volume}
  {24}},\ \bibinfo {pages} {L549} (\bibinfo {year} {1991})}\BibitemShut
  {NoStop}%
\bibitem [{\citenamefont {Chhajlany}\ \emph {et~al.}(2016)\citenamefont
  {Chhajlany}, \citenamefont {Grzybowski}, \citenamefont {Stasi{\'n}ska},
  \citenamefont {Lewenstein},\ and\ \citenamefont {Dutta}}]{Chhajlany2016}%
  \BibitemOpen
  \bibfield  {author} {\bibinfo {author} {\bibfnamefont {R.~W.}\ \bibnamefont
  {Chhajlany}}, \bibinfo {author} {\bibfnamefont {P.~R.}\ \bibnamefont
  {Grzybowski}}, \bibinfo {author} {\bibfnamefont {J.}~\bibnamefont
  {Stasi{\'n}ska}}, \bibinfo {author} {\bibfnamefont {M.}~\bibnamefont
  {Lewenstein}},\ and\ \bibinfo {author} {\bibfnamefont {O.}~\bibnamefont
  {Dutta}},\ }\bibfield  {title} {\bibinfo {title} {Hidden string order in a
  hole superconductor with extended correlated hopping},\ }\href@noop {}
  {\bibfield  {journal} {\bibinfo  {journal} {Phys. Rev. Lett.}\ }\textbf
  {\bibinfo {volume} {116}},\ \bibinfo {pages} {225303} (\bibinfo {year}
  {2016})}\BibitemShut {NoStop}%
\bibitem [{\citenamefont {Pozsgay}\ \emph {et~al.}(2021)\citenamefont
  {Pozsgay}, \citenamefont {Gombor},\ and\ \citenamefont
  {Hutsalyuk}}]{Pozsgay2021}%
  \BibitemOpen
  \bibfield  {author} {\bibinfo {author} {\bibfnamefont {B.}~\bibnamefont
  {Pozsgay}}, \bibinfo {author} {\bibfnamefont {T.}~\bibnamefont {Gombor}},\
  and\ \bibinfo {author} {\bibfnamefont {A.}~\bibnamefont {Hutsalyuk}},\
  }\bibfield  {title} {\bibinfo {title} {Integrable hard-rod deformation of the
  {Heisenberg} spin chains},\ }\href@noop {} {\bibfield  {journal} {\bibinfo
  {journal} {Phys. Rev. E}\ }\textbf {\bibinfo {volume} {104}},\ \bibinfo
  {pages} {064124} (\bibinfo {year} {2021})}\BibitemShut {NoStop}%
\bibitem [{\citenamefont {Borsi}\ \emph {et~al.}(2023)\citenamefont {Borsi},
  \citenamefont {Pristy{\'a}k},\ and\ \citenamefont {Pozsgay}}]{Borsi2023}%
  \BibitemOpen
  \bibfield  {author} {\bibinfo {author} {\bibfnamefont {M.}~\bibnamefont
  {Borsi}}, \bibinfo {author} {\bibfnamefont {L.}~\bibnamefont
  {Pristy{\'a}k}},\ and\ \bibinfo {author} {\bibfnamefont {B.}~\bibnamefont
  {Pozsgay}},\ }\bibfield  {title} {\bibinfo {title} {Matrix product symmetries
  and breakdown of thermalization from hard rod deformations},\ }\href@noop {}
  {\bibfield  {journal} {\bibinfo  {journal} {Phys. Rev. Lett.}\ }\textbf
  {\bibinfo {volume} {131}},\ \bibinfo {pages} {037101} (\bibinfo {year}
  {2023})}\BibitemShut {NoStop}%
\bibitem [{\citenamefont {Borsi}\ and\ \citenamefont
  {Pozsgay}(2025)}]{borsi2025volume}%
  \BibitemOpen
  \bibfield  {author} {\bibinfo {author} {\bibfnamefont {M.}~\bibnamefont
  {Borsi}}\ and\ \bibinfo {author} {\bibfnamefont {B.}~\bibnamefont
  {Pozsgay}},\ }\bibfield  {title} {\bibinfo {title} {Volume changing
  symmetries by matrix product operators},\ }\href@noop {} {\bibfield
  {journal} {\bibinfo  {journal} {J. Phys. A: Math. Theor.}\ }\textbf {\bibinfo
  {volume} {58}},\ \bibinfo {pages} {065203} (\bibinfo {year}
  {2025})}\BibitemShut {NoStop}%
\bibitem [{\citenamefont {Schollw{\"o}ck}(2011)}]{schollwock2011density}%
  \BibitemOpen
  \bibfield  {author} {\bibinfo {author} {\bibfnamefont {U.}~\bibnamefont
  {Schollw{\"o}ck}},\ }\bibfield  {title} {\bibinfo {title} {The density-matrix
  renormalization group in the age of matrix product states},\ }\href@noop {}
  {\bibfield  {journal} {\bibinfo  {journal} {Ann Phys.}\ }\textbf {\bibinfo
  {volume} {326}},\ \bibinfo {pages} {96} (\bibinfo {year} {2011})}\BibitemShut
  {NoStop}%
\bibitem [{\citenamefont {Hashimoto}\ \emph {et~al.}()\citenamefont
  {Hashimoto}, \citenamefont {Kunimi},\ and\ \citenamefont
  {Nikuni}}]{hashimotoSUN}%
  \BibitemOpen
  \bibfield  {author} {\bibinfo {author} {\bibfnamefont {D.}~\bibnamefont
  {Hashimoto}}, \bibinfo {author} {\bibfnamefont {M.}~\bibnamefont {Kunimi}},\
  and\ \bibinfo {author} {\bibfnamefont {T.}~\bibnamefont {Nikuni}},\
  }\href@noop {} {\bibinfo {title} {Construction of asymptotic quantum
  many-body scar states in the {SU({$\mathit N$})} {Hubbard} model}},\ \bibinfo
  {note} {unpublished}\BibitemShut {NoStop}%
\bibitem [{\citenamefont {Shiraishi}\ and\ \citenamefont
  {Mori}(2017)}]{Shiraishi2017}%
  \BibitemOpen
  \bibfield  {author} {\bibinfo {author} {\bibfnamefont {N.}~\bibnamefont
  {Shiraishi}}\ and\ \bibinfo {author} {\bibfnamefont {T.}~\bibnamefont
  {Mori}},\ }\bibfield  {title} {\bibinfo {title} {Systematic construction of
  counterexamples to the eigenstate thermalization hypothesis},\ }\href@noop {}
  {\bibfield  {journal} {\bibinfo  {journal} {Phys. Rev. Lett.}\ }\textbf
  {\bibinfo {volume} {119}},\ \bibinfo {pages} {030601} (\bibinfo {year}
  {2017})}\BibitemShut {NoStop}%
\bibitem [{\citenamefont {Kunimi}\ \emph {et~al.}(2025)\citenamefont {Kunimi},
  \citenamefont {Kato},\ and\ \citenamefont {Katsura}}]{kunimi_2025_16891505}%
  \BibitemOpen
  \bibfield  {author} {\bibinfo {author} {\bibfnamefont {M.}~\bibnamefont
  {Kunimi}}, \bibinfo {author} {\bibfnamefont {Y.}~\bibnamefont {Kato}},\ and\
  \bibinfo {author} {\bibfnamefont {H.}~\bibnamefont {Katsura}},\ }\bibfield
  {title} {\bibinfo {title} {Dataset for "{Systematic} construction of
  asymptotic quantum many-body scar states and their relation to supersymmetric
  quantum mechanics"},\ }\href {https://doi.org/10.5281/zenodo.16891505}
  {10.5281/zenodo.16891505} (\bibinfo {year} {2025})\BibitemShut {NoStop}%
\bibitem [{\citenamefont {Popkov}\ and\ \citenamefont
  {Salerno}(2005)}]{popkov2005logarithmic}%
  \BibitemOpen
  \bibfield  {author} {\bibinfo {author} {\bibfnamefont {V.}~\bibnamefont
  {Popkov}}\ and\ \bibinfo {author} {\bibfnamefont {M.}~\bibnamefont
  {Salerno}},\ }\bibfield  {title} {\bibinfo {title} {Logarithmic divergence of
  the block entanglement entropy for the ferromagnetic {Heisenberg} model},\
  }\href@noop {} {\bibfield  {journal} {\bibinfo  {journal} {Phys. Rev. A}\
  }\textbf {\bibinfo {volume} {71}},\ \bibinfo {pages} {012301} (\bibinfo
  {year} {2005})}\BibitemShut {NoStop}%
\bibitem [{\citenamefont {Ma}(2024)}]{ma2024lieb}%
  \BibitemOpen
  \bibfield  {author} {\bibinfo {author} {\bibfnamefont {R.}~\bibnamefont
  {Ma}},\ }\bibfield  {title} {\bibinfo {title} {{Lieb-Schultz-Mattis} theorem
  with long-range interactions},\ }\href@noop {} {\bibfield  {journal}
  {\bibinfo  {journal} {Phys. Rev. B}\ }\textbf {\bibinfo {volume} {110}},\
  \bibinfo {pages} {104412} (\bibinfo {year} {2024})}\BibitemShut {NoStop}%
\bibitem [{\citenamefont {Zhou}\ and\ \citenamefont
  {Li}(2024)}]{zhou2024validity}%
  \BibitemOpen
  \bibfield  {author} {\bibinfo {author} {\bibfnamefont {Y.-N.}\ \bibnamefont
  {Zhou}}\ and\ \bibinfo {author} {\bibfnamefont {X.}~\bibnamefont {Li}},\
  }\bibfield  {title} {\bibinfo {title} {Validity of the {Lieb-Schultz-Mattis}
  theorem in long-range interacting systems},\ }\href@noop {} {\bibfield
  {journal} {\bibinfo  {journal} {arXiv:2406.08948}\ } (\bibinfo {year}
  {2024})}\BibitemShut {NoStop}%
\bibitem [{\citenamefont {Rice}\ and\ \citenamefont {Mele}(1982)}]{Rice1982}%
  \BibitemOpen
  \bibfield  {author} {\bibinfo {author} {\bibfnamefont {M.}~\bibnamefont
  {Rice}}\ and\ \bibinfo {author} {\bibfnamefont {E.}~\bibnamefont {Mele}},\
  }\bibfield  {title} {\bibinfo {title} {Elementary excitations of a linearly
  conjugated diatomic polymer},\ }\href@noop {} {\bibfield  {journal} {\bibinfo
   {journal} {Phys. Rev. Lett.}\ }\textbf {\bibinfo {volume} {49}},\ \bibinfo
  {pages} {1455} (\bibinfo {year} {1982})}\BibitemShut {NoStop}%
\end{thebibliography}%
\end{document}